%% file: main.tex
\documentclass[aps,prd,preprint,nofootinbib]{revtex4-2}
\input{defs.tex}

\input{numbers.tex}

\begin{document}

\title{Loop-induced $gg\to ZH$ at NLO+PS for LHC simulations: exact, fast scale and PDF event weights,
  theory uncertainties, and the limits of rescaled LO+PS}

\author{L. Zhang}
\author{C. Palmer}
\affiliation{University of Maryland, College Park, MD, USA}

\date{\today}

\begin{abstract}
We study loop-induced $gg\to ZH$ production with charged-lepton and neutrino $Z$ decays
at next-to-leading order (NLO) in QCD matched to parton showers, using public software at
$\sqrt{s}=13$ and $13.6\TeV$.
An exact loop-amplitude cache reduces measured event-generation time by a factor of
\pScanTimeRatiosix{} with \pScanMaxsix{} scale and parton distribution function (PDF)
weights, while Born suppression improves boosted-event statistics without changing the
weighted prediction.  We present inclusive results for both decay channels and detailed
charged-lepton differential and Simplified Template Cross Section (STXS) Stage~1.3
predictions with theory uncertainties.  At 13.6\TeV{}, fixed-order top-quark mass-scheme
variations transferred to NLO+PS dominate the statistically precise central-Higgs STXS bins
from $\ptV=\pStxsMtFromsix\GeV$ onward, reaching \pStxsMtHighsix\%.
An inclusive $K$ factor does not reproduce the boosted and jet-dependent NLO+PS spectra
of the tested LO+PS samples, and the discrepancies depend strongly on the shower prescription.
For specified CMS- and ATLAS-style vetoed sources at 13.6\TeV{}, we construct separate
charged-lepton and neutrino corrections using ensembles of gradient-boosted classifiers,
with calibration, target-theory variations, and transfer uncertainties.
Independent evaluations show substantial improvement over inclusive rescaling in most
tested distributions, but not uniform closure.  In charged-lepton calibration comparisons,
the median STXS uncertainty from transported NLO scales and transfer terms is
\rbdRwStxs\% for the CMS-style correction and \rbaRwStxs\% for the ATLAS-style correction,
compared with original LO scale uncertainties of \rbdLoStxs\% and \rbaLoStxs\% at fixed $K$.
These comparisons do not include the complete target-theory budget.
The corrections do not restore the depleted hard-recoil population; the highest ATLAS-style
Higgs transverse-momentum bin remains below NLO+PS even at the upper edge of its assigned band.
The production setup, reference results, and source- and channel-specific models with their
application prescriptions support new NLO+PS generation and reuse of existing samples within
stated applicability limits.
\end{abstract}

\maketitle

\input{sec-intro.tex}
\input{sec-setup.tex}
\input{sec-events.tex}
\input{sec-shower.tex}
\input{sec-uncertainties.tex}
\input{sec-validation.tex}
\input{sec-results.tex}
\input{sec-lops.tex}
\input{sec-rwt.tex}

\input{sec-summary.tex}

\begin{acknowledgments}
This work was supported in part by the U.S. Department of Energy under Grant No.~DE-SC0010072. We thank the authors of \ggxyZH{} for reviewing and merging the amplitude cache.
We thank Weitao Wang (ATLAS) for providing the ATLAS generation configuration and
Sitian Qian (CMS) for fruitful discussions.
The authors used OpenAI Codex and Anthropic Claude Code under their direction for
manuscript language polishing, software-development assistance, monitoring computational
runs and recording their execution times. We acknowledge OpenAI for providing complimentary access to ChatGPT through the ChatGPT for Academic Researchers program.
The software used in this work was validated by the authors, with the \ggxy{} components
validated by the \ggxy{} authors.
The authors retain full responsibility for the scientific content and conclusions.
\end{acknowledgments}

\appendix
\input{sec-app-13tev.tex}
\input{sec-app-rwt-atlas.tex}
\input{sec-app-reproduction.tex}
\input{sec-app-conventions.tex}

\bibliography{refs}
\end{document}

%% file: defs.tex
\usepackage{amsmath,amssymb}
\usepackage{booktabs}
\usepackage{graphicx}
\usepackage{xcolor}
\usepackage[hidelinks]{hyperref}
\usepackage[capitalise]{cleveref}
\usepackage[htt]{hyphenat}  %
\newcommand{\ggllH}{\ensuremath{gg\to \ell^-\ell^+H}}

\newcommand{\mZH}{\ensuremath{m_{ZH}}}
\newcommand{\ptZ}{\ensuremath{p_{\mathrm{T}}(Z)}}
\newcommand{\ptV}{\ensuremath{p_{\mathrm{T}}^{V}}}
\newcommand{\ptH}{\ensuremath{p_{\mathrm{T}}(H)}}
\newcommand{\muR}{\ensuremath{\mu_{\mathrm{R}}}}
\newcommand{\muF}{\ensuremath{\mu_{\mathrm{F}}}}
\newcommand{\muz}{\ensuremath{\mu_0}}
\newcommand{\as}{\ensuremath{\alpha_{\mathrm{s}}}}
\newcommand{\hdamp}{\ensuremath{h_{\mathrm{damp}}}}
\newcommand{\GeV}{\ensuremath{\,\mathrm{GeV}}}
\newcommand{\TeV}{\ensuremath{\,\mathrm{TeV}}}
\newcommand{\fb}{\ensuremath{\,\mathrm{fb}}}
\newcommand{\code}[1]{\texttt{#1}}
\newcommand{\POWHEG}{\textsc{Powheg}}
\newcommand{\PBOX}{\textsc{Powheg-Box-V2}}
\newcommand{\ggxy}{\texttt{ggxy}}
\newcommand{\ggxyZH}{\texttt{ggxy\_ggZH}}
\newcommand{\PYTHIA}{\textsc{Pythia}\,8}

\newcommand{\RIVET}{\textsc{Rivet}}
\newcommand{\RECOLA}{\textsc{Recola}}
\newcommand{\COLLIER}{\textsc{Collier}}
\newcommand{\MCFM}{\textsc{MCFM}}
\newcommand{\LHAPDF}{\textsc{LHAPDF}}

%% file: numbers.tex
\newcommand{\nWeights}{155}            %
\newcommand{\bornsuppL}{250}           %
\newcommand{\nEvNLO}{$5\times10^{5}$}  %
\newcommand{\nEvLO}{$5\times10^{5}$}   %

\newcommand{\refLOthirteen}{6.366(4)}      \newcommand{\refLOthirteenband}{$^{+26.7\%}_{-20.0\%}$}
\newcommand{\refNLOthirteen}{11.80(1)}     \newcommand{\refNLOthirteenband}{$^{+16.6\%}_{-14.0\%}$}
\newcommand{\refLOsix}{7.026(4)}           \newcommand{\refLOsixband}{$^{+26.2\%}_{-19.7\%}$}
\newcommand{\refNLOsix}{13.02(1)}          \newcommand{\refNLOsixband}{$^{+16.4\%}_{-13.8\%}$}
\newcommand{\refLOthirteenVV}{12.581(8)}  \newcommand{\refLOthirteenbandVV}{$^{+26.7\%}_{-20.0\%}$}
\newcommand{\refNLOthirteenVV}{23.30(2)}  \newcommand{\refNLOthirteenbandVV}{$^{+16.5\%}_{-14.0\%}$}
\newcommand{\refLOsixVV}{13.883(8)}       \newcommand{\refLOsixbandVV}{$^{+26.2\%}_{-19.7\%}$}
\newcommand{\refNLOsixVV}{25.70(2)}       \newcommand{\refNLOsixbandVV}{$^{+16.4\%}_{-13.8\%}$}
\newcommand{\refStLOthirteen}{63.97(2)}    \newcommand{\refStNLOthirteen}{118.40(8)}
\newcommand{\refStLOsix}{70.59(2)}         \newcommand{\refStNLOsix}{130.50(9)}

\newcommand{\yrZHthirteen}{0.8839}         
\newcommand{\yrggZHthirteen}{0.1227}       \newcommand{\yrggZHthirteenscale}{$^{+25.1\%}_{-18.9\%}$}

\newcommand{\yrggZHfracthirteen}{13.9}     %
\newcommand{\yrZHsix}{0.9439}              
\newcommand{\yrggZHsix}{0.1360}
\newcommand{\yrggZHfracsix}{14.4}          %

\input{numbers-results.tex}

\input{numbers-rwt.tex}

\input{numbers-rwt-bdt17.tex}
\input{numbers-rwt-atlas.tex}

%% file: numbers-results.tex
\newcommand{\vLO}{7.0324(10)}                       %
\newcommand{\vLOband}{$^{+26.8\%}_{-20.0\%}$}       %
\newcommand{\vNLO}{13.043(4)}                       %
\newcommand{\vNLOband}{$^{+16.6\%}_{-14.0\%}$}      %
\newcommand{\vPowLO}{7.0306(19)}                    %
\newcommand{\vGrids}{13.031(7)}                     %
\newcommand{\vEvNLO}{13.05(6)}                      %
\newcommand{\vEvNLOband}{$^{+16.7\%}_{-14.0\%}$}    %
\newcommand{\vEvLO}{7.037(31)}                      %
\newcommand{\vStLOnn}{70.705(7)}                    %
\newcommand{\vMCFMnn}{71.111(18)}                   %
\newcommand{\vCacheEvents}{150}                     %
\newcommand{\vCacheWeights}{16\,800}                %
\newcommand{\vChecks}{66/66}                        %
\newcommand{\vEvNLObsPull}{$-0.4$}                  %
\newcommand{\vPartonNLOdip}{0.946(18)}              %
\newcommand{\vBSSTXSchi}{8.6/19}                    %
\newcommand{\vBSPtzchi}{38.1/33}                    %
\newcommand{\pLOsix}{7.0225(10)}                    %
\newcommand{\pLOsixband}{$^{+26.2\%}_{-19.7\%}$}    %
\newcommand{\pNLOsix}{13.014(4)}                    %
\newcommand{\pNLOsixband}{$^{+16.4\%}_{-13.8\%}$}   %
\newcommand{\pKsix}{1.8532(6)}                      %
\newcommand{\pStLOsix}{70.601(7)}                   %
\newcommand{\pStNLOsix}{130.599(31)}                %
\newcommand{\pPtzFallsix}{$2.4\times10^{3}$}        %
\newcommand{\pEvPDFsix}{1.77}                       %
\newcommand{\pEvAssix}{1.41}                        %
\newcommand{\pCacheEventssix}{150}                  %
\newcommand{\pCacheWeightssix}{23\,250}             %
\newcommand{\pMtLOsix}{$^{+0}_{-24}$}               %
\newcommand{\pMtNLOsix}{$^{+0}_{-13}$}              %
\newcommand{\pMtMzhBelowsix}{$0$ to $+5$}           %
\newcommand{\pMtMzhAbovesix}{$-33$ to $-11$}        %
\newcommand{\pMtPtzLowsix}{$-2$ to $-1$}            %
\newcommand{\pMtPtzHighsix}{$-44$ to $-18$}         %
\newcommand{\pMtPthHighsix}{$-38$ to $-14$}         %
\newcommand{\pMtMzhThrsix}{1.8}                     %
\newcommand{\pScanJobssix}{3}                       %
\newcommand{\pScanMaxsix}{450}                      %
\newcommand{\pScanCachedsix}{185}                   %
\newcommand{\pScanEvalRatiosix}{28}                 %
\newcommand{\pScanTimeRatiosix}{26}                 %
\newcommand{\pScanTimeUncachedsix}{40}              %
\newcommand{\pScanExtrapolatesix}{884}              %
\newcommand{\pScanEvalExtrasix}{53}                 %
\newcommand{\pScanTimeExtrasix}{41}                 %
\newcommand{\pPartonLOchisix}{10.0/16}              %
\newcommand{\pPartonNLOtailsix}{0.95--1.03}         %
\newcommand{\pPartonNLOtailErrsix}{4--17}           %
\newcommand{\pPartonNLOmidLosix}{$-5.5$}            %
\newcommand{\pPartonNLOdipsix}{0.945(7)}            %
\newcommand{\pPartonNLOmidHisix}{$+7.6$}            %
\newcommand{\pEvLOPDFsix}{1.79}                     %
\newcommand{\pEvLOAssix}{1.07}                      %
\newcommand{\pLOmcsix}{0.01}                        %
\newcommand{\pNLOmcsix}{0.03}                       %
\newcommand{\pNeffGainsix}{2.1}                     %
\newcommand{\pLopsDefTwoJCsix}{19.9(16)}            %
\newcommand{\pLopsDefZeroJsix}{0.73--0.97}          %
\newcommand{\pLopsRangeHalfsix}{0.75--1.50}         %
\newcommand{\pLopsRangeDampedsix}{0.85--4.56}       %
\newcommand{\pLopsCovDefsix}{5 of 16}               %
\newcommand{\pLopsCovHalfsix}{12 of 16}             %
\newcommand{\pLopsHighBinssix}{5}                   %
\newcommand{\pLopsKsix}{1.856}                      %
\newcommand{\pLopsDefTunesFoursix}{5.4--6.1}        %
\newcommand{\pLopsDefTunesSixsix}{14.2--17.4}       %
\newcommand{\pMzhShiftsix}{$<10^{-15}$}             %
\newcommand{\pStxsScalesix}{9--24}                  %
\newcommand{\pStxsPdfsix}{2.2--7.5}                 %
\newcommand{\pStxsMtsix}{2--50}                     %
\newcommand{\pStxsMtFromsix}{150}                   %
\newcommand{\pStxsMtHighsix}{32--50}                %
\newcommand{\pStxsPrecisesix}{16}                   %
\newcommand{\pStxsMatchsix}{3--18}                  %
\newcommand{\pLOthirteen}{6.3643(9)}                %
\newcommand{\pLOthirteenband}{$^{+26.7\%}_{-20.0\%}$}%
\newcommand{\pNLOthirteen}{11.8044(35)}             %
\newcommand{\pNLOthirteenband}{$^{+16.6\%}_{-14.0\%}$}%
\newcommand{\pStLOthirteen}{63.989(6)}              %
\newcommand{\pStNLOthirteen}{118.440(28)}           %
\newcommand{\pEvPDFthirteen}{1.80}                  %
\newcommand{\pEvAsthirteen}{1.43}                   %
\newcommand{\pMtLOthirteen}{$^{+0}_{-24}$}          %
\newcommand{\pMtNLOthirteen}{$^{+0}_{-13}$}         %
\newcommand{\pMtNoisythirteen}{250--400\GeV: 8\%}   %
\newcommand{\pPartonLOchithirteen}{9.9/16}          %
\newcommand{\pPartonNLOmidHithirteen}{$+28.5$}      %
\newcommand{\pEvLOPDFthirteen}{1.82}                %
\newcommand{\pEvLOAsthirteen}{1.08}                 %
\newcommand{\pLOmcthirteen}{0.01}                   %
\newcommand{\pNLOmcthirteen}{0.03}                  %
\newcommand{\pNeffGainthirteen}{2.9}                %
\newcommand{\pLopsKthirteen}{1.856}                 %
\newcommand{\pLopsDefTunesFourthirteen}{5.6--6.4}   %
\newcommand{\pLopsDefTunesSixthirteen}{15.7--18.5}  %
\newcommand{\pHdampPartonMidLosix}{$-6.3$}          %
\newcommand{\pHdampPartonMidHisix}{$+12.9$}         %
\newcommand{\pCpuSixBase}{1360}                     %
\newcommand{\pCpuThirteen}{1260}                    %
\newcommand{\pCpuSixExtra}{790}                     %
\newcommand{\pCpuVerify}{670}                       %
\newcommand{\pCpuShowerShareSix}{39}                %
\newcommand{\pCpuEvNLOSix}{90}                      %
\newcommand{\pMtChiEnergies}{24.2/21}               %
\newcommand{\pKMzhThrsix}{3.0}                      %
\newcommand{\pKMzhHighsix}{1.5--1.6}                %
\newcommand{\pKPtzLowsix}{1.7--2.1}                 %
\newcommand{\pKPtzSixsix}{2.6}                      %
\newcommand{\pKPtzEightsix}{4.0}                    %
\newcommand{\pKPthEightsix}{21}                     %
\newcommand{\pPtzhPsOverNLOsix}{1.6--1.9}           %
\newcommand{\pKChiPtz}{20.5/16}                     %
\newcommand{\pKChiPth}{7.7/16}                      %
\newcommand{\pKChiMzh}{6.7/17}                      %
\newcommand{\pMCFMoffset}{0.57}                     %
\newcommand{\pLOsixVV}{13.8800(20)}                 %
\newcommand{\pLOsixbandVV}{$^{+26.2\%}_{-19.7\%}$}  %
\newcommand{\pLOmcsixVV}{0.01}                      %
\newcommand{\pEvLOPDFsixVV}{1.79}                   %
\newcommand{\pEvLOAssixVV}{1.07}                    %
\newcommand{\pNLOsixVV}{25.704(8)}                  %
\newcommand{\pNLOsixbandVV}{$^{+16.4\%}_{-13.8\%}$} %
\newcommand{\pNLOmcsixVV}{0.03}                     %
\newcommand{\pEvPDFsixVV}{1.77}                     %
\newcommand{\pEvAssixVV}{1.41}                      %
\newcommand{\pMtCheckLOVV}{-23.7}                   %
\newcommand{\pMtCheckNLOVV}{-13.4}                  %
\newcommand{\pLOthirteenVV}{12.5791(18)}            %
\newcommand{\pLOthirteenbandVV}{$^{+26.7\%}_{-20.0\%}$}%
\newcommand{\pLOmcthirteenVV}{0.01}                 %
\newcommand{\pEvLOPDFthirteenVV}{1.82}              %
\newcommand{\pEvLOAsthirteenVV}{1.08}               %
\newcommand{\pNLOthirteenVV}{23.296(7)}             %
\newcommand{\pNLOthirteenbandVV}{$^{+16.5\%}_{-14.0\%}$}%
\newcommand{\pNLOmcthirteenVV}{0.03}                %
\newcommand{\pEvPDFthirteenVV}{1.80}                %
\newcommand{\pEvAsthirteenVV}{1.43}                 %

%% file: numbers-rwt.tex
\newcommand{\rwtR}[2]{\csname rwtR#1#2\endcsname}      %
\newcommand{\rwtChi}[2]{\csname rwtChi#1#2\endcsname}  %
\newcommand{\rwtHJ}[2]{\csname rwtHJ#1#2\endcsname}    %
\newcommand{\rwtHJraw}[1]{\csname rwtHJraw#1\endcsname}  %
\newcommand{\rwtNevLO}{$5\times10^{5}$}   %
\newcommand{\rwtNevNLO}{$5\times10^{5}$}   %
\expandafter\newcommand\csname rwtRshareptzA\endcsname{3.47\%}   %
\expandafter\newcommand\csname rwtRUnwptzA\endcsname{$1.91\pm0.01$}   %
\expandafter\newcommand\csname rwtRMapAptzA\endcsname{$1.00\pm0.01$}   %
\expandafter\newcommand\csname rwtRClfptzA\endcsname{$1.02\pm0.01$}   %
\expandafter\newcommand\csname rwtRshareptzB\endcsname{0.22\%}   %
\expandafter\newcommand\csname rwtRUnwptzB\endcsname{$3.28\pm0.09$}   %
\expandafter\newcommand\csname rwtRMapAptzB\endcsname{$1.01\pm0.03$}   %
\expandafter\newcommand\csname rwtRClfptzB\endcsname{$1.03\pm0.03$}   %
\expandafter\newcommand\csname rwtRshareptzC\endcsname{0.01\%}   %
\expandafter\newcommand\csname rwtRUnwptzC\endcsname{$4.61\pm0.39$}   %
\expandafter\newcommand\csname rwtRMapAptzC\endcsname{$1.02\pm0.09$}   %
\expandafter\newcommand\csname rwtRClfptzC\endcsname{$1.24\pm0.11$}   %
\expandafter\newcommand\csname rwtRsharepthA\endcsname{13.13\%}   %
\expandafter\newcommand\csname rwtRUnwpthA\endcsname{$1.31\pm0.01$}   %
\expandafter\newcommand\csname rwtRMapApthA\endcsname{$1.11\pm0.01$}   %
\expandafter\newcommand\csname rwtRClfpthA\endcsname{$0.97$}   %
\expandafter\newcommand\csname rwtRsharepthB\endcsname{0.35\%}   %
\expandafter\newcommand\csname rwtRUnwpthB\endcsname{$1.23\pm0.03$}   %
\expandafter\newcommand\csname rwtRMapApthB\endcsname{$0.87\pm0.03$}   %
\expandafter\newcommand\csname rwtRClfpthB\endcsname{$0.77\pm0.03$}   %
\expandafter\newcommand\csname rwtRsharejetA\endcsname{3.39\%}   %
\expandafter\newcommand\csname rwtRUnwjetA\endcsname{$2.31\pm0.03$}   %
\expandafter\newcommand\csname rwtRMapAjetA\endcsname{$1.86\pm0.02$}   %
\expandafter\newcommand\csname rwtRClfjetA\endcsname{$1.19\pm0.01$}   %
\expandafter\newcommand\csname rwtRsharejetB\endcsname{0.24\%}   %
\expandafter\newcommand\csname rwtRUnwjetB\endcsname{$1.35\pm0.05$}   %
\expandafter\newcommand\csname rwtRMapAjetB\endcsname{$0.77\pm0.03$}   %
\expandafter\newcommand\csname rwtRClfjetB\endcsname{$0.55\pm0.02$}   %
\expandafter\newcommand\csname rwtRsharedrA\endcsname{9.12\%}   %
\expandafter\newcommand\csname rwtRUnwdrA\endcsname{$1.45\pm0.01$}   %
\expandafter\newcommand\csname rwtRMapAdrA\endcsname{$1.31\pm0.01$}   %
\expandafter\newcommand\csname rwtRClfdrA\endcsname{$1.01\pm0.01$}   %
\expandafter\newcommand\csname rwtRshareptzhA\endcsname{9.11\%}   %
\expandafter\newcommand\csname rwtRUnwptzhA\endcsname{$1.60\pm0.01$}   %
\expandafter\newcommand\csname rwtRMapAptzhA\endcsname{$1.41\pm0.01$}   %
\expandafter\newcommand\csname rwtRClfptzhA\endcsname{$1.01\pm0.01$}   %
\expandafter\newcommand\csname rwtRsharemllA\endcsname{0.35\%}   %
\expandafter\newcommand\csname rwtRUnwmllA\endcsname{$0.71\pm0.04$}   %
\expandafter\newcommand\csname rwtRMapAmllA\endcsname{$0.72\pm0.04$}   %
\expandafter\newcommand\csname rwtRClfmllA\endcsname{$0.75\pm0.04$}   %
\expandafter\newcommand\csname rwtChiUnwptz\endcsname{$12417/31$}   %
\expandafter\newcommand\csname rwtChiUnwptzh\endcsname{$7793/29$}   %
\expandafter\newcommand\csname rwtChiUnwpth\endcsname{$4468/31$}   %
\expandafter\newcommand\csname rwtChiUnwjet\endcsname{$10813/28$}   %
\expandafter\newcommand\csname rwtChiUnwdr\endcsname{$3201/30$}   %
\expandafter\newcommand\csname rwtChiUnwmzh\endcsname{$7256/41$}   %
\expandafter\newcommand\csname rwtChiUnwyh\endcsname{$59/19$}   %
\expandafter\newcommand\csname rwtChiUnwnj\endcsname{$1220/3$}   %
\expandafter\newcommand\csname rwtChiUnwayzh\endcsname{$223/11$}   %
\expandafter\newcommand\csname rwtChiMapAptzh\endcsname{$4361/29$}   %
\expandafter\newcommand\csname rwtChiMapApth\endcsname{$733/31$}   %
\expandafter\newcommand\csname rwtChiMapAjet\endcsname{$6562/28$}   %
\expandafter\newcommand\csname rwtChiMapAdr\endcsname{$2291/30$}   %
\expandafter\newcommand\csname rwtChiMapAmzh\endcsname{$1435/41$}   %
\expandafter\newcommand\csname rwtChiMapAyh\endcsname{$30/19$}   %
\expandafter\newcommand\csname rwtChiClfptz\endcsname{$28/31$}   %
\expandafter\newcommand\csname rwtChiClfptzh\endcsname{$10/29$}   %
\expandafter\newcommand\csname rwtChiClfpth\endcsname{$214/31$}   %
\expandafter\newcommand\csname rwtChiClfjet\endcsname{$609/28$}   %
\expandafter\newcommand\csname rwtChiClfdr\endcsname{$42/30$}   %
\expandafter\newcommand\csname rwtChiClfmzh\endcsname{$39/41$}   %
\expandafter\newcommand\csname rwtChiClfyh\endcsname{$33/19$}   %
\expandafter\newcommand\csname rwtChiClfnj\endcsname{$1/3$}   %
\expandafter\newcommand\csname rwtChiClfayzh\endcsname{$6/11$}   %
\expandafter\newcommand\csname rwtHJLoAll\endcsname{0.21}   %
\expandafter\newcommand\csname rwtHJClfAll\endcsname{0.15}   %
\expandafter\newcommand\csname rwtHJNloAll\endcsname{0.50}   %
\expandafter\newcommand\csname rwtHJrawAll\endcsname{1067}   %
\expandafter\newcommand\csname rwtHJLoPtz\endcsname{1.2}   %
\expandafter\newcommand\csname rwtHJClfPtz\endcsname{0.8}   %
\expandafter\newcommand\csname rwtHJNloPtz\endcsname{9.9}   %
\expandafter\newcommand\csname rwtHJrawPtz\endcsname{45}   %
\expandafter\newcommand\csname rwtHJLoPth\endcsname{1.3}   %
\expandafter\newcommand\csname rwtHJClfPth\endcsname{2.1}   %
\expandafter\newcommand\csname rwtHJNloPth\endcsname{23.6}   %
\expandafter\newcommand\csname rwtHJrawPth\endcsname{28}   %
\expandafter\newcommand\csname rwtRDefUnwptzA\endcsname{$2.09\pm0.02$}   %
\expandafter\newcommand\csname rwtRDefUnwptzB\endcsname{$5.61\pm0.14$}   %
\expandafter\newcommand\csname rwtRDefUnwptzC\endcsname{$17.65\pm1.25$}   %
\expandafter\newcommand\csname rwtRDefUnwpthA\endcsname{$1.39\pm0.01$}   %
\expandafter\newcommand\csname rwtRDefUnwpthB\endcsname{$3.03\pm0.07$}   %
\expandafter\newcommand\csname rwtRDefUnwjetA\endcsname{$2.52\pm0.03$}   %
\expandafter\newcommand\csname rwtRDefUnwjetB\endcsname{$8.48\pm0.28$}   %
\expandafter\newcommand\csname rwtRDefUnwdrA\endcsname{$1.60\pm0.01$}   %
\expandafter\newcommand\csname rwtRDefUnwptzhA\endcsname{$1.56\pm0.01$}   %
\expandafter\newcommand\csname rwtRDefUnwmllA\endcsname{$0.71\pm0.04$}   %
\expandafter\newcommand\csname rwtChiDefUnwptz\endcsname{$18717/31$}   %
\expandafter\newcommand\csname rwtChiDefUnwptzh\endcsname{$7908/29$}   %
\expandafter\newcommand\csname rwtChiDefUnwpth\endcsname{$9141/36$}   %
\expandafter\newcommand\csname rwtChiDefUnwjet\endcsname{$21039/34$}   %
\expandafter\newcommand\csname rwtChiDefUnwdr\endcsname{$6270/30$}   %
\expandafter\newcommand\csname rwtChiDefUnwmzh\endcsname{$7256/41$}   %
\expandafter\newcommand\csname rwtChiDefUnwyh\endcsname{$157/19$}   %
\expandafter\newcommand\csname rwtChiDefUnwnj\endcsname{$2356/3$}   %
\expandafter\newcommand\csname rwtChiDefUnwayzh\endcsname{$487/11$}   %
\expandafter\newcommand\csname rwtRDefClfptzA\endcsname{$1.01\pm0.01$}   %
\expandafter\newcommand\csname rwtRDefClfptzB\endcsname{$1.02\pm0.03$}   %
\expandafter\newcommand\csname rwtRDefClfptzC\endcsname{$1.31\pm0.11$}   %
\expandafter\newcommand\csname rwtRDefClfpthA\endcsname{$0.97$}   %
\expandafter\newcommand\csname rwtRDefClfpthB\endcsname{$0.89\pm0.02$}   %
\expandafter\newcommand\csname rwtRDefClfjetA\endcsname{$1.16\pm0.01$}   %
\expandafter\newcommand\csname rwtRDefClfjetB\endcsname{$1.43\pm0.05$}   %
\expandafter\newcommand\csname rwtRDefClfdrA\endcsname{$1.01\pm0.01$}   %
\expandafter\newcommand\csname rwtRDefClfptzhA\endcsname{$1.01\pm0.01$}   %
\expandafter\newcommand\csname rwtRDefClfmllA\endcsname{$0.77\pm0.04$}   %
\expandafter\newcommand\csname rwtChiDefClfptz\endcsname{$44/31$}   %
\expandafter\newcommand\csname rwtChiDefClfptzh\endcsname{$14/29$}   %
\expandafter\newcommand\csname rwtChiDefClfpth\endcsname{$150/36$}   %
\expandafter\newcommand\csname rwtChiDefClfjet\endcsname{$507/34$}   %
\expandafter\newcommand\csname rwtChiDefClfdr\endcsname{$28/30$}   %
\expandafter\newcommand\csname rwtChiDefClfmzh\endcsname{$38/41$}   %
\expandafter\newcommand\csname rwtChiDefClfyh\endcsname{$23/19$}   %
\expandafter\newcommand\csname rwtChiDefClfnj\endcsname{$2/3$}   %
\expandafter\newcommand\csname rwtChiDefClfayzh\endcsname{$7/11$}   %
\expandafter\newcommand\csname rwtHJDefLoAll\endcsname{2.84}   %
\expandafter\newcommand\csname rwtHJDefClfAll\endcsname{0.65}   %
\expandafter\newcommand\csname rwtHJDefLoPtz\endcsname{43.8}   %
\expandafter\newcommand\csname rwtHJDefClfPtz\endcsname{11.4}   %
\expandafter\newcommand\csname rwtHJDefLoPth\endcsname{54.7}   %
\expandafter\newcommand\csname rwtHJDefClfPth\endcsname{33.1}   %
\expandafter\newcommand\csname rwtRDefJetUnwptzA\endcsname{$2.09\pm0.02$}   %
\expandafter\newcommand\csname rwtRDefJetUnwptzB\endcsname{$5.61\pm0.14$}   %
\expandafter\newcommand\csname rwtRDefJetUnwptzC\endcsname{$17.65\pm1.25$}   %
\expandafter\newcommand\csname rwtRDefJetUnwpthA\endcsname{$1.39\pm0.01$}   %
\expandafter\newcommand\csname rwtRDefJetUnwpthB\endcsname{$3.03\pm0.07$}   %
\expandafter\newcommand\csname rwtRDefJetUnwjetA\endcsname{$2.52\pm0.03$}   %
\expandafter\newcommand\csname rwtRDefJetUnwjetB\endcsname{$8.48\pm0.28$}   %
\expandafter\newcommand\csname rwtRDefJetUnwdrA\endcsname{$1.60\pm0.01$}   %
\expandafter\newcommand\csname rwtRDefJetUnwptzhA\endcsname{$1.56\pm0.01$}   %
\expandafter\newcommand\csname rwtRDefJetUnwmllA\endcsname{$0.71\pm0.04$}   %
\expandafter\newcommand\csname rwtChiDefJetUnwptz\endcsname{$18717/31$}   %
\expandafter\newcommand\csname rwtChiDefJetUnwptzh\endcsname{$7908/29$}   %
\expandafter\newcommand\csname rwtChiDefJetUnwpth\endcsname{$9141/36$}   %
\expandafter\newcommand\csname rwtChiDefJetUnwjet\endcsname{$21039/34$}   %
\expandafter\newcommand\csname rwtChiDefJetUnwdr\endcsname{$6270/30$}   %
\expandafter\newcommand\csname rwtChiDefJetUnwmzh\endcsname{$7256/41$}   %
\expandafter\newcommand\csname rwtChiDefJetUnwyh\endcsname{$157/19$}   %
\expandafter\newcommand\csname rwtChiDefJetUnwnj\endcsname{$2356/3$}   %
\expandafter\newcommand\csname rwtChiDefJetUnwayzh\endcsname{$487/11$}   %
\expandafter\newcommand\csname rwtRDefJetClfptzA\endcsname{$1.01\pm0.01$}   %
\expandafter\newcommand\csname rwtRDefJetClfptzB\endcsname{$1.02\pm0.03$}   %
\expandafter\newcommand\csname rwtRDefJetClfptzC\endcsname{$1.33\pm0.11$}   %
\expandafter\newcommand\csname rwtRDefJetClfpthA\endcsname{$0.97$}   %
\expandafter\newcommand\csname rwtRDefJetClfpthB\endcsname{$0.88\pm0.02$}   %
\expandafter\newcommand\csname rwtRDefJetClfjetA\endcsname{$1.00\pm0.01$}   %
\expandafter\newcommand\csname rwtRDefJetClfjetB\endcsname{$1.06\pm0.04$}   %
\expandafter\newcommand\csname rwtRDefJetClfdrA\endcsname{$1.01\pm0.01$}   %
\expandafter\newcommand\csname rwtRDefJetClfptzhA\endcsname{$1.01\pm0.01$}   %
\expandafter\newcommand\csname rwtRDefJetClfmllA\endcsname{$0.77\pm0.04$}   %
\expandafter\newcommand\csname rwtChiDefJetClfptz\endcsname{$44/31$}   %
\expandafter\newcommand\csname rwtChiDefJetClfptzh\endcsname{$16/29$}   %
\expandafter\newcommand\csname rwtChiDefJetClfpth\endcsname{$165/36$}   %
\expandafter\newcommand\csname rwtChiDefJetClfjet\endcsname{$40/34$}   %
\expandafter\newcommand\csname rwtChiDefJetClfdr\endcsname{$29/30$}   %
\expandafter\newcommand\csname rwtChiDefJetClfmzh\endcsname{$37/41$}   %
\expandafter\newcommand\csname rwtChiDefJetClfyh\endcsname{$25/19$}   %
\expandafter\newcommand\csname rwtChiDefJetClfnj\endcsname{$1/3$}   %
\expandafter\newcommand\csname rwtChiDefJetClfayzh\endcsname{$7/11$}   %
\expandafter\newcommand\csname rwtHJDefJetLoAll\endcsname{2.84}   %
\expandafter\newcommand\csname rwtHJDefJetClfAll\endcsname{0.51}   %
\expandafter\newcommand\csname rwtHJDefJetLoPtz\endcsname{43.8}   %
\expandafter\newcommand\csname rwtHJDefJetClfPtz\endcsname{9.9}   %
\expandafter\newcommand\csname rwtHJDefJetLoPth\endcsname{54.7}   %
\expandafter\newcommand\csname rwtHJDefJetClfPth\endcsname{29.2}   %
\expandafter\newcommand\csname rwtRCmsJetUnwptzA\endcsname{$1.91\pm0.01$}   %
\expandafter\newcommand\csname rwtRCmsJetUnwptzB\endcsname{$3.28\pm0.09$}   %
\expandafter\newcommand\csname rwtRCmsJetUnwptzC\endcsname{$4.61\pm0.39$}   %
\expandafter\newcommand\csname rwtRCmsJetUnwpthA\endcsname{$1.31\pm0.01$}   %
\expandafter\newcommand\csname rwtRCmsJetUnwpthB\endcsname{$1.23\pm0.03$}   %
\expandafter\newcommand\csname rwtRCmsJetUnwjetA\endcsname{$2.31\pm0.03$}   %
\expandafter\newcommand\csname rwtRCmsJetUnwjetB\endcsname{$1.35\pm0.05$}   %
\expandafter\newcommand\csname rwtRCmsJetUnwdrA\endcsname{$1.45\pm0.01$}   %
\expandafter\newcommand\csname rwtRCmsJetUnwptzhA\endcsname{$1.60\pm0.01$}   %
\expandafter\newcommand\csname rwtRCmsJetUnwmllA\endcsname{$0.71\pm0.04$}   %
\expandafter\newcommand\csname rwtChiCmsJetUnwptz\endcsname{$12417/31$}   %
\expandafter\newcommand\csname rwtChiCmsJetUnwptzh\endcsname{$7793/29$}   %
\expandafter\newcommand\csname rwtChiCmsJetUnwpth\endcsname{$4468/31$}   %
\expandafter\newcommand\csname rwtChiCmsJetUnwjet\endcsname{$10813/28$}   %
\expandafter\newcommand\csname rwtChiCmsJetUnwdr\endcsname{$3201/30$}   %
\expandafter\newcommand\csname rwtChiCmsJetUnwmzh\endcsname{$7256/41$}   %
\expandafter\newcommand\csname rwtChiCmsJetUnwyh\endcsname{$59/19$}   %
\expandafter\newcommand\csname rwtChiCmsJetUnwnj\endcsname{$1220/3$}   %
\expandafter\newcommand\csname rwtChiCmsJetUnwayzh\endcsname{$223/11$}   %
\expandafter\newcommand\csname rwtRCmsJetClfptzA\endcsname{$1.01\pm0.01$}   %
\expandafter\newcommand\csname rwtRCmsJetClfptzB\endcsname{$1.00\pm0.03$}   %
\expandafter\newcommand\csname rwtRCmsJetClfptzC\endcsname{$1.24\pm0.11$}   %
\expandafter\newcommand\csname rwtRCmsJetClfpthA\endcsname{$0.97\pm0.01$}   %
\expandafter\newcommand\csname rwtRCmsJetClfpthB\endcsname{$0.74\pm0.03$}   %
\expandafter\newcommand\csname rwtRCmsJetClfjetA\endcsname{$0.99\pm0.01$}   %
\expandafter\newcommand\csname rwtRCmsJetClfjetB\endcsname{$0.82\pm0.13$}   %
\expandafter\newcommand\csname rwtRCmsJetClfdrA\endcsname{$1.00\pm0.01$}   %
\expandafter\newcommand\csname rwtRCmsJetClfptzhA\endcsname{$1.01\pm0.01$}   %
\expandafter\newcommand\csname rwtRCmsJetClfmllA\endcsname{$0.75\pm0.04$}   %
\expandafter\newcommand\csname rwtChiCmsJetClfptz\endcsname{$27/31$}   %
\expandafter\newcommand\csname rwtChiCmsJetClfptzh\endcsname{$13/29$}   %
\expandafter\newcommand\csname rwtChiCmsJetClfpth\endcsname{$261/31$}   %
\expandafter\newcommand\csname rwtChiCmsJetClfjet\endcsname{$193/28$}   %
\expandafter\newcommand\csname rwtChiCmsJetClfdr\endcsname{$51/30$}   %
\expandafter\newcommand\csname rwtChiCmsJetClfmzh\endcsname{$40/41$}   %
\expandafter\newcommand\csname rwtChiCmsJetClfyh\endcsname{$36/19$}   %
\expandafter\newcommand\csname rwtChiCmsJetClfnj\endcsname{$4/3$}   %
\expandafter\newcommand\csname rwtChiCmsJetClfayzh\endcsname{$6/11$}   %
\expandafter\newcommand\csname rwtHJCmsJetLoAll\endcsname{0.21}   %
\expandafter\newcommand\csname rwtHJCmsJetClfAll\endcsname{0.26}   %
\expandafter\newcommand\csname rwtHJCmsJetLoPtz\endcsname{1.1}   %
\expandafter\newcommand\csname rwtHJCmsJetClfPtz\endcsname{1.2}   %
\expandafter\newcommand\csname rwtHJCmsJetLoPth\endcsname{1.3}   %
\expandafter\newcommand\csname rwtHJCmsJetClfPth\endcsname{7.9}   %

%% file: numbers-rwt-bdt17.tex
\newcommand{\rbdNeff}{90}   %

\newcommand{\rbdOutNlo}{0.49}   %
\newcommand{\rbdOutLo}{0.21}
\newcommand{\rbdHJNloPtz}{9}
\newcommand{\rbdHJLoPtz}{1.4}
\newcommand{\rbdHJNloPth}{25}
\newcommand{\rbdHJLoPth}{1.5}
\newcommand{\rbdKlo}{0.93}
\newcommand{\rbdKhi}{1.03}
\newcommand{\rbdKgroups}{23}
\newcommand{\rbdClo}{0.85}
\newcommand{\rbdChi}{1.06}
\newcommand{\rbdShareLow}{1.1}   %
\newcommand{\rbdShareHigh}{20}   %
\newcommand{\rbdLoPtz}{30}   %
\newcommand{\rbdRwPtz}{17}
\newcommand{\rbdMethPtz}{2.8}
\newcommand{\rbdSmallerPtz}{76}
\newcommand{\rbdLoPth}{29}   %
\newcommand{\rbdRwPth}{17}
\newcommand{\rbdMethPth}{2.6}

\newcommand{\rbdLoMzh}{30}   %
\newcommand{\rbdRwMzh}{13}
\newcommand{\rbdMethMzh}{1.4}
\newcommand{\rbdSmallerMzh}{100}
\newcommand{\rbdLoStxs}{28}   %
\newcommand{\rbdRwStxs}{17}
\newcommand{\rbdMethStxs}{2.1}

\newcommand{\rbdPthEdge}{525}   %
\newcommand{\rbdPthRw}{0.69}   %
\newcommand{\rbdPthUp}{1.04}   %
\newcommand{\rbdPthLoK}{1.07}   %
\newcommand{\rbdFullRw}{0.97}   %
\newcommand{\rbdFullLoK}{3.7}   %
\newcommand{\rbdFullLoKDn}{24}   %
\newcommand{\rbdFullLoKUp}{34}
\newcommand{\rbdSharedLo}{14}   %

\newcommand{\rbdPthInLoK}{1.5}   %
\newcommand{\rbdPthInRw}{0.97}
\newcommand{\rbdConfTwoFrac}{3.8}   %

\newcommand{\rbdConfMax}{2.9}
\newcommand{\rbdConfBins}{521}
\newcommand{\rbdExcessLo}{5}   %
\newcommand{\rbdExcessHi}{6}
\newcommand{\rbvPthRw}{0.69}   %
\newcommand{\rbvPthUp}{1.00}   %
\newcommand{\rbvPthLoK}{1.08}   %
\newcommand{\rbvFullRw}{0.99}   %
\newcommand{\rbvFullLoK}{3.7}   %
\newcommand{\rbvSharedHi}{24}
\newcommand{\rbvPthInLoK}{1.6}   %
\newcommand{\rbvPthInRw}{1.00}
\newcommand{\rbvConfTwoFrac}{4.0}   %

\newcommand{\rbvConfMax}{3.6}
\newcommand{\rbvConfBins}{424}
\newcommand{\rbvExcessLo}{10}   %
\newcommand{\rbvExcessHi}{12}
\newcommand{\rbaLoEv}{3.5}   %
\newcommand{\rbaNloEv}{3.4}
\newcommand{\rbaSharedLo}{66}   %
\newcommand{\rbaSharedHi}{71}
\newcommand{\rbaCovBins}{515}

\newcommand{\rbaCovTwoFrac}{3.5}   %

\newcommand{\rbaCovMax}{3.1}
\newcommand{\rbaPullMean}{$-$0.07}
\newcommand{\rbaPullRms}{0.92}

\newcommand{\rbaTailA}{$1.06\pm0.03$}   %
\newcommand{\rbaLoPtz}{26}   %
\newcommand{\rbaRwPtz}{17}
\newcommand{\rbaMethPtz}{2.1}

\newcommand{\rbaLoStxs}{26}   %
\newcommand{\rbaRwStxs}{17}

\newcommand{\rbaPthEdge}{525}   %
\newcommand{\rbaPthRw}{0.66}   %
\newcommand{\rbaPthUp}{0.92}   %
\newcommand{\rbaFullRw}{1.00}   %
\newcommand{\rbaFullLoK}{3.1}   %

\newcommand{\rbnLoEv}{3.3}   %
\newcommand{\rbnNloEv}{3.3}
\newcommand{\rbnSharedLo}{75}   %
\newcommand{\rbnSharedHi}{76}
\newcommand{\rbnCovBins}{421}

\newcommand{\rbnCovTwoFrac}{3.8}   %

\newcommand{\rbnCovMax}{3.4}
\newcommand{\rbnPullMean}{$-$0.21}
\newcommand{\rbnPullRms}{0.98}

\newcommand{\rbnTailA}{$1.03\pm0.03$}   %
\newcommand{\rbnLoPtz}{26}   %
\newcommand{\rbnRwPtz}{17}
\newcommand{\rbnMethPtz}{2.2}

\newcommand{\rbnPthRw}{0.64}   %
\newcommand{\rbnPthUp}{0.91}   %
\newcommand{\rbnFullRw}{0.96}   %
\newcommand{\rbnFullLoK}{3.0}   %

%% file: numbers-rwt-atlas.tex
\newcommand{\atlasR}[2]{\csname atlasR#1#2\endcsname}
\newcommand{\atlasChi}[2]{\csname atlasChi#1#2\endcsname}
\newcommand{\atlasPass}[2]{\csname atlasPass#1#2\endcsname}
\expandafter\newcommand\csname atlasNevLO\endcsname{$5\times10^{5}$} %
\expandafter\newcommand\csname atlasNevNLO\endcsname{$5\times10^{5}$} %
\expandafter\newcommand\csname atlasExcess\endcsname{0.58\%} %
\expandafter\newcommand\csname atlasSigLO\endcsname{6.006} %
\expandafter\newcommand\csname atlasSigNLO\endcsname{13.038} %
\expandafter\newcommand\csname atlasPullUnw\endcsname{64.52} %
\expandafter\newcommand\csname atlasRUnwptzA\endcsname{$1.68\pm0.01$} %
\expandafter\newcommand\csname atlasRUnwptzB\endcsname{$2.67\pm0.07$} %
\expandafter\newcommand\csname atlasRUnwptzC\endcsname{$3.23\pm0.29$} %
\expandafter\newcommand\csname atlasRUnwpthA\endcsname{$1.19\pm0.01$} %
\expandafter\newcommand\csname atlasRUnwpthB\endcsname{$1.03\pm0.03$} %
\expandafter\newcommand\csname atlasRUnwjetA\endcsname{$2.34\pm0.03$} %
\expandafter\newcommand\csname atlasRUnwjetB\endcsname{$1.19\pm0.05$} %
\expandafter\newcommand\csname atlasRUnwdrA\endcsname{$1.54\pm0.01$} %
\expandafter\newcommand\csname atlasRUnwptzhA\endcsname{$1.64\pm0.01$} %
\expandafter\newcommand\csname atlasRUnwmllA\endcsname{$0.75\pm0.04$} %
\expandafter\newcommand\csname atlasChiUnwptz\endcsname{$7554/29$} %
\expandafter\newcommand\csname atlasChiUnwptzh\endcsname{$9294/29$} %
\expandafter\newcommand\csname atlasChiUnwpth\endcsname{$2011/30$} %
\expandafter\newcommand\csname atlasChiUnwjet\endcsname{$10819/28$} %
\expandafter\newcommand\csname atlasChiUnwdr\endcsname{$4944/30$} %
\expandafter\newcommand\csname atlasChiUnwmzh\endcsname{$3060/41$} %
\expandafter\newcommand\csname atlasChiUnwyh\endcsname{$22/19$} %
\expandafter\newcommand\csname atlasChiUnwnj\endcsname{$2895/3$} %
\expandafter\newcommand\csname atlasChiUnwayzh\endcsname{$137/11$} %
\expandafter\newcommand\csname atlasPullMap\endcsname{0.36} %
\expandafter\newcommand\csname atlasNeffMap\endcsname{97\%} %
\expandafter\newcommand\csname atlasPassMap1\endcsname{pass} %
\expandafter\newcommand\csname atlasPassMap2\endcsname{fail} %
\expandafter\newcommand\csname atlasPassMap3\endcsname{fail} %
\expandafter\newcommand\csname atlasRMapptzA\endcsname{$1.00\pm0.01$} %
\expandafter\newcommand\csname atlasRMapptzB\endcsname{$1.01\pm0.03$} %
\expandafter\newcommand\csname atlasRMapptzC\endcsname{$0.97\pm0.09$} %
\expandafter\newcommand\csname atlasRMappthA\endcsname{$1.05\pm0.01$} %
\expandafter\newcommand\csname atlasRMappthB\endcsname{$0.79\pm0.02$} %
\expandafter\newcommand\csname atlasRMapjetA\endcsname{$1.85\pm0.02$} %
\expandafter\newcommand\csname atlasRMapjetB\endcsname{$0.72\pm0.03$} %
\expandafter\newcommand\csname atlasRMapdrA\endcsname{$1.34\pm0.01$} %
\expandafter\newcommand\csname atlasRMapptzhA\endcsname{$1.40\pm0.01$} %
\expandafter\newcommand\csname atlasRMapmllA\endcsname{$0.75\pm0.04$} %
\expandafter\newcommand\csname atlasChiMapptzh\endcsname{$4249/29$} %
\expandafter\newcommand\csname atlasChiMappth\endcsname{$363/30$} %
\expandafter\newcommand\csname atlasChiMapjet\endcsname{$7048/28$} %
\expandafter\newcommand\csname atlasChiMapdr\endcsname{$2531/30$} %
\expandafter\newcommand\csname atlasChiMapmzh\endcsname{$1096/41$} %
\expandafter\newcommand\csname atlasChiMapyh\endcsname{$25/19$} %
\expandafter\newcommand\csname atlasPullFive\endcsname{2.28} %
\expandafter\newcommand\csname atlasNeffFive\endcsname{92\%} %
\expandafter\newcommand\csname atlasPassFive1\endcsname{fail} %
\expandafter\newcommand\csname atlasPassFive2\endcsname{fail} %
\expandafter\newcommand\csname atlasPassFive2b\endcsname{fail} %
\expandafter\newcommand\csname atlasPassFive3\endcsname{fail} %
\expandafter\newcommand\csname atlasRFiveptzA\endcsname{$1.01\pm0.01$} %
\expandafter\newcommand\csname atlasRFiveptzB\endcsname{$1.05\pm0.03$} %
\expandafter\newcommand\csname atlasRFiveptzC\endcsname{$1.09\pm0.11$} %
\expandafter\newcommand\csname atlasRFivepthA\endcsname{$0.95\pm0.00$} %
\expandafter\newcommand\csname atlasRFivepthB\endcsname{$0.75\pm0.02$} %
\expandafter\newcommand\csname atlasRFivejetA\endcsname{$1.17\pm0.01$} %
\expandafter\newcommand\csname atlasRFivejetB\endcsname{$0.52\pm0.02$} %
\expandafter\newcommand\csname atlasRFivedrA\endcsname{$1.01\pm0.01$} %
\expandafter\newcommand\csname atlasRFiveptzhA\endcsname{$1.01\pm0.01$} %
\expandafter\newcommand\csname atlasRFivemllA\endcsname{$0.79\pm0.04$} %
\expandafter\newcommand\csname atlasChiFiveptz\endcsname{$29/29$} %
\expandafter\newcommand\csname atlasChiFiveptzh\endcsname{$18/29$} %
\expandafter\newcommand\csname atlasChiFivepth\endcsname{$395/30$} %
\expandafter\newcommand\csname atlasChiFivejet\endcsname{$620/28$} %
\expandafter\newcommand\csname atlasChiFivedr\endcsname{$82/30$} %
\expandafter\newcommand\csname atlasChiFivemzh\endcsname{$31/41$} %
\expandafter\newcommand\csname atlasChiFiveyh\endcsname{$27/19$} %
\expandafter\newcommand\csname atlasChiFivenj\endcsname{$3/3$} %
\expandafter\newcommand\csname atlasChiFiveayzh\endcsname{$11/11$} %
\expandafter\newcommand\csname atlasPullSix\endcsname{2.29} %
\expandafter\newcommand\csname atlasNeffSix\endcsname{91\%} %
\expandafter\newcommand\csname atlasPassSix1\endcsname{fail} %
\expandafter\newcommand\csname atlasPassSix2\endcsname{fail} %
\expandafter\newcommand\csname atlasPassSix2b\endcsname{fail} %
\expandafter\newcommand\csname atlasPassSix3\endcsname{fail} %
\expandafter\newcommand\csname atlasRSixptzA\endcsname{$1.01\pm0.01$} %
\expandafter\newcommand\csname atlasRSixptzB\endcsname{$1.02\pm0.03$} %
\expandafter\newcommand\csname atlasRSixptzC\endcsname{$1.06\pm0.10$} %
\expandafter\newcommand\csname atlasRSixpthA\endcsname{$0.94\pm0.00$} %
\expandafter\newcommand\csname atlasRSixpthB\endcsname{$0.74\pm0.04$} %
\expandafter\newcommand\csname atlasRSixjetA\endcsname{$0.99\pm0.01$} %
\expandafter\newcommand\csname atlasRSixjetB\endcsname{$0.64\pm0.04$} %
\expandafter\newcommand\csname atlasRSixdrA\endcsname{$1.00\pm0.01$} %
\expandafter\newcommand\csname atlasRSixptzhA\endcsname{$1.01\pm0.01$} %
\expandafter\newcommand\csname atlasRSixmllA\endcsname{$0.78\pm0.04$} %
\expandafter\newcommand\csname atlasChiSixptz\endcsname{$25/29$} %
\expandafter\newcommand\csname atlasChiSixptzh\endcsname{$23/29$} %
\expandafter\newcommand\csname atlasChiSixpth\endcsname{$447/30$} %
\expandafter\newcommand\csname atlasChiSixjet\endcsname{$185/28$} %
\expandafter\newcommand\csname atlasChiSixdr\endcsname{$91/30$} %
\expandafter\newcommand\csname atlasChiSixmzh\endcsname{$33/41$} %
\expandafter\newcommand\csname atlasChiSixyh\endcsname{$27/19$} %
\expandafter\newcommand\csname atlasChiSixnj\endcsname{$7/3$} %
\expandafter\newcommand\csname atlasChiSixayzh\endcsname{$11/11$} %

%% file: sec-intro.tex
\section{Introduction}
\label{sec:intro}

Higgs-boson production in association with a $Z$ boson receives contributions from quark-initiated
processes and from the loop-induced gluon-fusion channel, $gg\to ZH$.  Although the latter first
appears at $\mathcal{O}(\as^2)$, it constitutes a sizeable part of the inclusive $ZH$ cross section
and carries a much larger perturbative uncertainty than the quark-initiated contribution.  At
13\TeV{}, the YR4 reference prediction assigns $\sigma(gg\to ZH)=\yrggZHthirteen$ of
$\yrZHthirteen\,$pb (\yrggZHfracthirteen\%) to gluon fusion, with a scale uncertainty of
\yrggZHthirteenscale~\cite{deFlorian:2016spz,CampilloAveleira:2025ggzh}.  At 13.6\TeV{}, the
corresponding contribution is $\yrggZHsix$ of $\yrZHsix\,$pb
(\yrggZHfracsix\%)~\cite{Karlberg:2024zxx}.  The top-quark loop also shapes the kinematics.  The
relative $gg\to ZH$ contribution grows in the boosted regime~\cite{Englert:2013vua,Hespel:2015zea},
which is widely used in LHC analyses.

The calculation of $gg\to ZH$ beyond leading order (LO) is challenging because the NLO QCD
correction requires two-loop amplitudes with massive quarks.  Early NLO predictions therefore
relied on the heavy-top limit and on expansions in complementary kinematic regimes~\cite{Altenkamp:2012sx,Hasselhuhn:2016rqt,Davies:2020drs,Alasfar:2021ppe,Bonciani:2018omm}.
The full top-mass dependence was subsequently obtained numerically and through improved analytic
expansions, leading to complete NLO predictions and their comparison and combination~\cite{Chen:2020gae,Wang:2021rxu,Chen:2022rua,Degrassi:2022mro,CampilloAveleira:2025ggzh};
soft-gluon resummation is also available at NLL~\cite{Harlander:2014wda,Das:2025wbj,Banerjee:2026hzp}.  The amplitudes of
Refs.~\cite{Davies:2025out,Davies:2025otz} are now implemented in the public code
\ggxy{}~\cite{Davies:2026ggxy} and interfaced to \PBOX{}~\cite{Nason:2004rx,Frixione:2007vw,Alioli:2010xd}.  This provides NLO QCD predictions matched to
parton showers (\ggxyZH{}, Ref.~\cite{Davies:2026ggzhps}).

LHC simulations of this channel have so far commonly relied on LO samples showered with
\PYTHIA{} and normalized to a higher-order inclusive cross section, as in the VH($b\bar b$)
measurements of ATLAS~\cite{ATLAS:2020fcp} and CMS~\cite{CMS:2023vzh}.  The treatment of hard
radiation depends on the shower configuration.  For this colorless Born final state, the
unvetoed \PYTHIA{} default allows emissions up to the kinematic limit
(``power shower''~\cite{Plehn:2005cq,Corke:2010zj}).  The CMS- and ATLAS-style sources studied
here instead apply a \POWHEG{} veto using \code{SCALUP} $=\mZH$
(\cref{sec:rwt-samples,sec:rwt-atlas}); their predictions must be distinguished from the
unvetoed default.  Large NLO+PS samples present a separate practical challenge.  In the
unmodified implementation, scale and parton-distribution-function (PDF) weights repeat expensive
loop-amplitude evaluations.  Inclusive generation also provides few events in the boosted
Simplified Template Cross Section (STXS) bins.  We therefore ask whether NLO+PS samples with the
statistics and theory weights needed for LHC analyses can be generated efficiently, and how
reliably existing LO+PS samples can be corrected to reproduce their kinematics.

We address both questions using only public software.  We introduce an exact amplitude cache for
\ggxyZH{} to reduce the cost of scale and PDF event weights
(\cref{sec:cache}), together with a Born-suppression factor that enhances the sampling of the
high-$\ptV$ region without changing the physical prediction (\cref{sec:bornsupp}).  We validate the
resulting setup and present NLO and NLO+PS predictions at 13.6 and 13\TeV{} with a detailed theory
uncertainty budget, including predictions in the STXS Stage~1.3 bins
(\cref{sec:uncertainties,sec:validation,sec:results}).  We compare inclusively rescaled LO+PS
predictions with NLO+PS across several shower-start prescriptions to isolate their effect on
boosted and jet-dependent observables (\cref{sec:lops}).  Separately, we test generator-level
multivariate reweighting for the experiment-inspired sources and identify where the tested
corrections succeed or remain limited (\cref{sec:rwt}).  The complete run definitions, reference
results, and validation scripts are provided in the public package~\cite{ggxy-ggZH-ampcache}, with
the reproduction procedure summarized in \cref{app:reproduction}.

%% file: sec-setup.tex
\section{Calculation, parameters, and codes}
\label{sec:setup}

\subsection{Process and calculation}
\label{sec:process}

We consider loop-induced $pp\to ZH+X$ production through gluon fusion at NLO QCD with full
top-quark mass dependence, as implemented in \ggxy{}~\cite{Davies:2026ggxy} and in \PBOX{}
through \ggxyZH{}~\cite{Davies:2026ggzhps}.  The main results use
$Z\to\ell^-\ell^+$ with $\ell=e,\mu,\tau$, summed over the three flavors.  The $Z$ boson is
off shell and decays at amplitude level, preserving spin correlations and off-shell effects, with
$15<m_{\ell\ell}<150\GeV$.  The upper boundary follows the validity range of the two-loop virtual
amplitudes, whose expansion in external virtualities extends to approximately
$150\GeV$~\cite{Davies:2026ggzhps}.  The Higgs boson is on shell in the hard process and is decayed
by \PYTHIA{} for showered predictions (\cref{sec:shower}); none of the reported observables depends
on its decay mode.

Born and real-emission one-loop matrix elements are evaluated with
\RECOLA{}~\cite{Actis:2016mpe} and \COLLIER{}~\cite{Denner:2016kdg}.  The two-loop virtual
correction is provided by the \ggxy{} amplitudes, using Pad\'e-improved expansions~\cite{Davies:2025out,Davies:2026ggzhps}; their accuracy is discussed in \cref{sec:approx}.
The real contribution includes the $qg$ and $q\bar q$ channels and $Z$-boson radiation from
initial-state quarks (\code{extZ 1}).  Real-emission partons below $0.1\GeV$ in transverse momentum
are removed by the implementation-level infrared cutoff.

We repeat the calculation for $gg\to Z(\to\nu\bar\nu)H$, summed over the three neutrino flavors,
with the same settings apart from the decay mode (\code{vdecaymode 12}), the corresponding invariant-mass
window, and independent random seeds.  Inclusive results and reference comparisons for both decay
channels are given in \cref{tab:inclusive,tab:paper-comparison}; dedicated neutrino reweighting
results appear in \cref{sec:rwt-vv}.  The neutrino differential and STXS outputs and reweighting
models are included in the data record (\cref{app:reproduction}).  Unless stated otherwise, the
differential and STXS predictions displayed in the paper refer to the charged-lepton channel.

Although the Born contribution to $gg\to ZH$ is loop induced rather than tree level, we refer to it
as LO.  ``LO'' and ``NLO'' denote fixed-order predictions, while ``LO+PS'' and ``NLO+PS'' denote
predictions matched to a parton shower.  At LO, the $Z$ and Higgs bosons are back to back in the
transverse plane, so $\ptZ=\ptH$ and no QCD parton is present.  At NLO, one additional real-emission
parton is possible.  In LO+PS, all QCD radiation is generated by the shower.  In NLO+PS, \POWHEG{}
generates the hardest emission from the NLO real-emission kernel and the subsequent shower is vetoed
against harder emissions; the matching and shower settings are described in \cref{sec:shower}.

\subsection{Parameters}
\label{sec:parameters}

\begin{table}[h]
\centering
\caption{Parameters, identical at both energies except for the beam energy.}
\label{tab:parameters}
\small
\begin{tabular}{@{}lp{0.68\columnwidth}@{}}
\toprule
$\sqrt{s}$ & $13.6\TeV$ and $13\TeV$ ($pp$; beam energies 6800 and 6500\GeV) \\
$m_Z$, $\Gamma_Z$ & $91.1876\GeV$, $2.4955\GeV$ \\
$m_W$, $G_F$ & $80.385\GeV$, $1.16637\times10^{-5}\GeV^{-2}$ \\
$m_H$ & $125\GeV$, on shell \\
$m_t$ & $172.5\GeV$, on-shell scheme (\code{mtscheme 0}); $\overline{\mathrm{MS}}$ variations in \cref{sec:mtscheme} \\
$\kappa_{HZZ}$ & 1 \\
$m_{\ell\ell}$ & $15$--$150\GeV$ \\
$\muR=\muF=\muz$ & $\mZH/2$, $\mZH$ the invariant mass of the $\ell^-\ell^+H$ system \\
PDF & PDF4LHC21\_40\_pdfas~\cite{PDF4LHC:2022cjn} (\LHAPDF{}~\cite{Buckley:2014ana} 93300), also for LO \\
$\as(m_Z)$ & 0.118, from the PDF set (\code{alphas\_from\_pdf 1}) \\
$\hdamp$ & $250\GeV$ (\cref{sec:powheg}) \\
\bottomrule
\end{tabular}
\end{table}

Input parameters follow Ref.~\cite{Davies:2026ggzhps}, reproducing Tables~1 and~3 of that
work at both center-of-mass energies (\cref{sec:validation}).  Cross-code validation
(\cref{sec:validation}) was performed with NNPDF3.1 NNLO~\cite{Ball:2017nwa}
(\code{NNPDF31\_nnlo\_as\_0118\_mc\_hessian\_pdfas}, \LHAPDF{} ID 325300) at $13.6\TeV$;
NNPDF3.1 variations are also retained as event weights for comparison.  Using the same
PDF set at LO and NLO ensures that the differential $K$-factor,
$K=\sigma_{\mathrm{NLO}}/\sigma_{\mathrm{LO}}$, isolates higher-order ME corrections.

\subsection{Codes and versions}
\label{sec:codes}

The nominal NLO and NLO+PS calculations use the public \ggxy{} and \ggxyZH{} implementations in
\PBOX{}.  Showering and analysis use \PYTHIA{} and \RIVET{}, including the LHC Higgs Working Group
STXS classifier.  We use \MCFM{} for an independent LO cross-check.  The exact software commits,
library versions, patches, and build toolchains are listed in \cref{tab:software} and fixed by the
public reproduction package~\cite{ggxy-ggZH-ampcache}.

\subsection{Fixed-order calculations}
\label{sec:fo}

The nominal fixed-order LO and NLO predictions are obtained with \ggxy{}, including the seven-point
scale variations and the differential distributions used in this work.  Two independent LO
calculations provide cross-checks.  A Born-only \PBOX{} stage-2 integration tests the implementation
used for event generation, while \MCFM{} evaluates stable-$Z$ production with an independent
analytic one-loop amplitude.  The integration statistics, random-seed treatment, and binning used
for these calculations are given in \cref{app:reproduction}.

%% file: sec-events.tex
\section{Event generation}
\label{sec:events}

We generate the NLO+PS and LO+PS samples with \ggxyZH{} in \PBOX{}~\cite{Nason:2004rx,Frixione:2007vw,Alioli:2010xd}.  This section describes the choices that
affect the physics content or the efficiency of the samples; the complete run settings and event
statistics are given in \cref{app:reproduction}.

\subsection{\POWHEG{} matching}
\label{sec:powheg}

The real-emission matrix element $R$ is separated into singular and finite-remnant components,
$R=R_s+R_f$, with
\begin{equation}
  R_s=F_{\mathrm{damp}}R,\qquad
  F_{\mathrm{damp}}=\frac{\hdamp^2}{\hdamp^2+k_{\mathrm{T}}^2}.
\end{equation}
We use $\hdamp=250\GeV$, following Ref.~\cite{Davies:2026ggzhps}.  The singular component enters
the Sudakov exponent, while $R_f$ is generated as a finite, unexponentiated contribution.  In the
hard-emission limit, the Sudakov-generated term is proportional to $\bar B R_s/B$, where
$\bar B$ is the NLO underlying-Born weight and $B$ is the Born weight~\cite{Alioli:2010xd}.
The finite remnant carries no such $\bar B/B$ enhancement.  This local ratio can enhance hard
radiation through terms beyond fixed NLO; it need not equal the inclusive $K$ factor.
The choice of \hdamp{} controls this partition and constitutes a matching ambiguity.  We
probe it with dedicated samples, including the change in the generated radiation
(\cref{sec:matching}).

\subsection{Born suppression for the high-$\ptV$ tail}
\label{sec:bornsupp}

The physical spectrum falls steeply in the boosted region: at 13.6\TeV{}, the bin-averaged
fixed-order NLO differential cross section $d\sigma/d\ptZ$ decreases by a factor of
\pPtzFallsix{} between the $125\text{--}150$ and $500\text{--}600\GeV$ bins.  Inclusive
generation therefore provides few events in the highest STXS bins.  We enhance this region
with a Born-suppression factor~\cite{Alioli:2010xd},
\begin{equation}
  f(p_{\mathrm{T}}) = \left(\frac{p_{\mathrm{T}}^2 + (rL)^2}{p_{\mathrm{T}}^2+L^2}\right)^{n},
  \label{eq:bornsupp}
\end{equation}
using $L=\bornsuppL\GeV$, $n=1$, and $r=0.1$.  Here $p_{\mathrm{T}}$ is the Higgs-boson
transverse momentum in the underlying Born configuration, where it balances the lepton pair.
The generation density is multiplied by $f$, and the event receives the compensating factor
$1/f$.  The weighted sample therefore retains the physical cross section while preferentially
sampling large $\ptV$.

The suppression depends only on the underlying Born kinematics, so it does not alter the
real-emission matrix element or the scale and PDF variations.  Above $\ptZ=400\GeV$, a sample
generated with the suppression has \pNeffGainsix{} times the effective number of events
$N_{\mathrm{eff}}=(\sum w)^2/\sum w^2$ of an unweighted sample of the same size at 13.6\TeV{}
(\pNeffGainthirteen{} at 13\TeV{}).  Closure against unsuppressed generation and fixed-order NLO
is shown in \cref{sec:validation}.

\subsection{LO events}
\label{sec:loevents}

For the LO+PS reference, we generate Born events with the \POWHEG{} real emission disabled.
These events are unweighted and carry $\code{SCALUP}=\mZH$, the partonic center-of-mass energy.
This scale sets the shower limit when \code{pTmaxMatch = 1} is selected, or the veto threshold
when the \POWHEG{} hooks use \code{pThard = 0}.  For this colorless Born final state, the
unvetoed default shower (\code{pTmaxMatch = 0}) instead extends to the kinematic limit
(\cref{sec:lops-setup}).  The mean event weight reproduces the fixed-order LO cross section at
all scale points (\cref{sec:validation}); the corresponding run-card settings are listed in
\cref{app:reproduction}.

\subsection{Event weights}
\label{sec:weights}

Each production event carries \nWeights{} alternative weights evaluated at the same phase-space
point.  They cover renormalization- and factorization-scale variations, PDF4LHC21 members and
its $\as$ variations, and an NNPDF3.1 set retained for validation.  The complete weight list and
identifiers are given in \cref{app:conventions}.  Cross sections are obtained from the mean signed
event weight; the normalization and statistical convention are also specified there.

\subsection{Exact amplitude cache}
\label{sec:cache}

\begin{figure}[h]
\centering
\includegraphics[width=8.6cm]{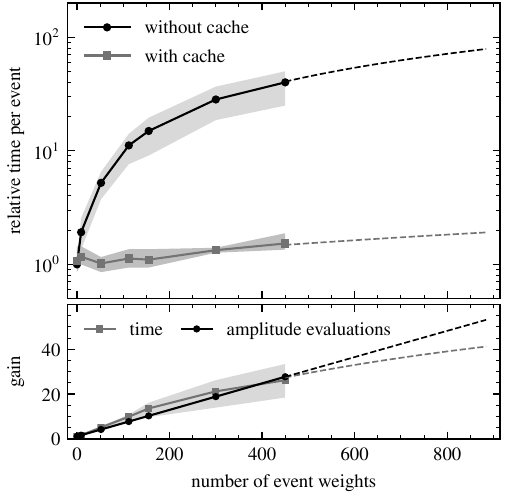}
\caption{Event-generation cost as a function of the number of alternative weights, measured with
the same 50 events at 13.6\TeV{}.  The upper panel shows the time per event relative to generation
without alternative weights or caching; bands give the range across \pScanJobssix{} runs.  The
lower panel shows the cache gain in generation time and one-loop amplitude evaluations.  Dashed
lines extrapolate the measured linear trends from \pScanMaxsix{} to \pScanExtrapolatesix{}
weights.}
\label{fig:cache}
\end{figure}

Alternative \POWHEG{} weights normally trigger a new evaluation of $\tilde B$ at the same
phase-space point.  For \ggxyZH{}, this repeats the one-loop Born and real amplitudes from
\RECOLA{} and the two-loop virtual amplitude from \ggxy{}, making large weight lists far more
expensive than the nominal event generation.  Without caching, an event with
\pScanMaxsix{} weights takes \pScanTimeUncachedsix{} times as long as an event without
alternative weights (\cref{fig:cache}).

The amplitudes do not depend on the PDF or on $\muF$, and their powers of \as{} can be factored
from the kinematic dependence.  We therefore developed an exact amplitude cache for \ggxyZH{},
which has been merged upstream (commit \code{1a6dbc6}).  The cache keys each amplitude by the
parton flavors, the bit representation of the momenta, and the mass-dependent inputs; the
two-loop virtual term also includes $\muR$.  The corresponding power of \as{} is restored for
the target scale, and the Born spin-correlation tensor is cached together with the Born
amplitude.  A cached value is reused only when all keyed inputs agree; otherwise the amplitude
is evaluated again.  Including the mass inputs prevents reuse of a nominal-mass result for a
different mass, but does not turn that result into a mass-scheme variation.  The closure tests
reported here cover PDF and scale weights in the on-shell setup; no
$\overline{\mathrm{MS}}$ NLO+PS sample is generated.  Implementation controls and memory settings
are documented in \cref{app:reproduction} and in the public package~\cite{ggxy-ggZH-ampcache}.

With identical random seeds, cached and uncached generation gives identical event kinematics and
alternative weights in the closure samples.  The verification covers \vCacheEvents{} events and
\vCacheWeights{} weights, with additional production checks of \pCacheEventssix{} events and
\pCacheWeightssix{} weights at each energy.

\Cref{fig:cache} shows the computational cost as the number of alternative weights is increased
from zero to \pScanMaxsix{}.  Without the cache, the number of one-loop amplitude evaluations
grows linearly with the weight list; with the cache it remains \pScanCachedsix{} per event.
At \pScanMaxsix{} weights, the cache reduces the number of amplitude evaluations by a factor
\pScanEvalRatiosix{} and the generation time by a factor \pScanTimeRatiosix{}.  A linear
extrapolation to \pScanExtrapolatesix{} weights, representative of larger LHC production weight
lists, gives factors of \pScanEvalExtrasix{} and \pScanTimeExtrasix{}, respectively.

%% file: sec-shower.tex
\section{Parton shower and analysis}
\label{sec:shower}

\subsection{\PYTHIA{} shower and \POWHEG{} veto}
\label{sec:pythia}

Generated LHE events are showered with \PYTHIA{}~\cite{Bierlich:2022pfr}, including multiparton
interactions and hadronization.  The Higgs boson is decayed to $b\bar b$ by \PYTHIA{}, and the
resulting events are passed through HepMC3~\cite{Buckley:2019xhk} to
\RIVET{}~\cite{Bierlich:2024vqo}.  LHE and shower-variation weights are propagated through the
same event record.

For NLO+PS, the shower is allowed to explore the full kinematic phase space, while the
\PYTHIA{} \POWHEG{} hooks veto QCD shower emissions above the event's reference hardness.
In the nominal setup, this hardness is \code{SCALUP} (\code{pThard = 0}), and shower emissions
are tested with the \POWHEG{} transverse-momentum definition (\code{pTdef = 1}).  This preserves
the separation between the \POWHEG{} emission and subsequent QCD shower radiation.  Alternative
hardness definitions enter the matching uncertainty in \cref{sec:matching}.  The complete
interface settings are listed in \cref{app:reproduction}.

For the shower-start comparison in \cref{sec:lops-setup}, Born-only LO events are showered
without a \POWHEG{} emission veto.  Those configurations vary the allowed shower phase space
directly.  The CMS- and ATLAS-style sources instead use the veto prescriptions specified in
\cref{sec:rwt-samples,sec:rwt-atlas}; they are separate from the unvetoed comparison samples.

\subsection{Tunes}
\label{sec:tunes}

We use the Monash tune~\cite{Skands:2014pea} as the nominal shower configuration and compare it
with the ATLAS A14~\cite{ATL-PHYS-PUB-2014-021} and CMS CP5~\cite{CMS:2019csb} tunes
(\cref{tab:tunes}).  Each tune uses its native PDF set for initial-state radiation and
multiparton interactions.  In this production-sample comparison, the hard matrix element uses
PDF4LHC21; the experiment-inspired sources have the PDF choices specified separately.
The tune comparison probes shower couplings and PDFs as well as nonperturbative parameters.

\begin{table}[h]
\centering
\caption{\PYTHIA{} tunes used for the shower comparison.}
\label{tab:tunes}
\small
\begin{tabular}{ll}
\toprule
Tune & Shower/MPI PDF \\
\midrule
Monash (nominal) & NNPDF2.3 QCD+QED LO \\
A14 & NNPDF2.3 LO \\
CP5 & NNPDF3.1 NNLO, $\as(m_Z)=0.118$ \\
\bottomrule
\end{tabular}
\end{table}

The exact tune settings used in the production, including the CP5 parameters and energy-dependent
cross-section inputs, are recorded in \cref{app:reproduction}.

\subsection{Shower uncertainties}
\label{sec:psvar}

We evaluate renormalization-scale variations in initial- and final-state radiation with
\PYTHIA{}'s automated variation framework~\cite{Mrenna:2016sih}.  The ISR and FSR scales are
varied independently by factors of $\tfrac12$ and $2$.  For each component, the upward and
downward shifts define an envelope; the ISR and FSR envelopes are then added in quadrature,
separately upwards and downwards.  Variations of the non-singular splitting-kernel terms are not
included.  We combine these scale variations with the Monash--A14--CP5 tune comparison when
constructing the matching and shower uncertainty in \cref{sec:matching}.

\subsection{Analysis and STXS classification}
\label{sec:rivet}

Events are classified into STXS Stage~1.3 bins using the official LHC Higgs WG \RIVET{} routine
\code{SimplifiedTemplateCrossSections}~\cite{LHCHWG:STXSrivet,LHCHWG:STXS13}.  After separating the forward-Higgs bin $|y_H|>2.5$,
the fine classification assigns the remaining events to 18 bins: six $\ptV$ intervals with
boundaries $0,75,150,250,400,600,\infty\GeV$, each split into
$N_{\mathrm{jets}}=0,1,\ge2$.  We also use the standard eight-bin grouping, which keeps separate
0- and $\ge1$-jet bins in the $150\text{--}250$ and $250\text{--}400\GeV$ intervals.

Jets are clustered with the anti-$k_{\mathrm{T}}$ algorithm~\cite{Cacciari:2008gp}, radius
$R=0.4$, and counted for $p_{\mathrm{T}}>30\GeV$.  Stable descendants of the Higgs boson and of
the decaying $Z$ boson are excluded from the jet inputs, as required by the STXS definition.  The
vector-boson transverse momentum is evaluated from the hard-process $Z$ after shower recoil, so
$\ptV=\ptZ$ for both charged-lepton and neutrino decays.

A companion \RIVET{} analysis records the differential observables used in this work, including
$\ptZ$, $\ptH$, $\mZH$, $p_{\mathrm{T}}(ZH)$, $y_{ZH}$, jet multiplicity, and leading-jet
$p_{\mathrm{T}}$.  The validation of the shower and STXS analysis chain is presented in
\cref{sec:validation}; detailed event-record conventions are collected in
\cref{app:conventions}.

%% file: sec-uncertainties.tex
\section{Uncertainties}
\label{sec:uncertainties}

We consider four uncertainty sources: perturbative scales, PDFs and $\as$, the top-quark mass
scheme, and matching and shower effects.  Fixed-order scale bands come from the integrations;
scale, PDF, and $\as$ variations of the generated samples use event weights.  The mass-scheme
study uses separate fixed-order calculations, while matching and shower variations use dedicated
samples, reshowers, or shower weights as specified below.

\subsection{Scales}
\label{sec:scale}

Perturbative QCD scale uncertainties are defined by the standard seven-point envelope
\begin{equation}
  \begin{split}
  (\muR,\muF)/\muz\in\{&(1,1),(\tfrac12,\tfrac12),(2,2),(2,1),\\
  &(\tfrac12,1),(1,2),(1,\tfrac12)\},
  \end{split}
\end{equation}
at fixed PDF, following Refs.~\cite{Davies:2026ggzhps,CampilloAveleira:2025ggzh}.  Generated
events contain the full nine scale combinations, from which this envelope is selected.

For rescaled LO+PS, the scale variation is separated into shape and normalization components.
Each LO+PS variation is normalized to its own inclusive cross section and therefore changes only
the shape.  This envelope is added in quadrature to the inclusive NLO scale uncertainty carried
by the normalization factor $K=\sigma_{\mathrm{NLO}}/\sigma_{\mathrm{LO}}$.  The two components
are treated as uncorrelated.

\subsection{PDF and $\as$}
\label{sec:pdf}

PDF and strong-coupling uncertainties follow the PDF4LHC21 prescription~\cite{PDF4LHC:2022cjn}:
\begin{equation}
  \begin{split}
  \delta_{\mathrm{PDF}}\sigma &= \left[\sum_{k=1}^{40}\left(\sigma^{(k)}-\sigma^{(0)}\right)^2\right]^{1/2},\\
  \delta_{\as}\sigma &= \frac{\left|\sigma(\as=0.119)-\sigma(\as=0.117)\right|}{2}.
  \end{split}
\end{equation}
Both are evaluated directly from event weights.  The quoted $\as$ variation corresponds to
$\delta\as(m_Z)=0.001$, matching Ref.~\cite{CampilloAveleira:2025ggzh}; the standard
PDF4LHC21 choice $\delta\as=0.0015$ is obtained by rescaling this component by 1.5.  PDF and
$\as$ uncertainties are added in quadrature.  Each $\as$ variation uses its associated PDF set,
so the change is applied consistently to the PDF evolution and the matrix element.

\subsection{Top-quark mass scheme}
\label{sec:mtscheme}

The baseline calculation uses the on-shell top-quark mass.  We estimate the mass-scheme
uncertainty by comparing it with predictions obtained using the running
$\overline{\mathrm{MS}}$ mass $\overline m_t(\mu_t)$ at
$\mu_t\in\{\mZH,\mZH/2,\mZH/4\}$, following
Refs.~\cite{Baglio:2020wgt,CampilloAveleira:2025ggzh}.  The envelope of these predictions relative
to the on-shell result defines the scheme uncertainty.  It estimates missing higher-order
dependence on the mass definition and is distinct from the experimental uncertainty on $m_t$.

We evaluate this variation at fixed order with \ggxy{} at LO and NLO, inclusively and
differentially, at both collision energies.  No NLO+PS samples are generated in the
$\overline{\mathrm{MS}}$ scheme.  For the central-Higgs STXS bins, we therefore transfer the
fixed-order NLO $\overline{\mathrm{MS}}$/on-shell ratio in each $\ptV$ interval to the
corresponding NLO+PS bins.  The same ratio is applied to all jet multiplicities within that
interval.  For the forward-Higgs bin $|y_H|>2.5$, we use the inclusive fixed-order ratio,
without a forward-region selection.  These are transfer prescriptions rather than direct
NLO+PS variations; they provide neither jet-dependent scheme information nor a separate
calculation of the forward-region dependence.

\subsection{ME--PS matching and shower uncertainties}
\label{sec:matching}

The separation of radiation between the NLO calculation and the parton shower is not unique.  We
probe three components of this ambiguity at 13.6\TeV{}:
\begin{itemize}
\item \textit{Matching:} $\hdamp=\infty$ is compared with the nominal
$\hdamp=250\GeV$ to vary how much of the real emission is exponentiated by \POWHEG{}.
We also vary the \PYTHIA{} \POWHEG-veto hardness definitions with
\code{pThard = 1} and \code{pTdef = 2} relative to the nominal
\code{pThard = 0}, \code{pTdef = 1} setup (\cref{sec:pythia}).
\item \textit{Shower scales:} ISR and FSR renormalization scales are varied as described in
\cref{sec:psvar}.
\item \textit{Tune:} CP5 and A14 are compared with the nominal Monash tune
(\cref{sec:tunes}).
\end{itemize}
The matching contribution is the envelope of the $\hdamp$ and veto variations.  We evaluate the
shower-scale component and tune spread separately, and treat the three components as uncorrelated
when adding them in quadrature, separately upwards and downwards, in each STXS bin.
\Cref{tab:matching} gives the decomposition and total uncertainty.

\begin{table*}[tbp]
\centering
\caption{Matching and shower uncertainties of the NLO+PS prediction in the STXS Stage~1.3 fine
bins at 13.6\TeV{}.  ``Matching'' is the envelope of the $\hdamp=\infty$ and \POWHEG-veto
variations; ``Shower scale'' is the ISR/FSR scale component; ``Tunes'' is the CP5/A14 spread
relative to Monash.  The total is their quadrature sum, evaluated separately upwards and
downwards.  MC gives the statistical precision of the comparison between the independent
$\hdamp=\infty$ and nominal samples.  The veto variants use the nominal events and shower seeds.
Variations are not resolved in bins where the nominal MC uncertainty exceeds 10\%.}
\label{tab:matching}
\ifdim\columnwidth<\textwidth\small\else\footnotesize\fi
\input{tab-matching.tex}
\end{table*}

\subsection{Accuracy and scope of the calculation}
\label{sec:approx}

The LO amplitudes retain the exact top-quark mass dependence.  The two-loop virtual amplitudes use
Pad\'e-improved expansions in the forward and high-energy limits~\cite{Davies:2025out,Davies:2026ggzhps}.  Ref.~\cite{CampilloAveleira:2025ggzh} finds agreement
with numerical virtual amplitudes containing the exact top-mass dependence~\cite{Chen:2022rua} at better than the per-mille level.  The expansion in external virtualities
sets the upper validity range $m_{\ell\ell}<150\GeV$ used throughout this work.

The calculation isolates the loop-induced gluon-fusion contribution.  The
$\mathcal{O}(\as^2)$ interference with Drell--Yan-type $q\bar q\to ZH$ amplitudes belongs to the
quark-initiated production channel and is not included.

%% file: tab-matching.tex
\begin{tabular}{lccccc}
\toprule
$\ptV$ (GeV), jets & Matching (\%) & MC (\%) & Shower scale (\%) & Tunes (\%) & Total (\%) \\
\midrule
$|y_H|>2.5$ & $^{+0.7}_{-0.0}$ & 1.7 & $^{+0.0}_{-3.2}$ & $^{+0.7}_{-0.0}$ & $^{+1.0}_{-3.2}$ \\
\midrule
0--75, 0 jets & $^{+2.0}_{-3.9}$ & 1.7 & $^{+0.3}_{-2.7}$ & $^{+0.0}_{-3.3}$ & $^{+2.1}_{-5.8}$ \\
75--150, 0 jets & $^{+1.4}_{-5.6}$ & 0.5 & $^{+1.8}_{-1.1}$ & $^{+0.0}_{-2.6}$ & $^{+2.3}_{-6.2}$ \\
150--250, 0 jets & $^{+1.0}_{-6.8}$ & 0.5 & $^{+3.4}_{-1.9}$ & $^{+0.0}_{-3.8}$ & $^{+3.6}_{-8.0}$ \\
250--400, 0 jets & $^{+0.6}_{-8.6}$ & 1.4 & $^{+1.7}_{-0.9}$ & $^{+0.0}_{-3.9}$ & $^{+1.8}_{-9.5}$ \\
400--600, 0 jets & $^{+0.0}_{-6.0}$ & 6.5 & $^{+0.7}_{-5.0}$ & $^{+0.0}_{-12.1}$ & $^{+0.7}_{-14.4}$ \\
$>600$, 0 jets & $^{+35.3}_{-0.0}$ & 36.8 & $^{+5.3}_{-58.8}$ & $^{+6.6}_{-6.5}$ & $^{+36.3}_{-59.1}$ \\
\midrule
0--75, 1 jet & $^{+4.5}_{-0.0}$ & 1.8 & $^{+7.2}_{-3.6}$ & $^{+8.6}_{-0.0}$ & $^{+12.1}_{-3.6}$ \\
75--150, 1 jet & $^{+5.7}_{-0.0}$ & 0.6 & $^{+2.6}_{-3.1}$ & $^{+8.2}_{-0.0}$ & $^{+10.4}_{-3.1}$ \\
150--250, 1 jet & $^{+5.2}_{-0.0}$ & 0.6 & $^{+3.2}_{-4.4}$ & $^{+8.2}_{-0.0}$ & $^{+10.2}_{-4.4}$ \\
250--400, 1 jet & $^{+6.9}_{-0.0}$ & 1.3 & $^{+4.5}_{-6.2}$ & $^{+11.5}_{-0.0}$ & $^{+14.2}_{-6.2}$ \\
400--600, 1 jet & $^{+6.1}_{-0.0}$ & 5.0 & $^{+9.4}_{-11.8}$ & $^{+14.3}_{-0.0}$ & $^{+18.2}_{-11.8}$ \\
$>600$, 1 jet & $^{+12.2}_{-21.8}$ & 19.4 & $^{+55.5}_{-10.3}$ & $^{+10.8}_{-2.0}$ & $^{+57.8}_{-24.2}$ \\
\midrule
0--75, $\ge2$ jets & $^{+10.4}_{-8.6}$ & 1.8 & $^{+4.3}_{-10.7}$ & $^{+0.0}_{-4.8}$ & $^{+11.3}_{-14.5}$ \\
75--150, $\ge2$ jets & $^{+6.5}_{-9.4}$ & 0.8 & $^{+5.5}_{-8.1}$ & $^{+0.0}_{-5.6}$ & $^{+8.5}_{-13.6}$ \\
150--250, $\ge2$ jets & $^{+7.3}_{-8.2}$ & 0.7 & $^{+4.9}_{-5.3}$ & $^{+0.0}_{-5.4}$ & $^{+8.8}_{-11.2}$ \\
250--400, $\ge2$ jets & $^{+11.0}_{-7.7}$ & 1.3 & $^{+7.4}_{-10.2}$ & $^{+0.0}_{-9.0}$ & $^{+13.3}_{-15.6}$ \\
400--600, $\ge2$ jets & $^{+13.2}_{-6.0}$ & 3.7 & $^{+4.3}_{-3.0}$ & $^{+0.0}_{-8.1}$ & $^{+13.9}_{-10.5}$ \\
$>600$, $\ge2$ jets & $^{+19.6}_{-0.2}$ & 12.2 & $^{+2.6}_{-10.4}$ & $^{+0.0}_{-8.8}$ & $^{+19.7}_{-13.6}$ \\
\bottomrule
\end{tabular}

%% file: sec-validation.tex
\section{Validation}
\label{sec:validation}

We validate the calculation at three levels: consistency among the fixed-order and event-generation
implementations, comparison with the published \ggxyZH{} results, and closure of the event-level
analysis and Born-suppressed production.

\subsection{Fixed-order and event-generation consistency}

\Cref{tab:verification} compares integration and event-generation results at 13.6\TeV{} with
NNPDF3.1.  Standalone \ggxy{}, the \POWHEG{} stage-2 integration, and the generated event weights
are statistically compatible at LO and NLO.  These routes share matrix-element ingredients and
test their integration and event-generation interfaces.  The LO event weights reproduce the
fixed-order prediction at all seven scale points.  At NLO, the standalone calculation covers
seven scale points; separate regression checks compare all nine stored scale weights and the
PDF and $\as$ variations with reference event-weight results.  The automated suite passes
\vChecks{} generator-level checks, which do not resolve the independent \MCFM{} comparison below.

\begin{table*}[tbp]
\centering
\caption{Cross sections in fb at 13.6\TeV{} with NNPDF3.1 and $\muz=\mZH/2$.
$\ell\ell H$ is summed over $\ell=e,\mu,\tau$ with $15<m_{\ell\ell}<150\GeV$.
The scale bands are the seven-point envelopes.}
\label{tab:verification}
\small
\begin{tabular}{lll}
\toprule
& $\ell\ell H$ & stable $Z$ \\
\midrule
LO, \ggxy{} & \vLO{} \vLOband & \vStLOnn{} \\
LO, \POWHEG{} stage 2 & \vPowLO{} & \\
LO, event weights & \vEvLO{} & \\
LO, \MCFM{} & & \vMCFMnn{} \\
NLO, \ggxy{} & \vNLO{} \vNLOband & \\
NLO, \POWHEG{} grids & \vGrids{} & \\
NLO, event weights & \vEvNLO{} \vEvNLOband & \\
\bottomrule
\end{tabular}
\end{table*}

The independent \MCFM{} stable-$Z$ calculation lies \pMCFMoffset\% above \ggxy{}, well beyond
the quoted integration uncertainties.  The same relative difference is observed with NNPDF3.1
and PDF4LHC21 and at both collision energies.  Its origin remains unresolved.  This comparison
therefore does not establish agreement between the independent LO implementations, and no
correction is applied to either prediction.

\subsection{Agreement with the published \ggxyZH{} calculation}

\Cref{tab:paper-comparison} compares our fixed-order results with Tables~1 and~3 of
Ref.~\cite{Davies:2026ggzhps}.  The charged-lepton, neutrino, and stable-$Z$ cross sections agree
within the quoted Monte Carlo precision at both 13 and 13.6\TeV{}.  The stable-$Z$ results are also
consistent with the updated NLO predictions of Ref.~\cite{CampilloAveleira:2025ggzh}.

\begin{table*}[tbp]
\centering
\caption{Fixed-order cross sections in fb with PDF4LHC21, compared with
Ref.~\cite{Davies:2026ggzhps}.}
\label{tab:paper-comparison}
\footnotesize\setlength{\tabcolsep}{4pt}
\begin{tabular}{llllll}
\toprule
& & \multicolumn{2}{c}{13.6\TeV} & \multicolumn{2}{c}{13\TeV} \\
& & this work & Ref.~\cite{Davies:2026ggzhps} & this work & Ref.~\cite{Davies:2026ggzhps} \\
\midrule
$\ell\ell H$ & LO  & \pLOsix  & \refLOsix{} \refLOsixband   & \pLOthirteen  & \refLOthirteen{} \refLOthirteenband \\
             & NLO & \pNLOsix & \refNLOsix{} \refNLOsixband & \pNLOthirteen & \refNLOthirteen{} \refNLOthirteenband \\
$\nu\bar\nu H$ & LO & \pLOsixVV & \refLOsixVV{} \refLOsixbandVV & \pLOthirteenVV & \refLOthirteenVV{} \refLOthirteenbandVV \\
             & NLO & \pNLOsixVV & \refNLOsixVV{} \refNLOsixbandVV & \pNLOthirteenVV & \refNLOthirteenVV{} \refNLOthirteenbandVV \\
stable $Z$   & LO  & \pStLOsix & \refStLOsix  & \pStLOthirteen  & \refStLOthirteen \\
             & NLO & \pStNLOsix & \refStNLOsix & \pStNLOthirteen & \refStNLOthirteen \\
\bottomrule
\end{tabular}
\end{table*}

\subsection{Event-level and Born-suppression closure}

At parton level, the production LO events reproduce the fixed-order LO $\ptZ$ spectrum bin by bin,
with $\chi^2/N_{\mathrm{bins}}=\pPartonLOchisix$ at 13.6\TeV{} and
\pPartonLOchithirteen{} at 13\TeV{}.  On the NNPDF3.1 validation sample, the integrated
\RIVET{} cross section after showering matches the LHE normalization for every scale and PDF
weight.  Every event is assigned an STXS bin, and the official Stage~1.3 implementation agrees
with the independent bin assignment used in our validation.

We test the Born suppression with 50,000 events generated with identical seeds, once without
suppression and once with $L=\bornsuppL\GeV$.  The inclusive cross sections agree with pull
\vEvNLObsPull; all nine scale variations and the PDF and $\as$ uncertainties close within the
validation tolerances.  After showering, the suppressed and unsuppressed samples agree in the
Stage~1.3 fine bins and in $\ptZ$, with
$\chi^2/N_{\mathrm{bins}}=\vBSSTXSchi$ and \vBSPtzchi{}, respectively.

The unsuppressed validation sample has limited statistics above about $\ptZ=550\GeV$.  We
therefore also compare the larger Born-suppressed production sample at parton level with
fixed-order NLO.  At 13.6\TeV{}, the ratio above $400\GeV$ spans \pPartonNLOtailsix{}, with
Monte Carlo uncertainties of \pPartonNLOtailErrsix\%.  Between 125 and $400\GeV$, the deviations
span \pPartonNLOmidLosix\% to \pPartonNLOmidHisix\%.  A similar structure appears without Born
suppression: the ratio in the $125\text{--}150\GeV$ bin is \vPartonNLOdip{} in the unsuppressed
validation sample and \pPartonNLOdipsix{} in the production sample.  With $\hdamp=\infty$, the
13.6\TeV{} deviations over 125--400\GeV{} span \pHdampPartonMidLosix\% to
\pHdampPartonMidHisix\%.  Their persistence without suppression and dependence on \hdamp{}
point to higher-order matching effects in the \POWHEG{} hardest emission, rather than a bias
introduced by the Born-suppression factor.  Agreement with fixed order in the tail is a
consistency check, not a substitute for suppressed-versus-unsuppressed closure.

%% file: sec-results.tex
\section{Predictions}
\label{sec:results}

\subsection{Inclusive rates}

\Cref{tab:inclusive} gives the fixed-order cross sections and their uncertainty budgets.  At
13.6\TeV{}, the charged-lepton channel increases from
$\sigma_{\mathrm{LO}}=\pLOsix\fb$ to $\sigma_{\mathrm{NLO}}=\pNLOsix\fb$, corresponding to
$K=\pKsix$.  The neutrino channel receives a nearly identical relative NLO correction.
Its cross section is approximately twice that of the charged-lepton channel, consistent with
the branching-fraction expectation for the two $Z$-boson decay modes.

The NLO correction reduces the seven-point scale band from \pLOsixband{} to \pNLOsixband{}.
At NLO, the PDF and $\as$ uncertainties are \pEvPDFsix\% and \pEvAssix\%, while the
top-quark mass-scheme band is \pMtNLOsix\%.  Scale variation remains the largest inclusive
uncertainty.  The mass-scheme band is comparable to its downward variation and substantially
larger than the PDF and $\as$ uncertainties.  The 13\TeV{} results show the same hierarchy.

\begin{table*}[tbp]
\centering
\caption{Inclusive fixed-order cross sections in fb with PDF4LHC21, summed over the three
charged-lepton or neutrino flavors.  Scale denotes the seven-point envelope; PDF and $\as$
($\delta\as=0.001$) are obtained from event weights; $m_t$ is the envelope of the
$\overline{\mathrm{MS}}$ predictions with $\mu_t\in\{\mZH,\mZH/2,\mZH/4\}$ relative to the
on-shell result; MC is the relative integration uncertainty.  For $\nu\bar\nu H$, the relative
$m_t$ band is transferred from the charged-lepton calculation; the explicit
$\mu_t=\mZH$ check gives shifts of $\pMtCheckLOVV\%$ at LO and $\pMtCheckNLOVV\%$ at NLO.}
\label{tab:inclusive}
\small
\begin{tabular}{lcccccc}
\toprule
& $\sigma$ (fb) & Scale & PDF (\%) & $\as$ (\%) & $m_t$ scheme & MC (\%) \\
\midrule
\multicolumn{7}{l}{$gg\to\ell^-\ell^+H$} \\
LO, 13.6\TeV  & \pLOsix & \pLOsixband & \pEvLOPDFsix & \pEvLOAssix & \pMtLOsix & \pLOmcsix \\
NLO, 13.6\TeV & \pNLOsix & \pNLOsixband & \pEvPDFsix & \pEvAssix & \pMtNLOsix & \pNLOmcsix \\
LO, 13\TeV    & \pLOthirteen & \pLOthirteenband & \pEvLOPDFthirteen & \pEvLOAsthirteen & \pMtLOthirteen & \pLOmcthirteen \\
NLO, 13\TeV   & \pNLOthirteen & \pNLOthirteenband & \pEvPDFthirteen & \pEvAsthirteen & \pMtNLOthirteen & \pNLOmcthirteen \\
\midrule
\multicolumn{7}{l}{$gg\to\nu\bar\nu H$} \\
LO, 13.6\TeV  & \pLOsixVV & \pLOsixbandVV & \pEvLOPDFsixVV & \pEvLOAssixVV & \pMtLOsix & \pLOmcsixVV \\
NLO, 13.6\TeV & \pNLOsixVV & \pNLOsixbandVV & \pEvPDFsixVV & \pEvAssixVV & \pMtNLOsix & \pNLOmcsixVV \\
LO, 13\TeV    & \pLOthirteenVV & \pLOthirteenbandVV & \pEvLOPDFthirteenVV & \pEvLOAsthirteenVV & \pMtLOthirteen & \pLOmcthirteenVV \\
NLO, 13\TeV   & \pNLOthirteenVV & \pNLOthirteenbandVV & \pEvPDFthirteenVV & \pEvAsthirteenVV & \pMtNLOthirteen & \pNLOmcthirteenVV \\
\bottomrule
\end{tabular}
\end{table*}

\subsection{Differential structure at NLO and NLO+PS}

\begin{figure*}
\centering
\includegraphics[width=\textwidth]{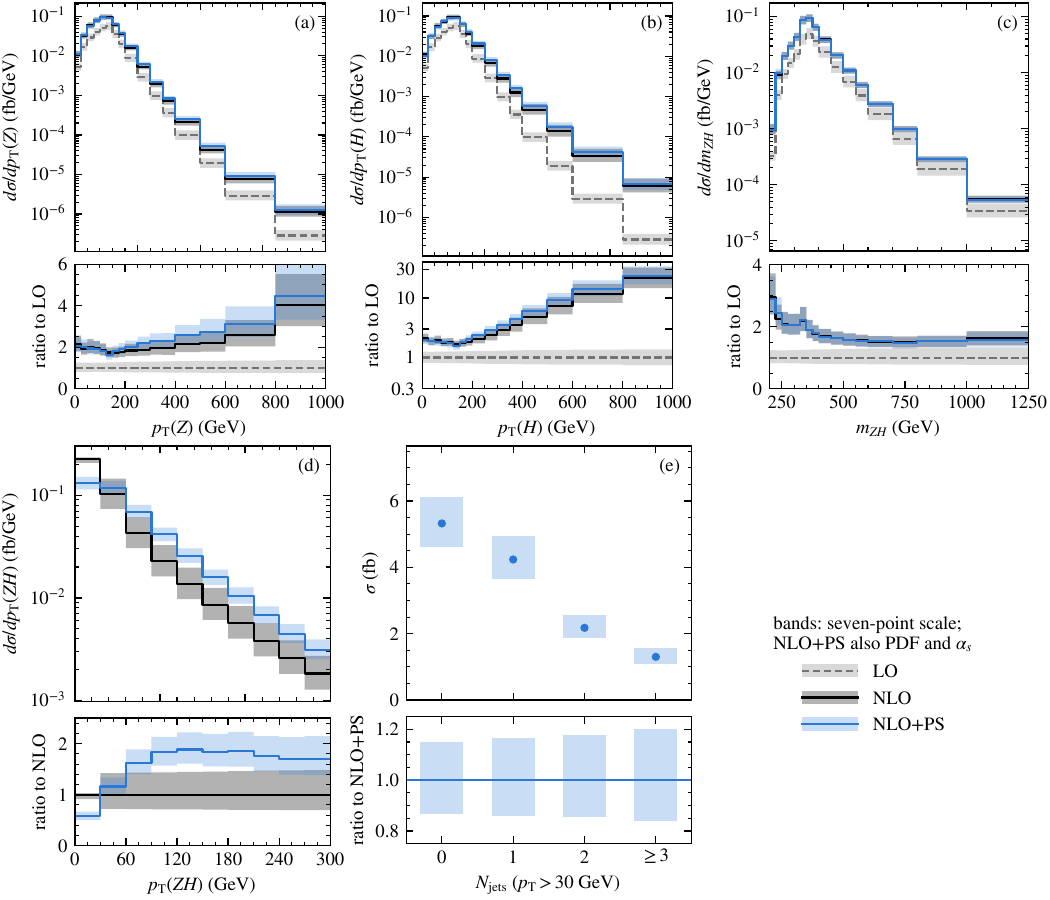}
\caption{Differential predictions for $gg\to\ell^-\ell^+H$ at 13.6\TeV{} with PDF4LHC21:
(a)~$\ptZ$, (b)~$\ptH$, (c)~$\mZH$, (d)~$p_{\mathrm{T}}(ZH)$, and (e)~jet multiplicity for
$p_{\mathrm{T}}>30\GeV$.  Fixed-order LO and NLO bands are seven-point scale envelopes.
NLO+PS uses the Monash tune; its bands combine the scale envelope, PDF, and $\as$ uncertainties
in quadrature, without the mass-scheme or matching variations.  The lower panels show ratios
to LO for $\ptZ$, $\ptH$ (logarithmic scale), and $\mZH$, to NLO for $p_{\mathrm{T}}(ZH)$,
and to the central NLO+PS prediction for jet multiplicity.}
\label{fig:dists}
\end{figure*}

\Cref{fig:dists} shows the main differential distributions at 13.6\TeV{}.  The NLO correction is
strongly kinematics dependent and cannot be represented by a single inclusive $K$ factor.  In
$\mZH$, the differential $K$ factor is \pKMzhThrsix{} in the low-mass
$200\text{--}225\GeV$ interval and decreases to \pKMzhHighsix{} above $600\GeV$.  The behavior is
opposite in the boosted transverse-momentum tails: $K(\ptZ)$ is \pKPtzLowsix{} below
$400\GeV$ and rises to \pKPtzSixsix{} at $600\text{--}800\GeV$ and
\pKPtzEightsix{} at $800\text{--}1000\GeV$.

The NLO real emission also separates the $Z$- and Higgs-boson transverse momenta.  At LO they are
back to back and satisfy $\ptZ=\ptH$; at NLO, a hard jet can recoil predominantly against the
Higgs boson while the $Z$ remains softer.  Consequently the $\ptH$ tail receives an even larger
correction, with $K$ reaching \pKPthEightsix{} at $800\text{--}1000\GeV$.  This hard-recoil
topology was also observed in merged $0+1$-jet LO studies~\cite{Amoroso:2020lgh}.

NLO+PS leaves the $\mZH$ spectrum essentially unchanged relative to fixed-order NLO and produces
additional hardening in the $\ptZ$ and $\ptH$ tails.  By contrast, the nonzero
$p_{\mathrm{T}}(ZH)$ spectrum is only LO accurate in the fixed-order NLO calculation and is more
sensitive to matching.  NLO+PS exceeds the fixed-order prediction by a factor of
\pPtzhPsOverNLOsix{} between 60 and $300\GeV$.  The Sudakov-generated real contribution can
receive the local $\bar B/B$ enhancement discussed in \cref{sec:powheg}, whereas the finite
remnant does not.  Varying \hdamp{} changes this partition and probes the resulting
higher-order matching dependence (\cref{sec:matching}).  The differential fixed-order $K$
factors at 13 and 13.6\TeV{} are statistically compatible; the energy comparison and its
Monte Carlo precision are discussed in \cref{app:dists13}.

\subsection{Top-quark mass-scheme dependence}

\begin{figure*}[tbp]
\centering
\includegraphics[width=\textwidth]{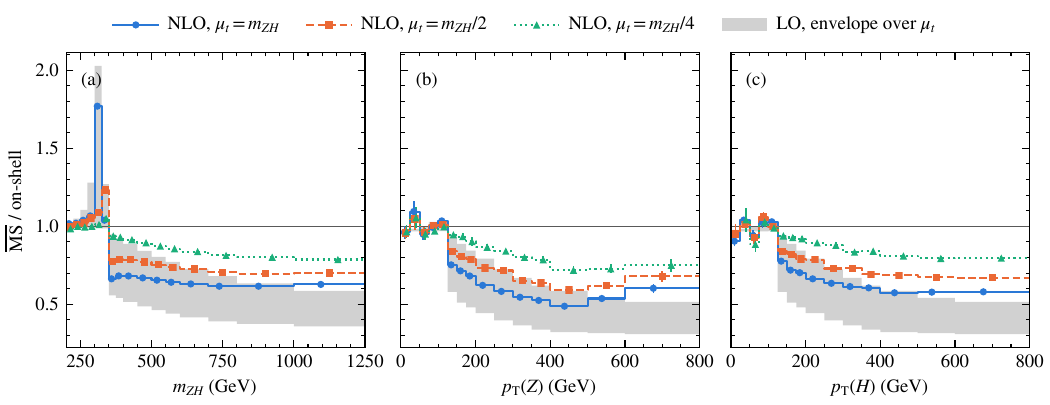}
\caption{Top-quark mass-scheme dependence at fixed order for $\ggllH$ at 13.6\TeV{} with
PDF4LHC21.  Shown is the $\overline{\mathrm{MS}}$/on-shell cross-section ratio for
$\mu_t=\mZH$, $\mZH/2$, and $\mZH/4$ in (a)~$\mZH$, (b)~$\ptZ$, and (c)~$\ptH$.
Markers show NLO results with Monte Carlo errors from the two integrations combined in
quadrature; the grey band is the envelope of the corresponding LO ratios.}
\label{fig:mtscheme}
\end{figure*}

\Cref{fig:mtscheme} shows the fixed-order NLO prediction in the $\overline{\mathrm{MS}}$ scheme
relative to the on-shell result for the three choices of $\mu_t$ defined in
\cref{sec:mtscheme}.  The inclusive NLO envelope is \pMtNLOsix\%, but the effect is much more
pronounced near the top-pair threshold and in the boosted region.  Percentage shifts quoted
below refer to cross sections integrated over each stated region; their ranges span the three
choices of $\mu_t$.  The shift is \pMtMzhBelowsix\,\% for
$200<\mZH<300\GeV$ and \pMtMzhAbovesix\,\% for $\mZH>400\GeV$.
Near the top-pair threshold, the finite-order mass-scheme predictions also differ in shape:
the largest single-bin $\overline{\mathrm{MS}}$/on-shell ratio between 300 and $350\GeV$ is
\pMtMzhThrsix{}, over the three $\mu_t$ choices.

For the cross section integrated over $\ptZ<100\GeV$, the shift is
\pMtPtzLowsix\,\%, whereas for $\ptZ>250\GeV$ it is \pMtPtzHighsix\,\%.
The corresponding shift for $\ptH>250\GeV$ is \pMtPthHighsix\,\%.
The mass-scheme dependence therefore becomes a leading uncertainty in the boosted phase space
targeted by the high-$\ptV$ STXS bins.

\subsection{STXS Stage~1.3 predictions and uncertainties}
\label{sec:stxs}

\Cref{tab:stxs} gives the NLO+PS Stage~1.3 fine-bin predictions at 13.6\TeV{}.  Scale, PDF, and
$\as$ uncertainties are evaluated directly from the event weights, while matching and shower
uncertainties use the variations of \cref{sec:matching}.  No $\overline{\mathrm{MS}}$ NLO+PS
sample is generated.  For the central-Higgs bins, the fixed-order NLO mass-scheme ratio in each
$\ptV$ interval is transferred to all jet bins in that interval.  The forward-Higgs bin instead
receives the inclusive fixed-order ratio, without a forward-region selection
(\cref{sec:mtscheme}).

In the \pStxsPrecisesix{} bins with Monte Carlo uncertainties below 5\%, scale uncertainties span
\pStxsScalesix\%, PDF+$\as$ uncertainties \pStxsPdfsix\%, top-mass scheme bands
\pStxsMtsix\%, and matching+shower uncertainties \pStxsMatchsix\%.  For asymmetric bands,
these ranges refer to the larger absolute excursion.  Scale variation is largest in the
low-$\ptV$ bins and the forward-Higgs region.  With the stated transfer prescription, the
mass-scheme band is the largest uncertainty in every statistically precise central-Higgs jet
bin from $\ptV=\pStxsMtFromsix\GeV$ onward, reaching \pStxsMtHighsix\%.
The common mass-scheme band across jet multiplicities is imposed by this prescription; its
jet dependence has not been calculated.

\begin{table*}[tbp]
\centering
\caption{NLO+PS (Monash) cross sections at 13.6\TeV{} in the STXS Stage~1.3 fine bins of
$gg\to Z(\to\ell\ell)H$, summed over $\ell=e,\mu,\tau$.  The table gives the relative Monte
Carlo uncertainty, seven-point scale envelope, PDF+$\as$ uncertainty ($\delta\as=0.001$),
top-mass scheme band from the fixed-order transfer of \cref{sec:mtscheme}, and the matching+shower
uncertainty of \cref{sec:matching}.  The mass-scheme transfer uses the $\ptV$-interval ratio
for central-Higgs bins and the inclusive ratio for the forward-Higgs bin.  The 13\TeV{}
predictions are given in \cref{tab:stxs13}.}
\label{tab:stxs}
\ifdim\columnwidth<\textwidth\small\else\footnotesize\fi
\input{tab-stxs.tex}
\end{table*}

%% file: tab-stxs.tex
\begin{tabular}{lcccccc}
\toprule
$\ptV$ (GeV), jets & $\sigma$ (fb) & MC (\%) & Scale (\%) & PDF+$\as$ (\%) & $m_t$ (\%) & Match.+PS (\%) \\
\midrule
$|y_H|>2.5$ & 0.476 & 1.2 & $^{+18}_{-15}$ & 7.5 & $^{+0}_{-13}$ & $^{+1}_{-3}$ \\
\midrule
0--75, 0 jets & 0.985 & 1.2 & $^{+17}_{-14}$ & 2.2 & $^{+0}_{-2}$ & $^{+2}_{-6}$ \\
75--150, 0 jets & 2.84 & 0.3 & $^{+16}_{-13}$ & 2.2 & $^{+0}_{-8}$ & $^{+2}_{-6}$ \\
150--250, 0 jets & 1.17 & 0.4 & $^{+12}_{-12}$ & 2.2 & $^{+0}_{-32}$ & $^{+4}_{-8}$ \\
250--400, 0 jets & 0.0975 & 1.0 & $^{+7}_{-10}$ & 2.7 & $^{+0}_{-43}$ & $^{+2}_{-9}$ \\
400--600, 0 jets & 0.00358 & 4.5 & $^{+5}_{-9}$ & 3.6 & $^{+0}_{-50}$ & $^{+1}_{-14}$ \\
$>600$, 0 jets & 9.21e-05 & 29.2 & $^{+3}_{-22}$ & 6.1 & $^{+0}_{-39}$ & $^{+36}_{-59}$ \\
\midrule
0--75, 1 jet & 0.778 & 1.3 & $^{+18}_{-14}$ & 2.2 & $^{+0}_{-2}$ & $^{+12}_{-4}$ \\
75--150, 1 jet & 2.03 & 0.4 & $^{+17}_{-14}$ & 2.2 & $^{+0}_{-8}$ & $^{+10}_{-3}$ \\
150--250, 1 jet & 1.13 & 0.4 & $^{+15}_{-13}$ & 2.2 & $^{+0}_{-32}$ & $^{+10}_{-4}$ \\
250--400, 1 jet & 0.137 & 0.9 & $^{+13}_{-12}$ & 2.6 & $^{+0}_{-43}$ & $^{+14}_{-6}$ \\
400--600, 1 jet & 0.00712 & 3.4 & $^{+12}_{-13}$ & 3.3 & $^{+0}_{-50}$ & $^{+18}_{-12}$ \\
$>600$, 1 jet & 0.000455 & 13.1 & $^{+29}_{-21}$ & 5.1 & $^{+0}_{-39}$ & $^{+58}_{-24}$ \\
\midrule
0--75, $\ge2$ jets & 0.61 & 1.3 & $^{+20}_{-16}$ & 2.2 & $^{+0}_{-2}$ & $^{+11}_{-15}$ \\
75--150, $\ge2$ jets & 1.47 & 0.6 & $^{+19}_{-15}$ & 2.2 & $^{+0}_{-8}$ & $^{+8}_{-14}$ \\
150--250, $\ge2$ jets & 1.05 & 0.5 & $^{+17}_{-14}$ & 2.3 & $^{+0}_{-32}$ & $^{+9}_{-11}$ \\
250--400, $\ge2$ jets & 0.226 & 0.9 & $^{+19}_{-15}$ & 2.6 & $^{+0}_{-43}$ & $^{+13}_{-16}$ \\
400--600, $\ge2$ jets & 0.0196 & 2.8 & $^{+24}_{-18}$ & 3.5 & $^{+0}_{-50}$ & $^{+14}_{-11}$ \\
$>600$, $\ge2$ jets & 0.00163 & 7.8 & $^{+31}_{-22}$ & 5.4 & $^{+0}_{-39}$ & $^{+20}_{-14}$ \\
\bottomrule
\end{tabular}

%% file: sec-lops.tex
\section{Rescaled LO+PS versus NLO+PS}
\label{sec:lops}

An inclusive $K$ factor corrects the normalization of an LO+PS sample but leaves its kinematic
shapes unchanged.  We test whether this rescaling reproduces the NLO+PS prediction in boosted
and jet-dependent observables, and isolate the dependence on the shower starting scale.
Differential reweighting of experiment-inspired samples is studied separately in \cref{sec:rwt}.

Earlier $gg\to ZH$ studies compared Born-only showers with merged zero- and one-jet matrix
elements and found sizeable shower-start dependence in hard-recoil observables~\cite{Amoroso:2020lgh}.  The two \POWHEG{}+\PYTHIA{} configurations in that study used a
restricted start or a vetoed power shower, rather than the unvetoed power shower considered below.
Related studies of $gg\to HH$ and $gg\to ZZ$ examined the sensitivity of pair-recoil spectra to
showering and matching~\cite{Jones:2017giv,Alioli:2016xab}.  Here we use the NLO+PS calculation
of \cref{sec:setup} as the reference and quantify the differences in STXS Stage~1.3 bins.

\subsection{Comparison setup and shower-start prescriptions}
\label{sec:lops-setup}

The comparison uses $5\cdot10^5$ charged-lepton events at each order and collision energy,
with PDF4LHC21 and the hard-process scales of \cref{tab:parameters}.  We use the Monash tune
as the baseline and repeat the comparison with CP5 and A14.  The NLO events are showered with
the \POWHEG{} veto of \cref{sec:pythia}.  For each LO shower-start prescription, the same Born
events are showered without that veto.  These variations isolate shower-start effects rather than reproduce the
experiment-inspired vetoed sources of \cref{sec:rwt-samples,sec:rwt-atlas}.

For the colorless Born final state, the default ISR setting
\code{SpaceShower:\allowbreak pTmaxMatch = 0} allows evolution up to the kinematic limit~\cite{Bierlich:2022pfr,Corke:2010zj}.  This is the unvetoed ``power shower''~\cite{Plehn:2005cq}; it is not restricted by the event's \code{SCALUP} value.
Setting \code{pTmaxMatch = 1} instead uses \code{SCALUP} $=\mZH$ as the evolution limit,
optionally reduced by \code{pTmaxFudge}.  We also test smooth damping of hard emissions without
restricting the available phase space~\cite{Corke:2010zj}.  The settings applied to ISR and FSR
are given in \cref{tab:lops-variants}.  The limits refer to the \PYTHIA{} evolution variable,
not directly to the reconstructed leading-jet transverse momentum.

\begin{table*}[tbp]
\centering
\caption{LO+PS shower-start prescriptions, applied to the same Born events without a
\POWHEG{} veto.  The listed settings are applied in both the \code{SpaceShower} and
\code{TimeShower} namespaces.  The damped configuration uses
\code{pTdampFudge = 1} and is evaluated only at 13.6\TeV{} with Monash.}
\label{tab:lops-variants}
\small
\begin{tabular}{lll}
\toprule
Label & Settings & Shower evolution limit \\
\midrule
default & \code{pTmaxMatch = 0} & kinematic limit (colorless Born state) \\
power & \code{pTmaxMatch = 2} & kinematic limit \\
$\mZH$ & \code{pTmaxMatch = 1} & \code{SCALUP} $=\mZH$ \\
$\mZH/2$ & \code{pTmaxMatch = 1}, \code{pTmaxFudge = 0.5} & $\mZH/2=\muF$ \\
$\mZH/4$ & \code{pTmaxMatch = 1}, \code{pTmaxFudge = 0.25} & $\mZH/4$ \\
damped & \code{pTmaxMatch = 0}, \code{pTdampMatch = 1} & kinematic limit, damping scale $\mZH$ \\
\bottomrule
\end{tabular}
\end{table*}

We multiply LO+PS by $K=\sigma_{\mathrm{NLO}}/\sigma_{\mathrm{LO}}$, using the inclusive
cross sections of the corresponding event samples: $K=\pLopsKsix$ at 13.6\TeV{} and
\pLopsKthirteen{} at 13\TeV{}.  With this common normalization, ratios of rescaled LO+PS to
NLO+PS cross sections are also ratios of the corresponding cross-section fractions.
The rescaled LO+PS scale band combines its seven-point shape variations with the inclusive
NLO normalization uncertainty in quadrature, as defined in \cref{sec:scale}.
The shower-start dependence is studied separately from these hard-process scale variations.

\subsection{Boosted spectra and STXS bins}
\label{sec:lops-results}

\begin{figure*}
\centering
\includegraphics[width=\textwidth]{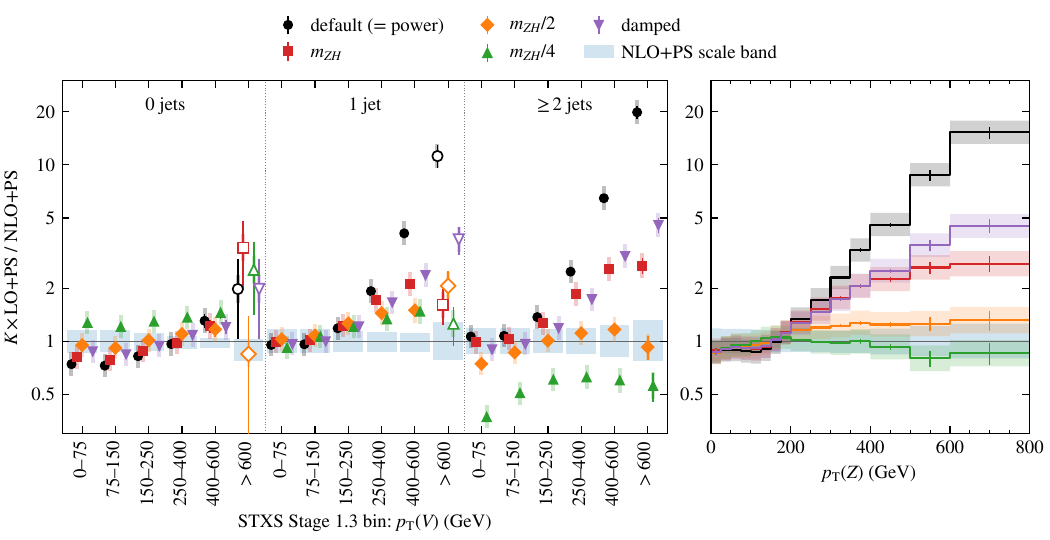}
\caption{Rescaled LO+PS divided by NLO+PS at 13.6\TeV{} with Monash for the unvetoed
prescriptions of \cref{tab:lops-variants}.  Left: the 18 central STXS Stage~1.3 fine bins,
excluding the forward-Higgs bin.  Markers show the ratios with Monte Carlo errors; boxes show
the rescaled LO+PS scale bands of \cref{sec:scale}.  Open markers identify bins with an NLO+PS
Monte Carlo uncertainty above 10\%.  Right: the same comparison per $\ptZ$ bin, with scale
bands and Monte Carlo errors.  The blue band is the NLO+PS seven-point scale envelope relative
to its central value.  The power configuration coincides with the default and is not shown
separately.}
\label{fig:lops}
\end{figure*}

The unvetoed default start produces a growing excess in the boosted region
(\cref{fig:lops,tab:lops}).  Across Monash, CP5, and A14, the ratio of cross-section fractions
above $\ptZ=400\GeV$ is \pLopsDefTunesFoursix{} at 13.6\TeV{}, rising to
\pLopsDefTunesSixsix{} above $600\GeV$.  The corresponding ranges at 13\TeV{} are
\pLopsDefTunesFourthirteen{} and \pLopsDefTunesSixthirteen{}.
In the 18 central STXS fine bins, the excess is concentrated in bins with jets and increases
with $\ptV$.  For Monash it reaches \pLopsDefTwoJCsix{} above $600\GeV$ with at least two
jets, while the zero-jet bins below $400\GeV$ lie below NLO+PS
(ratios \pLopsDefZeroJsix).

Reducing or damping the shower start improves some distributions but does not reproduce all
STXS bins.  Among bins with an NLO+PS Monte Carlo uncertainty below 10\%, the ratios span
\pLopsRangeHalfsix{} for the $\mZH/2$ start and \pLopsRangeDampedsix{} for the damped start.
For the default start, the rescaled LO+PS scale band contains the NLO+PS central prediction
in \pLopsCovDefsix{} bins; the $\mZH/2$ start gives the largest count, \pLopsCovHalfsix{}.
For the default start, none of the \pLopsHighBinssix{} statistically precise bins with jets
and $\ptV>250\GeV$ is covered.  This remains true for that start when using either $K$
times the full LO+PS scale envelope or correlated variations of the LO+PS shape and NLO
normalization.  The hard-process scale band therefore does not account for the observed
shower-start dependence.

\begin{table*}[tbp]
\centering
\caption{Ratio of LO+PS to NLO+PS cross-section fractions above each $\ptZ$ threshold at
13.6\TeV{} with Monash.  With the inclusive rescaling used here, these are also ratios of
rescaled LO+PS to NLO+PS cross sections in the selected regions.  The shower-start prescriptions
are those of \cref{tab:lops-variants}; the fractions include all Higgs rapidities.}
\label{tab:lops}
\small
\input{tab-lops.tex}
\end{table*}

\subsection{Origin of the excess and implications for corrections}

\Cref{tab:effect} separates three stages: the independent fixed-order calculation, the generated
events before showering, and the same events after showering.  The LO events reproduce the
fixed-order spectrum.  The NLO events already include the \POWHEG{} emission and differ from
fixed-order NLO by the matching effects discussed in \cref{sec:validation}.
Comparing each event sample before and after showering isolates the additional shower effect.

\begin{figure}[tbp]
\centering
\includegraphics{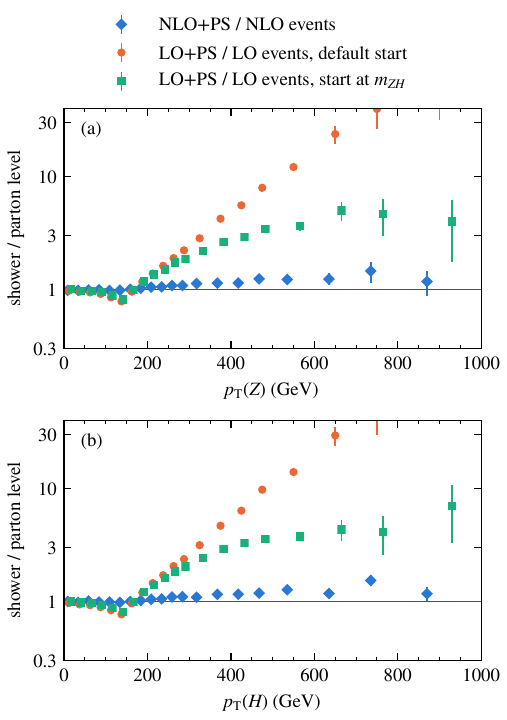}%
\caption{Normalized (a)~$\ptZ$ and (b)~$\ptH$ distributions after showering divided by those
before showering, using the same events at 13.6\TeV{} with Monash.  Curves show NLO events
with the matched shower and LO events with the unvetoed default and $\mZH$ starts.
The plotted errors combine the two histogram Monte Carlo errors in quadrature; the covariance
between the shared-event distributions is not included.}
\label{fig:shower-effect}
\end{figure}

Above $\ptZ=400\GeV$, showering raises the NLO event fraction by about a fifth, whereas the
unvetoed default LO shower increases it by more than a factor of eight (\cref{tab:effect}).
The $\mZH/4$ start reduces the LO shower enhancement to a level closer to the NLO one, but
agreement in this projection does not imply agreement in the jet bins.
\Cref{fig:shower-effect} shows the same contrast in the normalized $\ptZ$ and $\ptH$ spectra:
the default LO shower hardens the tails much more than the matched NLO shower.

The large default-start excess is thus generated when showering the Born events; it is not
removed by correcting the inclusive normalization.  This behavior is consistent with
extrapolating soft and collinear radiation kernels into hard-recoil regions, where they do not
replace the process-specific real-emission matrix element~\cite{Amoroso:2020lgh,Jones:2017giv}.
The NLO+PS reference includes that matrix element for the hardest emission, although its
matching and shower uncertainties remain relevant (\cref{sec:matching}).

\begin{table*}[tbp]
\centering
\caption{Cross-section fractions (in \%) above each $\ptZ$ threshold at 13.6\TeV{} with Monash,
including all Higgs rapidities.  Upper part: fixed-order predictions, generated events before
showering, and showered predictions.  Lower part: after-shower divided by before-shower
fractions, using the same underlying event sample in each ratio.}
\label{tab:effect}
\small
\input{tab-effect.tex}
\end{table*}

\paragraph{Invariant mass and recoil.}  For the tested configurations at 13.6\TeV{}, the
largest relative change in the normalized $\mZH$ distribution after showering is
\pMzhShiftsix{}, evaluated in bins containing more than 1\% of the before-shower cross section.  This does not by itself exclude an improvement in $\ptZ$
from a weight depending only on $\mZH$: correlations between the two variables allow such a
weight to change the marginal $\ptZ$ distribution.  A mass-only weight does, however, leave
the conditional recoil distribution at fixed $\mZH$ unchanged.  It therefore cannot repair a
discrepancy in that conditional distribution.  This distinction motivates the tests of
radiation-sensitive reweighting variables in \cref{sec:rwt}.

%% file: tab-lops.tex
\begin{tabular}{lccccc}
\toprule
LO shower start & $>75$ & $>150$ & $>250$ & $>400$ & $>600\GeV$ \\
\midrule
default $=$ power & 1.03 & 1.26 & 2.27 & 6.11 & 17.35 \\
$\mZH$ & 1.02 & 1.19 & 1.68 & 2.33 & 2.52 \\
$\mZH/2$ & 1.02 & 1.11 & 1.22 & 1.24 & 1.19 \\
$\mZH/4$ & 1.02 & 1.04 & 0.99 & 0.90 & 0.79 \\
damped & 1.02 & 1.18 & 1.66 & 2.79 & 4.33 \\
\bottomrule
\end{tabular}

%% file: tab-effect.tex
\begin{tabular}{lccccc}
\toprule
& $>75$ & $>150$ & $>250$ & $>400$ & $>600\GeV$ \\
\midrule
LO, fixed order & 81.8 & 29.0 & 3.15 & 0.175 & 0.00909 \\
LO events, parton level & 81.9 & 28.9 & 3.16 & 0.179 & 0.00859 \\
NLO, fixed order & 80.4 & 27.8 & 3.26 & 0.206 & 0.0132 \\
NLO events, parton level & 80.4 & 29.2 & 3.45 & 0.211 & 0.0132 \\
NLO+PS & 80.7 & 30.1 & 3.82 & 0.250 & 0.0167 \\
LO+PS, start default $=$ power & 82.7 & 38.1 & 8.67 & 1.53 & 0.289 \\
LO+PS, start $\mZH/4$ & 81.9 & 31.3 & 3.76 & 0.225 & 0.0132 \\
LO+PS, start damped & 82.3 & 35.6 & 6.34 & 0.698 & 0.0723 \\
\midrule
NLO+PS / NLO events & 1.00 & 1.03 & 1.10 & 1.19 & 1.26 \\
LO+PS / LO events, start default $=$ power & 1.01 & 1.32 & 2.74 & 8.52 & 33.67 \\
LO+PS / LO events, start $\mZH$ & 1.01 & 1.24 & 2.02 & 3.25 & 4.89 \\
LO+PS / LO events, start $\mZH/2$ & 1.00 & 1.16 & 1.47 & 1.74 & 2.30 \\
LO+PS / LO events, start $\mZH/4$ & 1.00 & 1.08 & 1.19 & 1.26 & 1.53 \\
LO+PS / LO events, start damped & 1.01 & 1.23 & 2.01 & 3.89 & 8.41 \\
\bottomrule
\end{tabular}

%% file: sec-rwt.tex
\section{Reweighting LO+PS to NLO+PS}
\label{sec:rwt}
\providecommand{\ptZH}{\ensuremath{p_{\mathrm{T}}(ZH)}}
\providecommand{\ptjone}{\ensuremath{p_{\mathrm{T}}(j_1)}}
\providecommand{\ptjtwo}{\ensuremath{p_{\mathrm{T}}(j_2)}}
\providecommand{\decision}[1]{\textcolor{teal}{[Decision: #1]}}
\providecommand{\rwtfullfig}[1]{\ifdim\columnwidth<\textwidth\includegraphics[width=\textwidth]{#1}%
\else\includegraphics[width=\textwidth,height=0.7\textheight,keepaspectratio]{#1}\fi}

An event-dependent correction can transfer NLO+PS information to existing LO+PS samples without
repeating their detector simulation.  We construct corrections for specified CMS- and ATLAS-style
$gg\to ZH$ samples in the charged-lepton and neutrino channels.  Each correction has its own
trained ensemble and calibration, together with target-theory variations and uncertainties
in the transfer.
Independent samples test the corrected boson distributions and STXS yields, while the
source population in hard-recoil regions determines the principal limitation.

\subsection{CMS-style source and target samples}
\label{sec:rwt-samples}

We first define the CMS-style construction; the ATLAS-style application is presented in
\cref{sec:rwt-atlas}.  The CMS-style source and target each contain $10^7$ events per
decay channel at $13.6\TeV$.
The CMS-style LO source follows the public production card~\cite{genproductions-gghz}:
Born-only \ggxyZH{} generation with NNPDF3.1 NNLO PDFs~\cite{Ball:2017nwa},
$\muR=\muF=m_Z+m_H$, and \code{SCALUP} $=\mZH$.
The charged-lepton mass window is $15<m_{\ell\ell}<150\GeV$.
Showering uses \PYTHIA{} with CP5~\cite{CMS:2019csb} and the CMS \POWHEG{} interface
~\cite{cmssw-powhegveto}: \code{veto = 1}, \code{pThard = 0}, \code{pTdef = 1},
\code{vetoCount = 100}, and \code{pTmaxMatch = 2} for ISR and FSR.
The shower explores the full phase space, but emissions harder than \code{SCALUP} in the
\POWHEG{} measure are vetoed.  This is distinct from the unvetoed shower-start comparisons
of \cref{sec:lops}.

The NLO+PS target uses the production setup of \cref{sec:events}, including Born suppression
with $L=\bornsuppL\GeV$ and the hard-process scale $\muR=\muF=\mZH/2$.
We select its NNPDF3.1 central weight and use the same CP5 tune and shower-interface settings
as for the source.  Both the perturbative order and the hard-process scale therefore change between source and target.

The $Z$ and Higgs momenta are taken from their final event-record copies.  The $Z$ decay
products are ordered in transverse momentum before final-state radiation.
Jets use anti-$k_{\mathrm{T}}$ clustering~\cite{Cacciari:2008gp} with $R=0.4$ and
$p_{\mathrm{T}}>30\GeV$, excluding all $Z$ and Higgs decay descendants.
The STXS jet count has no rapidity restriction; $j_1$ and $j_2$ are instead the leading jets
within $|\eta|<4.7$.  This latter definition is used for the jet inputs and
$\rho=\ptjone/\mZH$.

Whole event files are assigned in a fixed pseudo-random order to disjoint training,
calibration, and evaluation samples.  Calibration and evaluation each use $10^6$ events
per class; the remaining files form the training pool.  A further confirmation sample,
described in \cref{sec:rwt-results}, tests the residual-uncertainty prescription.

\subsection{Phase-space domain}
\label{sec:rwt-hole}

The source shower veto suppresses hard recoil, although it does not impose a strict bound
on reconstructed jet transverse momentum.  In the charged-lepton samples, events with
$\rho>1$ account for \rbdHJNloPtz\% of the NLO+PS cross section selected by
$\ptZ>400\GeV$, but only \rbdHJLoPtz\% of LO+PS.  For $\ptH>450\GeV$, the shares are
\rbdHJNloPth\% and \rbdHJLoPth\%, respectively.  The source is severely depleted, but not empty, in this region.

We restrict the learned correction to $\rho<1$ or no jet within the acceptance used to
define $j_1$.  The same restriction is applied to both training classes.
Only \rbdOutNlo\% of the inclusive NLO+PS cross section and \rbdOutLo\% of LO+PS lie
outside this domain in the charged-lepton evaluation sample, but the boosted selections
above enhance their importance.  Outside the domain, the nominal additional weight remains
one, retaining LO+PS$\times K$; a separate variation describes the population difference
(\cref{sec:rwt-unc}).  This separates the correction in well-populated regions from the
uncertainty associated with sparse hard recoil.

\subsection{Nominal correction}
\label{sec:rwt-methods}

We estimate the NLO+PS-to-LO+PS density ratio with gradient-boosted classifiers
~\cite{Cranmer:2015bka,Rogozhnikov:2016bdp,Andreassen:2019nnm}, using the 17 inputs in
\cref{tab:rwt-inputs}.  With equal total training weight for the two classes, the classifier
odds $p_i(x)/[1-p_i(x)]$ estimate the density ratio in the training domain.
Twenty classifiers use separately drawn subsets of $10^6$ events per class from the training
pool and different random seeds; their subsets may overlap.  LO events carry equal training
weights, while NLO generator weights enter the loss with negative values set to zero.
This positive-weight training approximation is assessed through closure against the NLO+PS
reference.

The classifiers use \code{HistGradientBoosting}\allowbreak\code{Classifier} in
\code{scikit-learn}~1.2.2~\cite{Pedregosa:2011sk}, with binary cross entropy, learning rate
0.05, at most 63 leaves, at least 500 events per leaf, and $L_2$ regularization 1.0.
Training stops after at most 1000 iterations or 20 iterations without improvement on a
20\% internal validation subset.  These validation events belong to the training pool,
not to the independent evaluation sample.

The nominal ensemble combines the odds geometrically,
\begin{equation}
 w_{\mathrm{ens}}(x)=
 \exp\!\left[\frac{1}{20}\sum_{i=1}^{20}
 \ln\frac{p_i(x)}{1-p_i(x)}\right].
\end{equation}
On the calibration sample, the ratio of normalized target and reweighted-source
$\mZH$ distributions defines a multiplicative factor $k(\mZH)$.
Starting from $25\GeV$ bins, adjacent bins are merged until both distributions have
relative statistical uncertainties below 2\%.  For the charged-lepton channel, the
\rbdKgroups{} factors range from \rbdKlo{} to \rbdKhi{}.
The product $k(\mZH)w_{\mathrm{ens}}$ is normalized on the calibration sample to give
unit-mean shape weights, with weight one outside the domain.
The overall cross section is set separately to the NLO value.  Thus the learned correction
changes the kinematics of LO+PS$\times K$, not its nominal inclusive normalization.

\begin{table}[tbp]
\centering
\caption{The 17 classifier inputs, evaluated with the definitions of \cref{sec:rwt-samples}.
Missing-jet inputs take fixed values outside their physical ranges.  In the neutrino channel,
the decay-product inputs refer to neutrinos and $m_{\ell\ell}$ is replaced by $m_{\nu\bar\nu}$.}
\label{tab:rwt-inputs}
\small
\begin{tabular}{@{}lp{0.36\columnwidth}@{}}
\toprule
Inputs & Kinematic information \\
\midrule
$\mZH$, $\ptZH$, $|y_{ZH}|$ & mass, recoil, longitudinal boost \\
$\ptZ$, $N_{\mathrm{jets}}$ ($0,1,2,\geq3$) & STXS kinematics \\
$\Delta\phi(Z,ZH)$ & azimuthal recoil geometry \\
$m_{\ell\ell}$ & $Z$ virtuality \\
$\ptjone$, $\eta(j_1)$, $\Delta R(j_1,ZH)$ & leading jet \\
$\ptjtwo$, $\eta(j_2)$, $\Delta R(j_2,ZH)$ & second jet \\
$p_{\mathrm{T}}(\ell_{1,2})$, $\eta(\ell_{1,2})$ & $Z$ decay products \\
\bottomrule
\end{tabular}
\end{table}

\subsection{Uncertainty prescription}
\label{sec:rwt-unc}

We distinguish uncertainty in the transfer from theory uncertainty in the target prediction.
The transfer prescription combines classifier variation, residual leading-jet dependence,
and a binwise residual term.  These components are treated as separate nuisance sources
and combined in quadrature, not as a product of simultaneous weight shifts.

\paragraph{Classifier ensemble.}
Each calibrated classifier supplies alternative event weights that are propagated through
the analysis selection.  Following Ref.~\cite{Palmer:2025wrq}, the resulting histogram bins
are standardized across the twenty predictions before principal-component analysis.
Components accounting for 99\% of the standardized variance are retained and transformed
back to bin contents.  Each defines a correlated shape variation; the compression is
performed at histogram level, rather than assigning independent classifier errors to events.

\paragraph{Leading-jet dependence.}
Inside the domain, the calibration-sample closure ratio $c(\rho)$ defines the alternatives
$w/c$ and $w\,c$.  Its bins are merged until the relative statistical uncertainty is below
5\%; the charged-lepton values range from \rbdClo{} to \rbdChi{}.
Outside the domain, the alternatives multiply the nominal unit weight by the target-to-source
ratio of cross-section shares, or its inverse, in bins of $\max(\ptZ,\ptH)$.
This ratio ranges from \rbdShareLow{} below $150\GeV$ to \rbdShareHigh{} above $600\GeV$
in the charged-lepton calibration sample.  It describes the depleted hard-recoil population
in these bins; it does not establish recovery of its multidimensional kinematics.

\paragraph{Residual differences.}
The ensemble spread does not capture biases shared by its members.  Residual differences
on the evaluation sample motivate an additional uncertainty of 1\% of each final histogram
bin content, treated as uncorrelated between bins.  This empirical term is added after
histogramming, not supplied as an event weight.  Its validation on a separate confirmation
sample and the remaining exceptions are reported in \cref{sec:rwt-results}.

\paragraph{Target-theory variations.}
For each seven-point NLO scale variation, the calibration target defines
$T_k=\sum w_k/\sum w$ in cells of $\mZH$ and $\rho$, where $w_k$ and $w$ are the varied
and nominal generator weights.  These factors multiply the corrected source weights.
PDF4LHC21 PDF and $\as$ variations~\cite{PDF4LHC:2022cjn} are transferred in the same cells
from the $5\times10^5$-event NLO+PS sample carrying those weights.
The improved perturbative scale description therefore comes from the NLO+PS reference;
the transfer uncertainties quantify the additional approximation made by reweighting.

\subsection{Validation and improvement over inclusive rescaling}
\label{sec:rwt-results}

\begin{figure*}[tbp]
\centering
\rwtfullfig{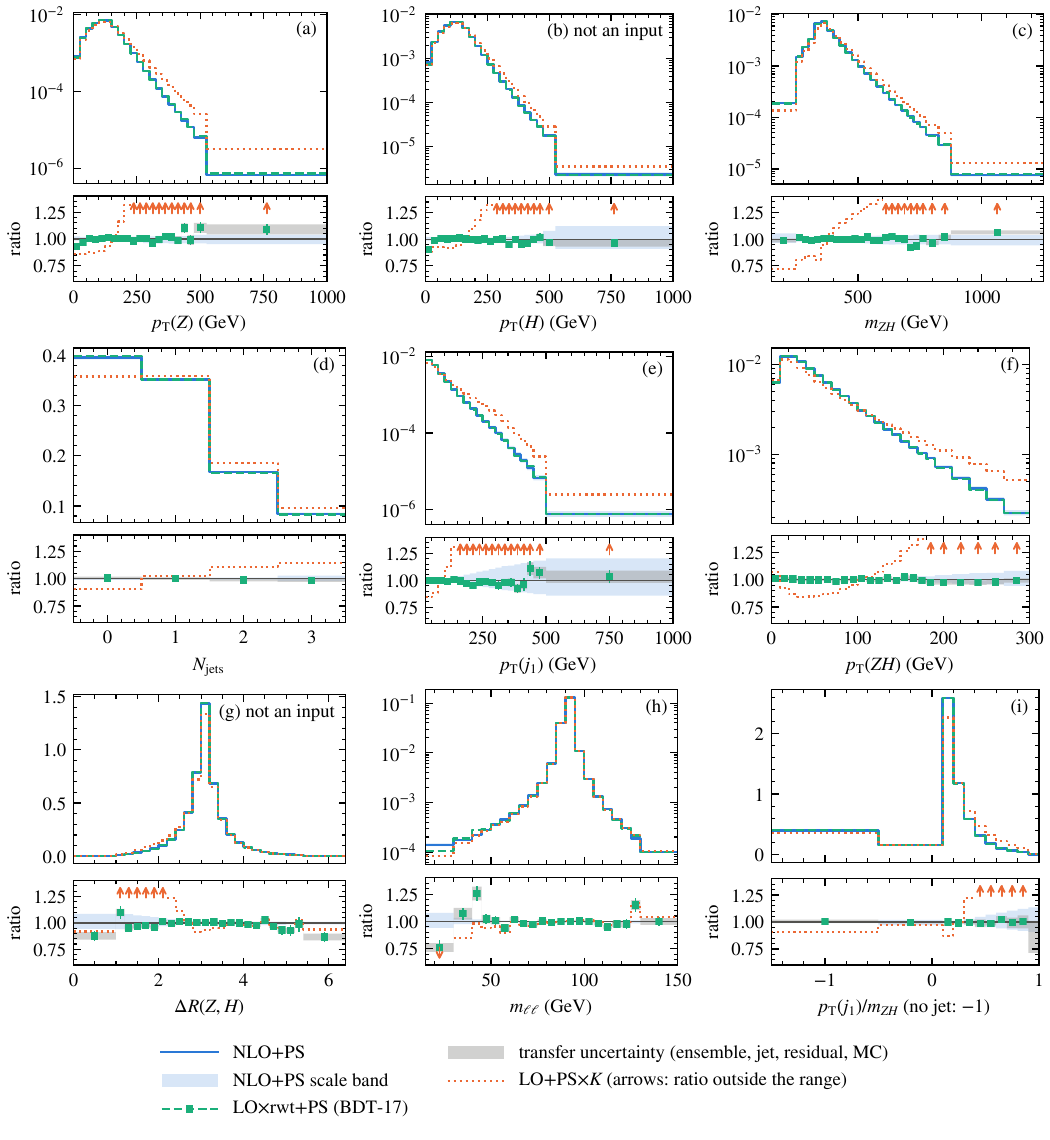}
\caption{CMS-style charged-lepton evaluation sample inside the training domain.
Upper panels: normalized NLO+PS (solid), LO+PS$\times K$ (dotted), and reweighted LO+PS
(dashed) distributions.  The NLO+PS band shows shape-only scale variations.
Lower panels: ratios to NLO+PS; bars combine source and target statistical errors, while
the gray band combines ensemble, leading-jet, residual, and reweighted-source statistical
uncertainties.  Arrows indicate LO+PS$\times K$ ratios outside the displayed range.
Adjacent presentation bins are merged to contain at least 1000 source events.  Panels (b) and (g) are not direct classifier inputs.}
\label{fig:rwt-closure}
\end{figure*}

\begin{figure*}[tbp]
\centering
\rwtfullfig{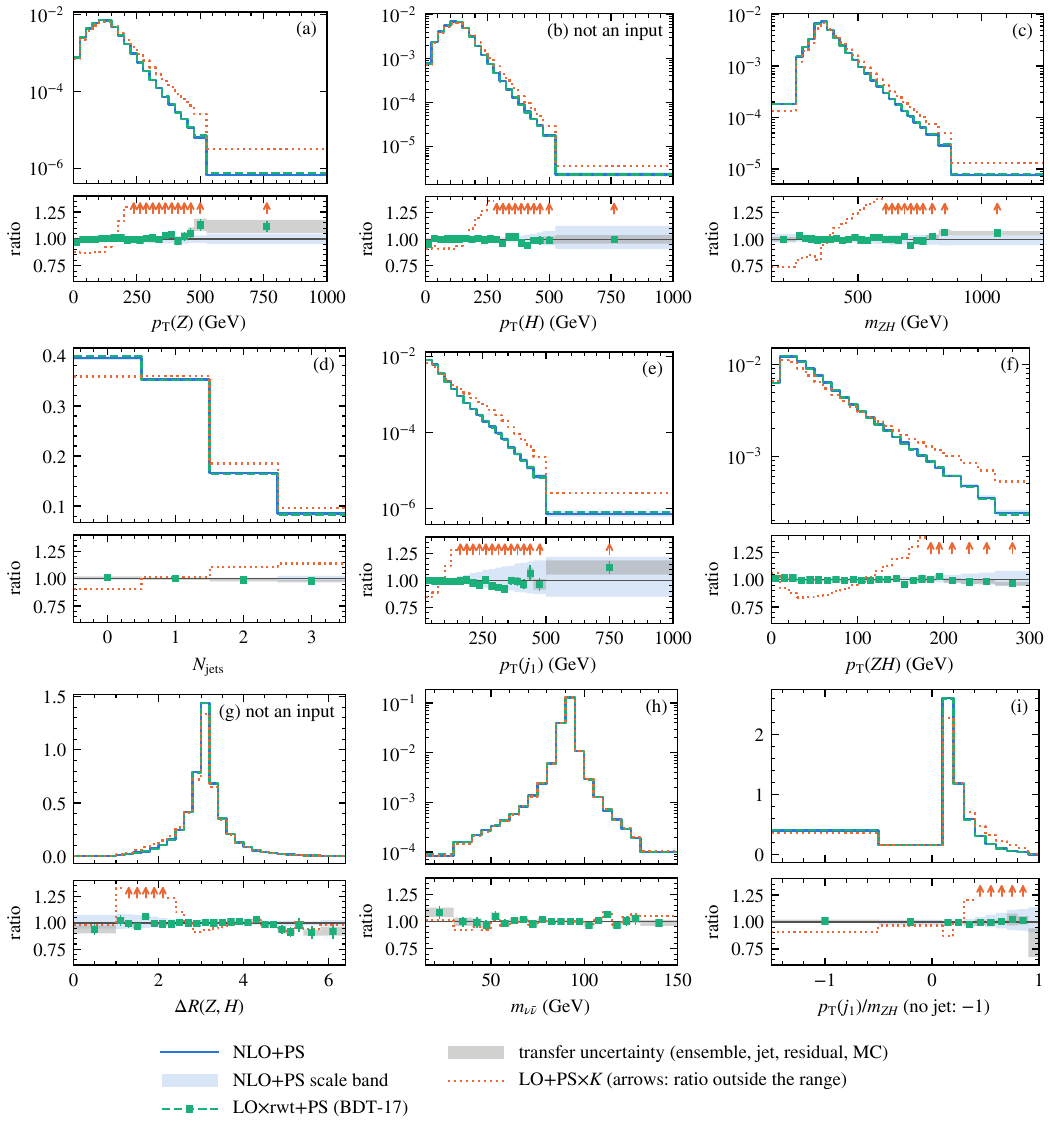}
\caption{As \cref{fig:rwt-closure}, for the independently trained CMS-style neutrino correction;
panel (h) shows $m_{\nu\bar\nu}$.}
\label{fig:rwt-closure-vv}
\end{figure*}

On the independent evaluation samples, reweighting brings the tested boson and jet
observables substantially closer to NLO+PS than inclusive rescaling alone
(\cref{fig:rwt-closure,fig:rwt-closure-vv}).  These shape comparisons are restricted to the
training domain.  Agreement within the transfer uncertainty holds over most of the tested
bins, including $\ptH$ and $\Delta R(Z,H)$, which are not direct classifier inputs.
The effective sample size after reweighting is about \rbdNeff\% of the event count in
both decay channels.

A further $10^6$ events per class, disjoint from training, calibration, and the sample used
to choose the residual term, test the uncertainty prescription.
For bins with at least 100 source events, pulls combine the source and target Monte Carlo
errors with the transfer uncertainty; target scale uncertainty is excluded.
The diagnostic requires at most 5\% of bins beyond two standard deviations and none beyond
three.  The charged-lepton correction satisfies it: \rbdConfTwoFrac\% of \rbdConfBins{}
bins exceed two, with a maximum pull of \rbdConfMax{}.
For neutrinos, \rbvConfTwoFrac\% of \rbvConfBins{} bins exceed two, but one reaches
\rbvConfMax{}; the strict criterion is therefore not satisfied in that channel.
These checks support the uncertainty prescription for most tested bins, rather than
unconditional closure under arbitrary selections.

\begin{figure*}[tbp]
\centering
\rwtfullfig{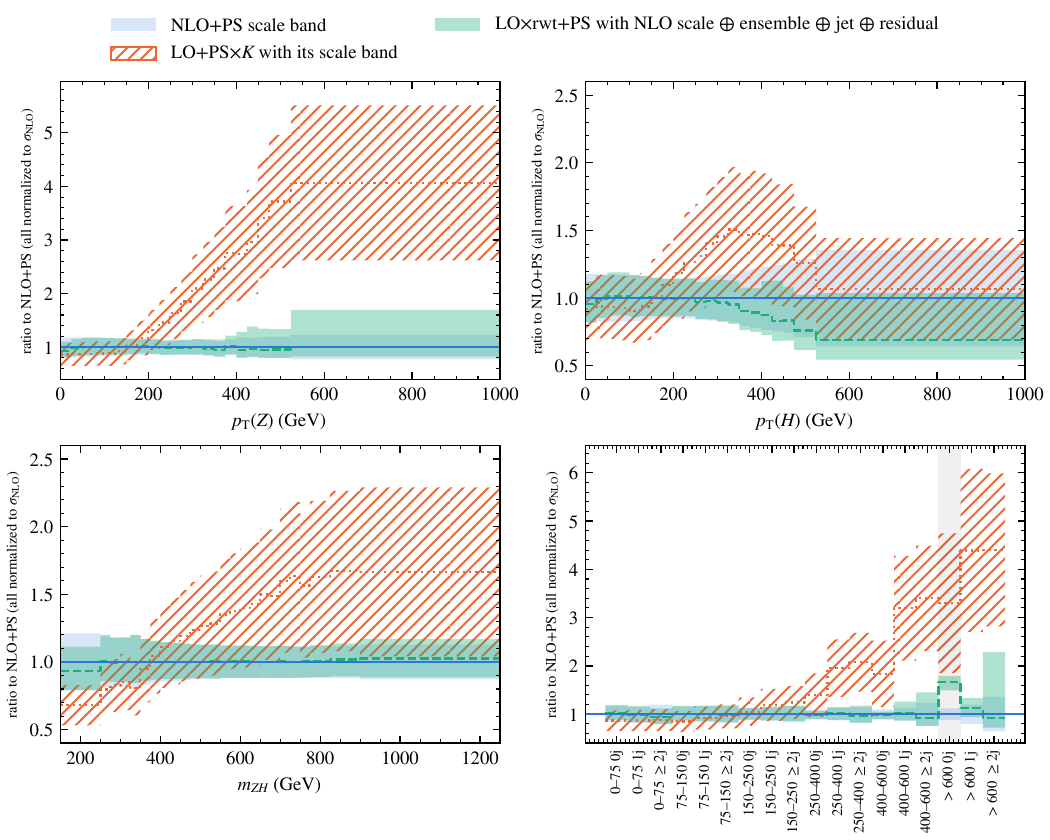}
\caption{CMS-style charged-lepton calibration sample in the full phase space, with all
nominal predictions normalized to $\sigma_{\mathrm{NLO}}$ and divided by NLO+PS.
Variations retain their changes in the total rate.  Hatched: LO+PS$\times K$ with the
original LO seven-point scale envelope, symmetrized to its larger excursion.
Shaded: reweighted LO+PS with transported NLO scale, ensemble, leading-jet, and 1\% residual
uncertainties; outside the domain its nominal correction is one.
The reweighted band retains separate upward and downward excursions.
Light: the symmetrized NLO+PS scale envelope.  No Monte Carlo statistical uncertainty is included.}
\label{fig:rwt-scaleband}
\end{figure*}

\begin{figure*}[tbp]
\centering
\rwtfullfig{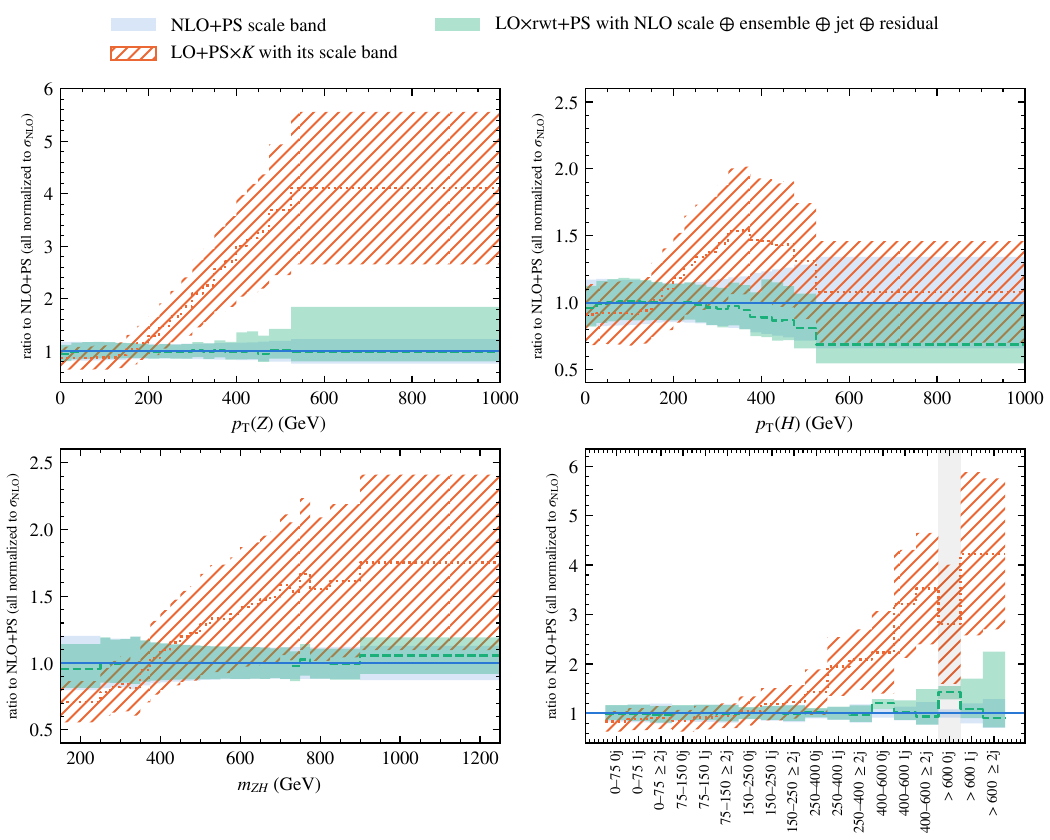}
\caption{As \cref{fig:rwt-scaleband}, for CMS-style $Z\to\nu\bar\nu$.}
\label{fig:rwt-scaleband-vv}
\end{figure*}

\Cref{fig:rwt-scaleband,fig:rwt-scaleband-vv} compare the full-phase-space predictions and
uncertainty sizes on the calibration samples, retaining rate changes in every variation.
Here the LO+PS$\times K$ band varies the original LO scale weights at fixed nominal $K$;
it is not the shape-plus-NLO-normalization prescription of \cref{sec:scale}.
The reweighted band combines transported NLO scale variations with the three transfer
components.  Neither band includes Monte Carlo statistics.  PDF, $\as$, mass-scheme,
and matching variations are not included in this comparison, so these bands are not the
complete theory budget of \cref{sec:uncertainties}.

The median uncertainties in \cref{tab:rwt-unc-summary} are smaller after reweighting,
while the central predictions move towards NLO+PS, except in the highest $\ptH$ bins discussed below.  The transferred prediction is narrower
in \rbdSmallerPtz--\rbdSmallerMzh\% of the bins across these observables; the large
outside-domain variation limits the improvement in the boosted bins above about $400\GeV$.
The neutrino results give similar median sizes, within one percentage point.

\begin{table}[tbp]
\centering
\caption{Median relative uncertainty sizes (\%) in the charged-lepton calibration sample.
For each bin, a component's size is its larger absolute excursion; components are combined
in quadrature.  ``Transfer'' includes ensemble, leading-jet, and residual terms and is
already included in the reweighted column.  These medians summarize the bands in
\cref{fig:rwt-scaleband}, not symmetric confidence intervals.}
\label{tab:rwt-unc-summary}
\small\setlength{\tabcolsep}{4pt}
\begin{tabular}{lccc}
\toprule
Observable & \shortstack{LO+PS$\times K$\\LO scale} &
\shortstack{Reweighted\\NLO scale\\+ transfer} & Transfer \\
\midrule
$\ptZ$ & \rbdLoPtz & \rbdRwPtz & \rbdMethPtz \\
$\ptH$ & \rbdLoPth & \rbdRwPth & \rbdMethPth \\
$\mZH$ & \rbdLoMzh & \rbdRwMzh & \rbdMethMzh \\
STXS & \rbdLoStxs & \rbdRwStxs & \rbdMethStxs \\
\bottomrule
\end{tabular}
\end{table}

The high-$\ptZ$ tail remains a limitation.  Above $450\GeV$, the in-domain reweighted
cross section exceeds NLO+PS by \rbdExcessLo--\rbdExcessHi\% for charged leptons and
\rbvExcessLo--\rbvExcessHi\% for neutrinos on the evaluation and confirmation samples.
For the full-phase-space calibration comparison, the corresponding ratios are
\rbdFullRw{} and \rbvFullRw{}, whereas LO+PS$\times K$ gives \rbdFullLoK{} and
\rbvFullLoK{}, respectively.  The LO discrepancy lies far outside its scale band
($-\rbdFullLoKDn/{+}\rbdFullLoKUp\%$ for charged leptons).
The near-unity full-phase-space ratios partly reflect compensation between an in-domain
excess and an outside-domain deficit.  They demonstrate improvement in this projection,
not closure of its separate recoil components.

The $\ptH$ tail shows the missing hard-recoil population directly.  Inside the training
domain, the reweighted cross section above $\ptH=\rbdPthEdge\GeV$ is \rbdPthInRw{} of NLO+PS
for charged leptons (\rbvPthInRw{} for neutrinos), whereas LO+PS$\times K$ is \rbdPthInLoK{}
(\rbvPthInLoK) times NLO+PS.  In the full phase space, however, the reweighted prediction falls
below NLO+PS above about $350\GeV$ and reaches \rbdPthRw{} (\rbvPthRw) above
$\rbdPthEdge\GeV$.  Above $\ptH=450\GeV$, \rbdHJNloPth\% of the charged-lepton NLO+PS cross
section has $\rho>1$: the Higgs boson recoils against a jet harder than $\mZH$, while the $Z$
boson remains softer.  The vetoed source contains only \rbdHJLoPth\% of such events, and their
weight remains one outside the domain, so reweighting cannot restore this contribution.  The
outside-domain variation, binned in $\max(\ptZ,\ptH)$, reaches NLO+PS in this bin (upper edges
\rbdPthUp{} and \rbvPthUp).  LO+PS$\times K$ lies closer to NLO+PS there (\rbdPthLoK{} and
\rbvPthLoK) only because its excess inside the domain offsets the same missing events.

\subsection{CMS-style neutrino channel}
\label{sec:rwt-vv}

The neutrino correction uses independent $10^7$-event source and target samples, with
$15<m_{\nu\bar\nu}<150\GeV$.  Training, calibration, and uncertainty construction are
performed separately with the same prescription, except for the PDF and $\as$ ratios, which
are taken from the charged-lepton sample.  Its inputs use generator-level neutrino
momenta; it is not obtained by substituting reconstructed missing transverse momentum
into the charged-lepton correction.  The parallel results above test the method in the
two decay channels while retaining their distinct residual limitations.

\subsection{ATLAS-style sources}
\label{sec:rwt-atlas}

We test the same reweighting prescription with ATLAS-style sources and separately trained
corrections for both decay channels.  The source follows the public ATLAS job options
~\cite{atlas-ggzh-joboptions}: Born-only \ggxyZH{} generation with PDF4LHC21,
$\muR=\muF=H_{\mathrm{T}}$, and \code{SCALUP} $=\mZH$, followed by the A14 shower
~\cite{ATL-PHYS-PUB-2014-021} with the ATLAS \POWHEG{} veto.
The target uses the NLO production setup, with $\muR=\muF=\mZH/2$, the PDF4LHC21
central weight, and A14 showering.  This tests the prescription under different
hard-process and shower settings, rather than transferring the CMS-trained models.

Training, calibration, the domain restriction, and uncertainty construction follow
\cref{sec:rwt-methods,sec:rwt-unc}.  The 1\% residual is retained without adjustment
and was fixed before examining the ATLAS-style evaluation samples.
The charged-lepton samples contain \rbaLoEv{} million source and \rbaNloEv{} million
target events; the neutrino samples contain \rbnLoEv{} and \rbnNloEv{} million.
The smaller training pools imply expected pairwise overlaps of
\rbaSharedLo--\rbaSharedHi\% and \rbnSharedLo--\rbnSharedHi\%, respectively,
compared with \rbdSharedLo--\rbvSharedHi\% for the CMS-style ensembles.

On the independent evaluation samples, fewer than 5\% of the tested bins have pulls
larger than two standard deviations in magnitude (\cref{tab:rwt-atlas}).
Each channel nevertheless has one bin beyond three, with maximum absolute pulls of
\rbaCovMax{} and \rbnCovMax{}; neither satisfies the strict diagnostic of
\cref{sec:rwt-results}.  The pull means are \rbaPullMean{} and \rbnPullMean{},
with standard deviations \rbaPullRms{} and \rbnPullRms{}.
These diagnostics test the combined transfer prescription, including the leading-jet
and residual terms, not the ensemble spread alone.
The in-domain tail ratios in \cref{tab:rwt-atlas} and the distributions in
\cref{app:rwt-atlas} show the performance on events excluded from training and calibration.

\begin{table}[tbp]
\centering
\caption{ATLAS-style corrections in the charged-lepton and neutrino channels.
The upper block uses independent evaluation samples inside the training domain;
pulls are evaluated in bins with at least 100 source events.
$R_{450}$ is the reweighted-to-NLO+PS ratio for $\ptZ>450\GeV$, with Monte Carlo errors.
The lower block gives median relative uncertainty sizes (\%) in $\ptZ$ on the
full-phase-space calibration samples, defined as in \cref{tab:rwt-unc-summary}.}
\label{tab:rwt-atlas}
\small\setlength{\tabcolsep}{4pt}
\begin{tabular}{@{}lcc@{}}
\toprule
Quantity & $Z\to\ell^-\ell^+$ & $Z\to\nu\bar\nu$ \\
\midrule
Tested bins & \rbaCovBins & \rbnCovBins \\
$|\mathrm{pull}|>2$ (\%) & \rbaCovTwoFrac & \rbnCovTwoFrac \\
Maximum $|\mathrm{pull}|$ & \rbaCovMax & \rbnCovMax \\
$R_{450}$ & \rbaTailA & \rbnTailA \\
\midrule
LO+PS$\times K$: LO scale & \rbaLoPtz & \rbnLoPtz \\
Reweighted: NLO scale + transfer & \rbaRwPtz & \rbnRwPtz \\
\bottomrule
\end{tabular}
\end{table}

\begin{figure*}[tbp]
\centering
\rwtfullfig{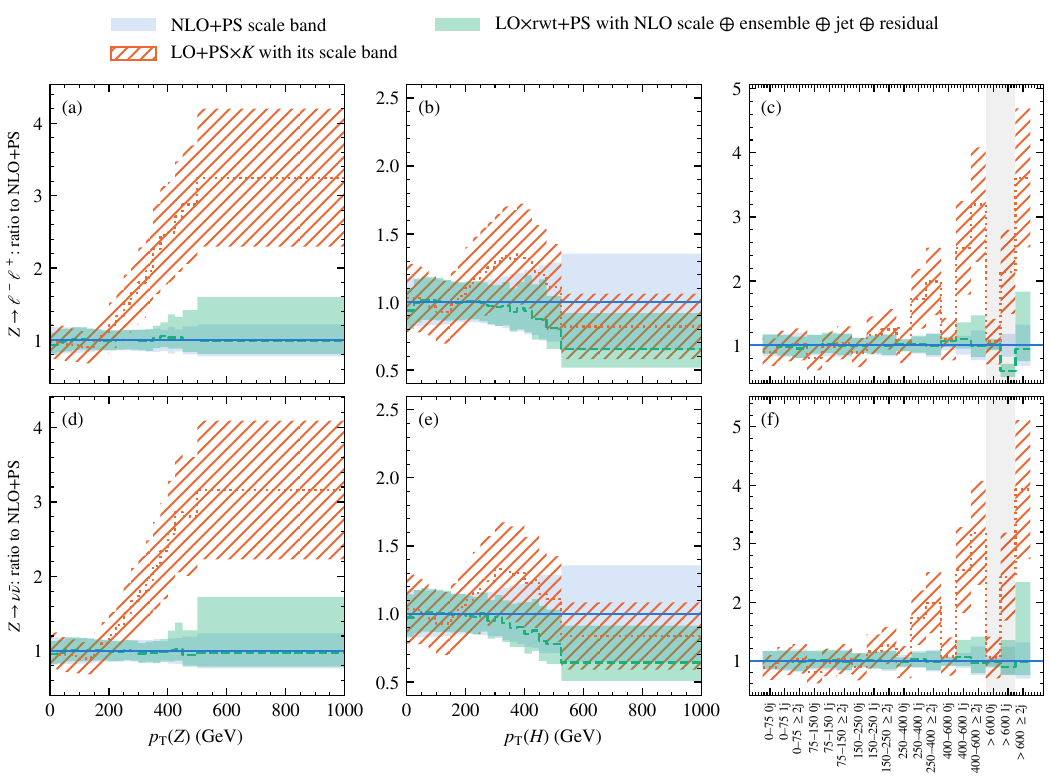}
\caption{ATLAS-style calibration samples in the full phase space:
(a--c) charged leptons and (d--f) neutrinos, for $\ptZ$, $\ptH$, and the STXS bins.
Normalization and uncertainty definitions follow \cref{fig:rwt-scaleband}.
The reweighted bands combine transported NLO scale variations with transfer uncertainties;
Monte Carlo statistics and the other target-theory uncertainties are not included.
The highest $\ptH$ bin retains an uncovered deficit in both channels.}
\label{fig:rwt-atlas}
\end{figure*}

In the full-phase-space calibration comparison (\cref{fig:rwt-atlas}), the median
$\ptZ$ uncertainty decreases from \rbaLoPtz\% to \rbaRwPtz\% for charged leptons
and from \rbnLoPtz\% to \rbnRwPtz\% for neutrinos.
The transfer contribution has median size \rbaMethPtz\% and \rbnMethPtz\%, respectively.
These compare the original LO scale band at fixed $K$ with transported NLO scale
variations plus transfer uncertainties, not the complete theory budget.
Above $\ptZ=450\GeV$, the full-phase-space reweighted-to-NLO+PS ratios are
\rbaFullRw{} and \rbnFullRw{}, compared with \rbaFullLoK{} and \rbnFullLoK{} for
inclusive rescaling alone.  As in the CMS-style comparison, agreement in this projection
does not establish closure of its separate recoil components.

The high-$\ptH$ deficit is more restrictive.  In the last displayed bin, beginning at
$\rbaPthEdge\GeV$, the full-phase-space reweighted predictions are only
\rbaPthRw{} and \rbnPthRw{} of NLO+PS for charged leptons and neutrinos.
Even the upper edges of the combined scale-plus-transfer bands reach only
\rbaPthUp{} and \rbnPthUp{}.  Thus the present prescription does not cover the
NLO+PS central prediction in that bin.  The outside-domain variation must not be
interpreted as a general correction for the depleted hard-recoil population.

\subsection{Application and model release}
\label{sec:rwt-application}

For samples matching one of the stated source configurations, its channel-specific correction and alternative
weights can be propagated to reconstructed analysis histograms using their generator-level
inputs.  The ensemble variations are then compressed for the chosen histograms, and the
binwise residual is added separately.  The calibration normalization is retained when
applying analysis selections, so that the correction can change acceptance as well as shape.

The method improves the tested boson distributions and STXS yields relative to inclusive
rescaling, but does not replace NLO+PS generation for analyses sensitive to the depleted
hard-recoil configurations.  In particular, agreement in a jet-inclusive projection does
not validate a selection that isolates or vetoes those configurations.  The uncovered
ATLAS-style Higgs-tail deficit provides an explicit example.
Applications with different source settings, selections, or binning require their own
closure checks; the empirical residual is not a universal error bound.

For each source configuration and decay channel, the accompanying data record (\cref{app:reproduction}) contains the
twenty ONNX classifiers, calibration and theory-transfer tables, and an inference script
for the nominal and alternative event weights.  It also provides the histogram-level
principal-component procedure, reference events with expected weights, and a model card
specifying the source, target, inputs, domain, normalization, and residual prescription.

%% file: sec-summary.tex
\section{Recommendations and conclusions}
\label{sec:conclusions}

For the tested LO+PS samples, an inclusive $K$ factor does not reproduce the NLO+PS
prediction in boosted and jet-dependent $gg\to ZH$ production.  Using the public \ggxyZH{} NLO+PS implementation, we have evaluated
inclusive charged-lepton and neutrino predictions at 13 and 13.6\TeV{}, together with
differential and STXS Stage~1.3 predictions for the charged-lepton channel.
At 13.6\TeV{}, the NLO correction reduces the inclusive seven-point scale band from
\pLOsixband{} to \pNLOsixband{}.  Its kinematic dependence and the contribution of hard
real radiation also change the distributions used in LHC measurements.

Scale variation remains the largest inclusive uncertainty, but the hierarchy changes in
the boosted region.  At 13.6\TeV{}, the top-quark mass-scheme band is largest in the
central-Higgs STXS bins from $\ptV=\pStxsMtFromsix\GeV$ onward with Monte Carlo errors
below 5\%, reaching \pStxsMtHighsix\% (\cref{sec:stxs}).  These bands use fixed-order
mass-scheme ratios transferred to NLO+PS, with the same ratio for every jet multiplicity
in a given $\ptV$ interval.  They are not direct $\overline{\mathrm{MS}}$ NLO+PS variations.
Scale, PDF+$\as$, mass-scheme, and matching and shower effects must therefore be assessed
separately rather than represented by the inclusive scale band alone.

The LO+PS comparison isolates a strong dependence on the treatment of shower radiation.
With the unvetoed default start at the kinematic limit, inclusive rescaling overestimates
the cross section above $\ptZ=600\GeV$ by factors of \pLopsDefTunesSixsix{} at
13.6\TeV{} across the three tunes.  Restricting or damping the shower start reduces the
excess, but none of the tested prescriptions describes all statistically precise STXS
bins within its scale band (\cref{sec:lops}).  This controlled comparison is distinct
from the experiment-inspired vetoed sources used for reweighting.

For the specified CMS- and ATLAS-style sources at 13.6\TeV{}, we provide separately trained
charged-lepton and neutrino corrections with calibration and transfer uncertainties.
They improve the tested boson distributions and STXS yields over inclusive rescaling
in most of the populated phase space.  In the charged-lepton calibration comparisons,
the median STXS uncertainty changes from \rbdLoStxs\% to \rbdRwStxs\% for CMS-style
samples and from \rbaLoStxs\% to \rbaRwStxs\% for ATLAS-style samples.
These compare the original LO scale band at fixed $K$ with transported NLO scales plus
transfer uncertainties, not the complete theory budget (\cref{sec:rwt-results,sec:rwt-atlas}).
The improvement therefore combines the NLO reference prediction with an additional,
separately evaluated uncertainty in the transfer.

Independent evaluation and confirmation samples test this prescription beyond calibration.
The charged-lepton CMS-style correction satisfies the stated pull diagnostic on its confirmation sample, while
CMS-style neutrinos and both ATLAS-style channels retain bins beyond its strict threshold.
For ATLAS-style samples, the 1\% residual was fixed before evaluation.  The hard-recoil limitation is also visible directly in the
full-phase-space Higgs spectrum.  In the last ATLAS-style $\ptH$ bin, beginning at
$\rbaPthEdge\GeV$, the corrected charged-lepton and neutrino predictions are
\rbaPthRw{} and \rbnPthRw{} of NLO+PS; even the upper band edges reach only
\rbaPthUp{} and \rbnPthUp{}.  Near-unity ratios in other projections do not remove
this deficit or establish closure of the separate recoil components.

For new simulation, we recommend NLO+PS generation, with Born suppression to improve
statistics in the boosted region and the uncertainty components of \cref{sec:uncertainties}.
The amplitude cache makes large theory-weight lists practical: at \pScanMaxsix{} weights,
it reduces the measured generation time by a factor of \pScanTimeRatiosix{}, while
cached and uncached samples retain identical event kinematics and alternative weights
in the closure tests (\cref{sec:cache}).  At 13.6\TeV{}, Born suppression increases the
effective statistics above $\ptZ=400\GeV$ by a factor of \pNeffGainsix{} at fixed event
count, without changing the weighted prediction (\cref{sec:bornsupp}).

For existing samples matching one of the specified source configurations, its calibrated correction
provides an alternative to inclusive rescaling without repeating detector simulation.
Its use requires the documented event-weight variations, histogram-level ensemble treatment,
and residual uncertainty, with closure checked for the analysis selection and binning.
Agreement in a jet-inclusive projection does not validate selections that isolate the
source's depleted hard-recoil region.  Such applications should use NLO+PS events rather
than rely on large weights or compensating discrepancies in different recoil regions.

The public production package provides the generation setup, reference results, and
validation scripts~\cite{ggxy-ggZH-ampcache}.  The reweighting deliverable consists of
source- and channel-specific ONNX ensembles, calibration and theory-transfer tables, inference and
uncertainty procedures, reference events, and model cards defining the source, target,
and applicability (\cref{sec:rwt-application,app:reproduction}).  Together, these provide
an NLO+PS reference for new production and a calibrated correction with stated limitations
for reuse of existing LO+PS samples.

%% file: sec-app-13tev.tex
\section{Distributions at 13 TeV}
\label{app:dists13}

\begin{figure*}[tbp]
\centering
\includegraphics[width=\textwidth]{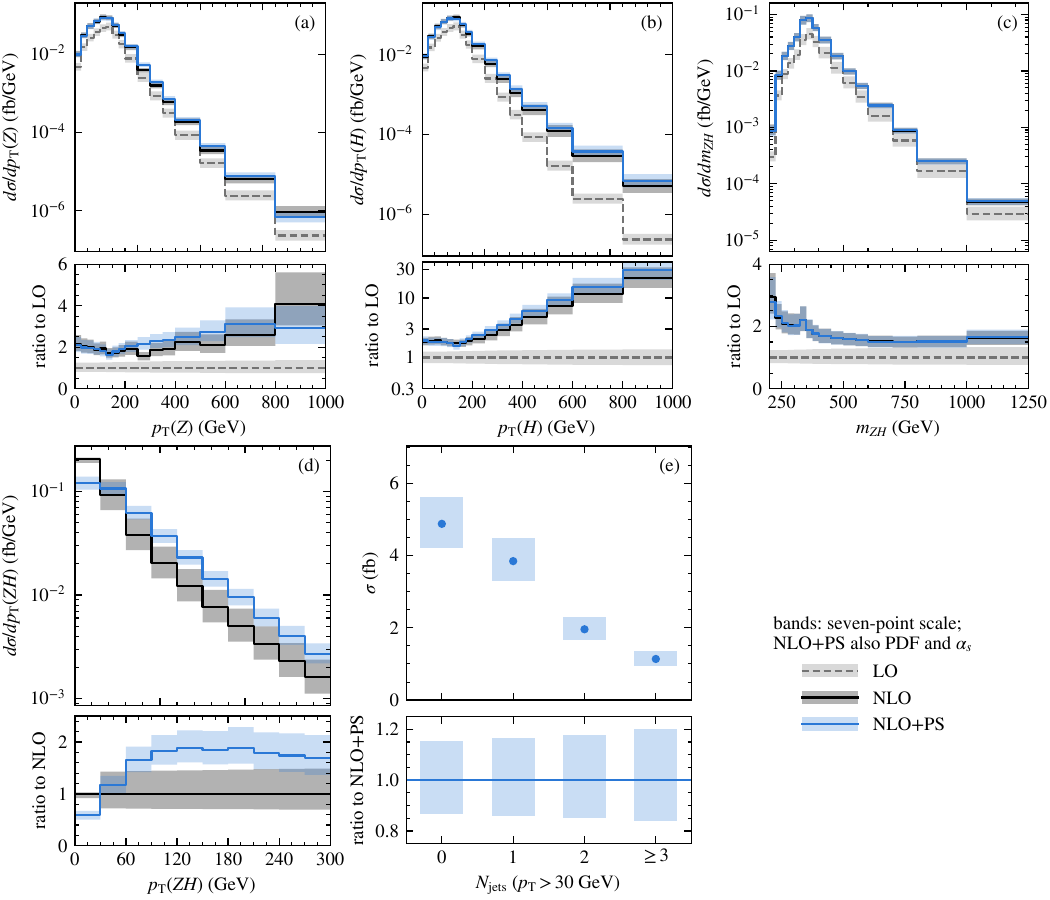}
\caption{As \cref{fig:dists}, at 13\TeV: differential distributions of $gg\to\ell^-\ell^+H$
(PDF4LHC21) in (a)~$\ptZ$, (b)~$\ptH$, (c)~$\mZH$, (d)~$p_{\mathrm{T}}(ZH)$ and (e)~the number of
jets; fixed-order LO and NLO with their seven-point scale bands, NLO+PS (Monash) with the scale
band combined in quadrature with the PDF and $\as$ uncertainties.}
\label{fig:dists13}
\end{figure*}

\Cref{fig:dists13} shows the distributions of \cref{fig:dists} at $\sqrt s=13\TeV$.
The differential fixed-order $K$ factors are statistically compatible with those at 13.6\TeV{}:
$\chi^2/N_{\mathrm{bins}}=\pKChiPtz$, \pKChiPth{}, and \pKChiMzh{} for
$\ptZ$, $\ptH$, and $\mZH$, respectively.  This comparison concerns the fixed-order corrections,
not equality of the matched distributions at the two energies.

The qualitative matching and shower trends are similar, but the precision differs between bins.
The largest central deviation of the parton-level \POWHEG{} spectrum from fixed-order NLO over
$125<\ptZ<400\GeV$ is \pPartonNLOmidHithirteen\%, in the $250\text{--}300\GeV$ bin.
The fixed-order integration error dominates the uncertainty of this ratio; that bin is
statistically compatible with its 13.6\TeV{} counterpart and does not establish an
energy-dependent matching effect.

\Cref{tab:stxs13} gives the STXS Stage~1.3 fine bins at 13\TeV{}; the matching
variations were made at 13.6\TeV{} only (\cref{tab:matching}).  The top-mass scheme ratios at
13\TeV{} are compatible with those at 13.6\TeV{} ($\chi^2/N=\pMtChiEnergies$ over the inclusive
cross section and the six $\ptV$ ranges for the three $\mu_t$).

\begin{table}[tbp]
\centering
\caption{As \cref{tab:stxs}, at 13\TeV: NLO+PS (Monash) cross sections in the STXS Stage~1.3 fine
bins in fb, with the relative MC error, the seven-point scale envelope, the PDF+$\as$ uncertainty and
the top-mass scheme band.  $^\dagger$: the MC error of this band exceeds 5\%
(\pMtNoisythirteen); the 13.6\TeV{} value of \cref{tab:stxs} is compatible with it.}
\label{tab:stxs13}
\ifdim\columnwidth<\textwidth\small\else\footnotesize\fi
\input{tab-stxs13.tex}
\end{table}

%% file: tab-stxs13.tex
\begin{tabular}{lccccc}
\toprule
$\ptV$ (GeV), jets & $\sigma$ (fb) & MC (\%) & Scale (\%) & PDF+$\as$ (\%) & $m_t$ (\%) \\
\midrule
$|y_H|>2.5$ & 0.407 & 1.3 & $^{+18}_{-15}$ & 7.9 & $^{+0}_{-13}$ \\
\midrule
0--75, 0 jets & 0.912 & 1.2 & $^{+17}_{-14}$ & 2.2 & $^{+0}_{-3}$ \\
75--150, 0 jets & 2.61 & 0.3 & $^{+16}_{-14}$ & 2.2 & $^{+0}_{-6}$ \\
150--250, 0 jets & 1.07 & 0.4 & $^{+12}_{-12}$ & 2.3 & $^{+0}_{-34}$ \\
250--400, 0 jets & 0.0882 & 1.0 & $^{+8}_{-10}$ & 2.8 & $^{+0}_{-35}$$^\dagger$ \\
400--600, 0 jets & 0.00342 & 4.4 & $^{+4}_{-8}$ & 4.0 & $^{+0}_{-50}$ \\
$>600$, 0 jets & 0.000151 & 20.0 & $^{+1}_{-13}$ & 5.0 & $^{+0}_{-39}$ \\
\midrule
0--75, 1 jet & 0.718 & 1.3 & $^{+18}_{-14}$ & 2.2 & $^{+0}_{-3}$ \\
75--150, 1 jet & 1.85 & 0.4 & $^{+17}_{-14}$ & 2.2 & $^{+0}_{-6}$ \\
150--250, 1 jet & 1.02 & 0.4 & $^{+15}_{-13}$ & 2.3 & $^{+0}_{-34}$ \\
250--400, 1 jet & 0.121 & 0.9 & $^{+13}_{-12}$ & 2.7 & $^{+0}_{-35}$$^\dagger$ \\
400--600, 1 jet & 0.00599 & 3.5 & $^{+12}_{-13}$ & 3.7 & $^{+0}_{-50}$ \\
$>600$, 1 jet & 0.000279 & 15.1 & $^{+21}_{-18}$ & 5.4 & $^{+0}_{-39}$ \\
\midrule
0--75, $\ge2$ jets & 0.55 & 1.4 & $^{+20}_{-16}$ & 2.2 & $^{+0}_{-3}$ \\
75--150, $\ge2$ jets & 1.31 & 0.6 & $^{+19}_{-15}$ & 2.2 & $^{+0}_{-6}$ \\
150--250, $\ge2$ jets & 0.936 & 0.5 & $^{+18}_{-15}$ & 2.3 & $^{+0}_{-34}$ \\
250--400, $\ge2$ jets & 0.197 & 0.9 & $^{+19}_{-15}$ & 2.7 & $^{+0}_{-35}$$^\dagger$ \\
400--600, $\ge2$ jets & 0.0163 & 2.5 & $^{+23}_{-18}$ & 3.7 & $^{+0}_{-50}$ \\
$>600$, $\ge2$ jets & 0.00128 & 8.6 & $^{+33}_{-23}$ & 5.3 & $^{+0}_{-39}$ \\
\bottomrule
\end{tabular}

%% file: sec-app-rwt-atlas.tex
\section{ATLAS-style closure}
\label{app:rwt-atlas}

\begin{figure*}[tbp]
\centering
\rwtfullfig{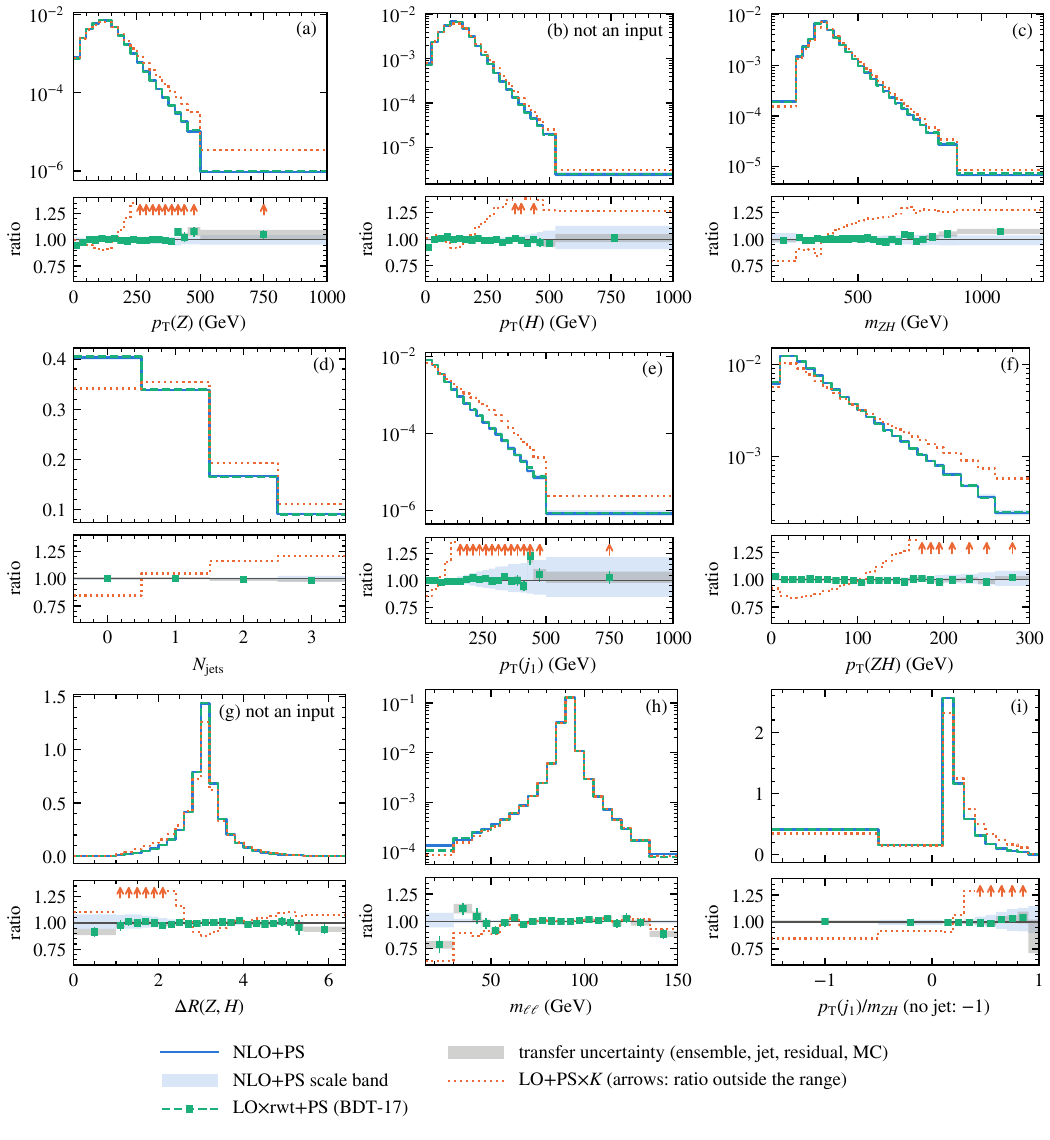}
\caption{As \cref{fig:rwt-closure}, for the ATLAS-style charged-lepton correction.}
\label{fig:rwt-atlas-closure}
\end{figure*}

\begin{figure*}[tbp]
\centering
\rwtfullfig{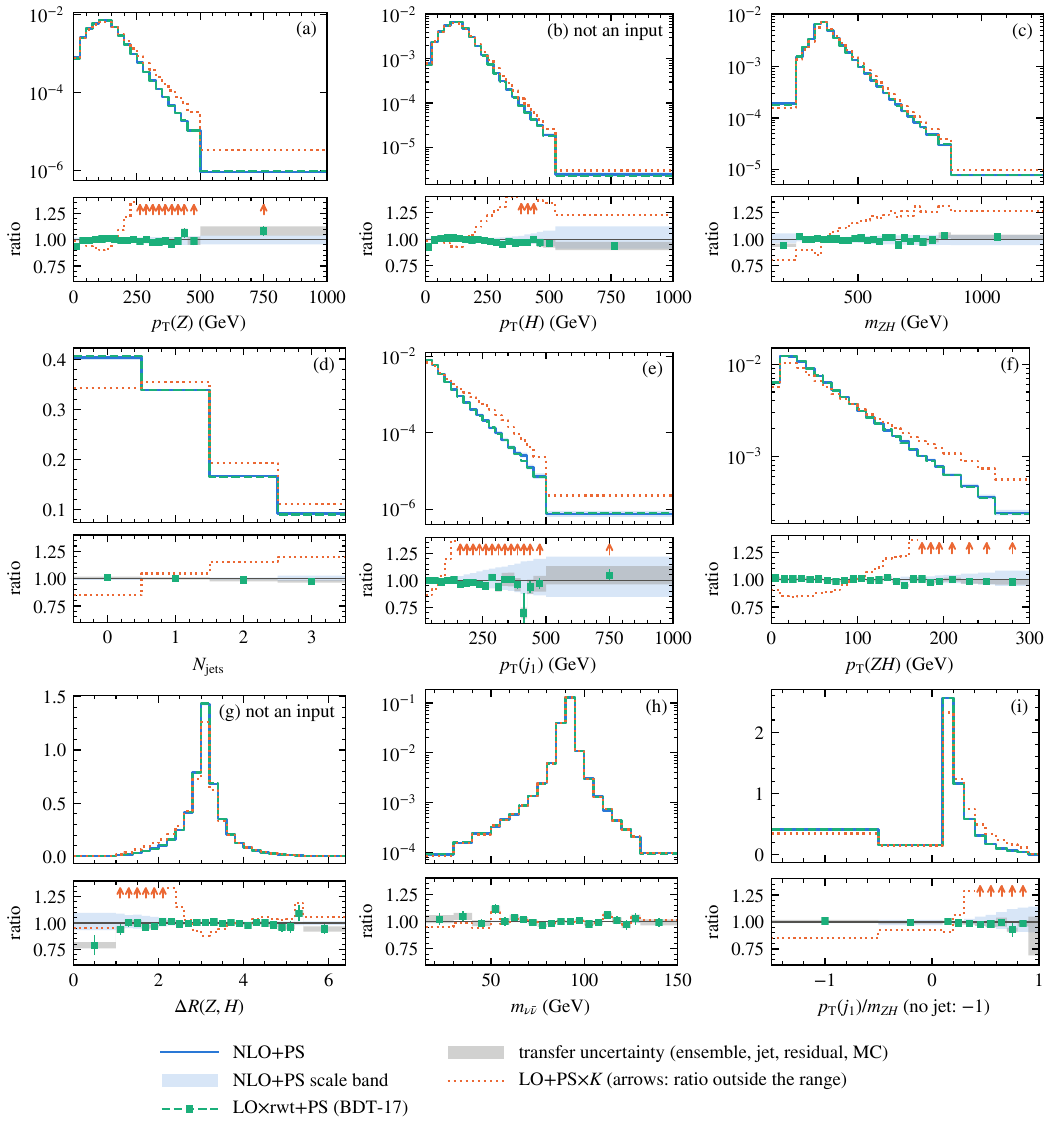}
\caption{As \cref{fig:rwt-closure}, for the ATLAS-style neutrino correction; panel (h) shows $m_{\nu\bar\nu}$.}
\label{fig:rwt-atlas-closure-vv}
\end{figure*}

\Cref{fig:rwt-atlas-closure,fig:rwt-atlas-closure-vv} show the closure of the ATLAS-style corrections on their
evaluation samples, in the layout of \cref{fig:rwt-closure}.

%% file: sec-app-reproduction.tex
\section{Reproduction}
\label{app:reproduction}
\sloppy

All results in this paper are generated with the public
package~\cite{ggxy-ggZH-ampcache}, release \code{v3.0.0} (commit \code{91a9aa8}, DOI~\href{https://doi.org/10.5281/zenodo.24316545}{10.5281/zenodo.24316545}).  The package specifies the software
versions and commits listed in \cref{tab:software} and includes workflow scripts, reference
data tables, and automated checks.  The driver script \code{production/paper.sh}
defines the run parameters, job allocations, event statistics, random seeds, and input configurations
for seven datasets: \code{13600} and \code{13000} (charged-lepton production at the two
energies), \code{vv13600} and \code{vv13000} (the corresponding neutrino channels),
\code{verify} (validation suite using NNPDF3.1 at $13.6\TeV$ and Born-suppression tests),
\code{atlas13600} (ATLAS-style charged-lepton reweighting at $13.6\TeV$),
and \code{common} (software build recipes).  The CMS-style charged-lepton source and its CP5
NLO+PS target reshower belong to \code{13600}, rather than a separate dataset.
The command \code{production/paper.sh list <set>} lists the runs, while the subcommands
\code{show}, \code{inputs}, and \code{outputs} document job configurations.  Each execution
records the originating Git commit hash.  Random seeds are deterministic functions of the job
index, and shower seeds are mapped directly from corresponding LHE seeds, making the runs
reproducible across batch systems.

\subsection{Software and fixed-order integration settings}

The exact software environment used for the production is summarized in \cref{tab:software}.
Build scripts fix the repository commits, apply the required patches, and ensure that the modified
libraries bundled with \ggxy{} take precedence over libraries in the host environment.  The
standalone \ggxy{} LO calculation evaluates scalar one-loop integrals with
\textsc{OneLOop}~\cite{vanHameren:2010cp}.

\begin{table*}[tbp]
\centering
\caption{Software.  Compilers and libraries are those of the LCG software releases
LCG\_107 and LCG\_108a; any equivalent installation can be used.}
\label{tab:software}
\small
\begin{tabular}{lp{7.4cm}l}
\toprule
Component & Version & Reference \\
\midrule
\ggxy{} & commit \code{9edd1d5} (bundles \RECOLA{} 1.4.1, \COLLIER{} 1.2.8, & \cite{Davies:2026ggxy} \\
        & CutTools, OneLOop) & \\
\PBOX{} & commit \code{6654ea3} & \cite{Alioli:2010xd} \\
\ggxyZH{} & commit \code{0fbaccf} (includes the amplitude cache, \cref{sec:cache}) & \cite{Davies:2026ggzhps,ggxy-ggZH-ampcache} \\
\quad Patch & Born suppression in $\ptH$ (\cref{sec:bornsupp}) & \cite{ggxy-ggZH-ampcache} \\
Generator toolchain & LCG\_107 (gcc/gfortran 11.3, \LHAPDF{} 6.5.5, Boost 1.87) & \\
\MCFM{} (LO check) & 10.3.1 with a patch selecting $gg\to ZH$ only; LCG\_106 (gcc 13) & \cite{Campbell:2019dru} \\
\PYTHIA{} & 8.315 & \cite{Bierlich:2022pfr} \\
\RIVET{}, YODA & 4.1.2, 2.1.2 & \cite{Bierlich:2024vqo} \\
HepMC3, FastJet & 3.3.1, 3.5.0 & \cite{Buckley:2019xhk,Cacciari:2011ma} \\
Shower toolchain & LCG\_108a (gcc 13) & \\
STXS classification & LHC Higgs WG \RIVET{} routine \code{SimplifiedTemplateCrossSections}, & \cite{LHCHWG:STXSrivet} \\
                    & commit \code{38eea57}: Stage~1.3~\cite{LHCHWG:STXS13}; framework & \\
                    & and Stages 1.1, 1.2~\cite{Berger:2019wnu,deFlorian:2016spz} & \\
\bottomrule
\end{tabular}
\end{table*}

\paragraph{Fixed-order integration.}
The fixed-order calculations use the following independent integration setups; these details are
recorded here to make the numerical results reproducible.
\begin{description}
\item[\ggxy{} Monte Carlo.] Integration programs based on \ggxy{} examples for $\ggllH$ and
  on-shell $Z$ production, configured with the parameters in \cref{tab:parameters}.  A single
  integration run evaluates the seven scale variations of \cref{sec:scale}.  Differential
  distributions in $\ptZ$, $\ptH$, $\mZH$, and auxiliary observables use $25\GeV$ binning aligned
  with STXS bin boundaries.  Sampling statistics per energy comprise $8\times2\cdot10^6$ points
  at LO (after $2\cdot10^5$ warmup points for phase-space grid optimization) and $40\times10^6$
  points at NLO ($1.5\cdot10^5$ warmup points); independent seeds are combined via unweighted
  averaging.
\item[\POWHEG{} stage~2.] Fixed-order integration in \PBOX{} evaluated with Born-only matrix
  elements (\code{bornonly 1}), with one run per scale point for each of 8 independent seeds and
  $3\times2\cdot10^5$ calls on a common importance grid ($3\times5\cdot10^4$ calls across three
  grid iterations).  Results are combined via variance-weighted averaging.
\item[\MCFM{}.] Standalone LO calculation of $gg\to ZH$ with stable $Z$ bosons using \MCFM{}'s
  analytic one-loop amplitudes (\code{gg\_zh}), providing a cross-check independent
  of \RECOLA{} and \ggxy{} expansions.  A dedicated patch adds an input switch to isolate the
  gluon-fusion contribution.
\end{description}

\subsection{Parton-shower settings}

For NLO+PS, \code{SpaceShower:pTmaxMatch = 2} and \code{TimeShower:pTmaxMatch = 2} open the
shower to the full phase space, while the \PYTHIA{} \POWHEG{} hooks veto harder emissions.  The
nominal \code{POWHEG:*} settings are \code{nFinal = 3}, \code{veto = 1},
\code{pThard = 0}, \code{pTemt = 0}, \code{pTdef = 1}, \code{emitted = 0}, and
\code{vetoCount = 100}.  Other \code{POWHEG:*} settings keep their defaults; MPI and QED
emissions are not vetoed.  The matching variations replace \code{pThard = 0} or
\code{pTdef = 1} as described in \cref{sec:matching}.

The nominal tune is Monash (\code{Tune:pp = 14}).  A14 uses \code{Tune:pp = 21}.  CP5 is
initialized from \code{Tune:pp = 14} and uses the parameters of Table~3 of
Ref.~\cite{CMS:2019csb}: MPI \code{ecmPow = 0.03344}, \code{bProfile = 2},
\code{pT0Ref = 1.41}, \code{coreRadius = 0.7634}, \code{coreFraction = 0.63},
\code{ecmRef = 7000}, \code{ColourReconnection:range = 5.176}, and
\code{SpaceShower:rapidityOrder = on}.  The CP5 shower/MPI PDF is selected with
\code{PDF:pSet = LHAPDF6:NNPDF31\_nnlo\_as\_0118}.
In the CP5 card, ISR, FSR, MPI, and \code{SigmaProcess} use $\as(m_Z)=0.118$ with two-loop
running; these \PYTHIA{} settings do not replace the hard-event weights in the LHE file.
The CP5 total and elastic $pp$ cross-section inputs are
$101.037/22.08\,\mathrm{mb}$ at 13.6\TeV{} and $100.309/21.89\,\mathrm{mb}$ at 13\TeV{}.

\subsection{Event-generation settings}

The \POWHEG{} grids are integrated with 16 independent seeds.  Per seed, the importance grid uses
$\code{ncall1}=2\cdot10^4$ and $\code{itmx1}=3$, the $\tilde B$ integration uses
$\code{ncall2}=4\cdot10^4$ and $\code{itmx2}=2$, and the radiation upper-bound step uses
$\code{nubound}=2\cdot10^4$.  The folding parameters are
$(\code{foldcsi},\code{foldy},\code{foldphi})=(1,1,1)$ and negative weights are retained
(\code{withnegweights 1}).  The grid-level cross section is the error-weighted mean of the 16
stage-2 integrations.  Events are generated seed by seed from the merged grids.

LO samples use the same process with \code{bornonly 1} and \code{LOevents 1}; eight integration
seeds are used before generating unweighted Born events.  The Born-suppression patch is inactive
when \code{bornsuppfact} is omitted.

The amplitude cache can be disabled with \code{GGXY\_AMPCACHE=0}.  Its entry limit is controlled
with \code{GGXY\_AMPCACHE\_MAX}; the default is $4\cdot10^5$ entries, corresponding to about
$100\,\mathrm{MB}$.  Closure tests compare cached and uncached jobs with identical random seeds.
The stored event kinematics and alternative weights are identical.  The only observed difference
is an internal \code{\#rwgt} comment at the $10^{-15}$ relative level from removing and restoring
the power of \as{}; this affects only a later \code{rwl\_add} re-reweighting of the LHE file.
The runs use the following steps:
\begin{enumerate}
\item Builds: \code{generator/build.sh <prefix>} (\ggxy{}, \PBOX{} with \ggxyZH{} and the
  Born-suppression patch, the \ggxy{} fixed-order programs) and \code{shower/build.sh <prefix>}
  (\PYTHIA{} driver, \RIVET{} analyses including the STXS routine).
\item Fixed order: \code{generator/run/fo.sh <lo|nlo> <out> <first seed> <n seeds> <points>
  <optimization points> <PDF> <fo-ggllh|fo-ggvvh|fo-ggzh> <sqrt(s)>}; combination with
  \code{analysis/fo\_combine.py}.
\item \POWHEG{} LO at the seven scale points: \code{run/powheg\_lo.sh}; combination with
  \code{analysis/powheg\_stat.py}.
\item Grids: \code{run/powheg\_grids.sh} with the card settings of the run (PDF, beam energy,
  \code{bornsuppfact}, or \code{bornonly}/\code{LOevents} for LO).
\item Events, one seed per job: \code{run/powheg\_events.sh} with the weight list
  \code{cards/pwg-rwl-pdf4lhc21.dat}; cross sections and variations with
  \code{analysis/lhe\_xsec.py} (per job, then combined).
\item Cache closure: \code{run/cache\_ab.sh} and \code{analysis/compare\_lhe.py}.
\item \MCFM{}: \code{mcfm/run\_mcfm.sh} and \code{analysis/mcfm\_xsec.py}.
\item Showers: \code{shower/run/shower\_batch.sh} (\code{shower/run/shower.sh} per event file);
  exact sums over files and jobs with \code{analysis/yoda\_summary.py};
  \code{analysis/compare\_orders.py} for \cref{sec:lops}.
\item Checks: \code{generator/check.py} (with \code{--paper-energy 13600|13000} against
  Ref.~\cite{Davies:2026ggzhps}) and \code{shower/check\_shower.py}.
\end{enumerate}

\Cref{tab:samples} lists the runs, with their statistics and use.
The \code{vv} sets repeat the decayed-$Z$ fixed-order, grid, event, and shower runs for
$Z\to\nu\bar\nu$ with independent seeds and the same statistics, including the 13.6\TeV{}
matching variations and reweighting chain; the $m_t$ scan is replaced by one check at
$\mu_t=\mZH$, at 13.6\TeV{} only.  Stable-$Z$ and code-validation runs are shared.
The CMS-style charged-lepton source uses NNPDF3.1, $\muR=\muF=m_Z+m_H$, and CP5
(\cref{sec:rwt-samples}).  Its \code{grids-lo-cms}, \code{events-lo-cms}, and
\code{table-lo-cms} runs are recorded under \code{13600}.  The \code{table-nlo-cp5} target
reuses the production NLO events with the NNPDF3.1 central weight and a CP5 reshower; it is not
an additional independently generated NLO sample.
The \code{atlas13600} set generates \atlasNevLO{} LO events with the $H_{\mathrm{T}}$ scale, checks
the scale weights against independent LO integrations, and reshowers \atlasNevNLO{} production NLO
events with A14, for earlier five- and six-input classifiers.  The \code{tables-atlas-zll-v1} and
\code{tables-atlas-zvv-v1} sets produce the ATLAS-style samples of \cref{sec:rwt-atlas}, and
\code{ensemble-atlas-zll-v2} and \code{ensemble-atlas-zvv-v2} their classifiers, tests, and models.

\begin{table*}[tbp]
\centering
\caption{Production and validation samples.  The first block is generated at both 13.6 and
13\TeV{} with PDF4LHC21.  The second block is specific to 13.6\TeV{}, with PDF choices shown
where they differ.  The CMS-style target reuses the nominal NLO events rather than adding a
new independent sample.}
\label{tab:samples}
\small
\begin{tabular}{lll}
\toprule
Run & Statistics & Use \\
\midrule
\ggxy{} $\ell\ell H$ LO / NLO & $8\times2\cdot10^6$ / $40\times10^6$ points & fixed order, 7 points, $K$ \\
\ggxy{} stable $Z$ LO / NLO & $2\times2\cdot10^6$ / $40\times10^6$ points & comparison with \cite{Davies:2026ggzhps}, Table~1 \\
\POWHEG{} LO (stage 2) & 7 points $\times$ 8 seeds & code comparison \\
NLO grids & 16 seeds, $L=\bornsuppL\GeV$ & NLO events \\
LO grids & 8 seeds & LO events \\
NLO events & \nEvNLO{} weighted, \nWeights{} weights & NLO+PS \\
LO events & \nEvLO{} unweighted, \nWeights{} weights & LO+PS \\
Cache closure & $3\times50$ events, cache on and off & \cref{sec:cache} \\
\MCFM{} LO, stable $Z$ & 7 points & independent LO check \\
\ggxy{} $\ell\ell H$ LO / NLO, $\overline{\mathrm{MS}}$ & as on-shell, $\mu_t=\mZH/\{1,2,4\}$ & $m_t$ scheme (\cref{sec:mtscheme}) \\
\midrule
\multicolumn{3}{l}{At 13.6\TeV{} only:} \\
NLO, NNPDF3.1 & 16 seeds; $5\cdot10^4$ events, & Born suppression \\
 & \quad unweighted and $L=\bornsuppL\GeV$ & (\cref{sec:validation}) \\
NLO, $\hdamp=\infty$ & 16 seeds; $5\cdot10^5$ events & matching (\cref{sec:matching}) \\
\ggxy{} stable $Z$ LO / NLO, $\overline{\mathrm{MS}}$ & as on-shell, $\mu_t=\mZH/\{1,2,4\}$ & comparison with \cite{CampilloAveleira:2025ggzh} \\
CMS-style LO+PS, NNPDF3.1, CP5 & \rwtNevLO{} events & source (\cref{sec:rwt-samples}) \\
NLO+PS, NNPDF3.1 weight, CP5 & \rwtNevNLO{} reused events & target (\cref{sec:rwt-samples}) \\
\bottomrule
\end{tabular}
\end{table*}

The regression suite compares fixed-order integrations and generated event weights with stored
reference results.  The corresponding numerical comparisons and their statistical precision
are reported in \cref{sec:validation}; they do not identify the compiler as the cause of any
run-to-run difference.

\paragraph{Computing cost.}  The base charged-lepton production runs took about \pCpuSixBase{} CPU hours at
13.6\TeV{} and \pCpuThirteen{} at 13\TeV{} (user and system time of the successful jobs), with \pCpuShowerShareSix\,\%
spent on the parton-shower runs (three tunes, five LO+PS starts).  The \nEvNLO{} NLO events with
\nWeights{} weights took \pCpuEvNLOSix{} CPU hours with the amplitude cache (\cref{sec:cache}).  The
variations made at 13.6\TeV{} only (PDF members, cache scan, $\hdamp=\infty$, \POWHEG{}-veto
variants, damped LO+PS start, stable-$Z$ $\overline{\mathrm{MS}}$ runs) added \pCpuSixExtra{} CPU hours, and the NNPDF3.1 validation set
\pCpuVerify{}.  These accounting categories exclude the experiment-inspired reweighting sample
preparation added to \cref{tab:samples}.

\paragraph{Data.}  The data record~\cite{ggzh-data-record} (DOI~\href{https://doi.org/10.5281/zenodo.24315535}{10.5281/zenodo.24315535}) holds the LHE files of the generated
samples of \cref{tab:samples} with their event weights (\nWeights{} per nominal production event:
scale, PDF4LHC21 and NNPDF3.1 members), the \POWHEG{} grids, with which more events of any sample can be generated without a
new integration, the per-event tables, weights, and reusable models of the reweighting study
(\cref{sec:rwt}), including the neutrino samples and their own \code{ggzh-bdt17-zvv-r5.tgz} (charged leptons:
\code{ggzh-bdt17-zll-r5.tgz}), the ATLAS-style samples with their own
\code{ggzh-bdt17-atlas-zll-r5.tgz} and \code{ggzh-bdt17-atlas-zvv-r5.tgz},
and the analysis outputs: the \RIVET{} histograms of every shower and parton-level sample
with all weight variations, and every result file behind the numbers of this paper.  Each file has a
checksum; \code{production/data\_release.sh} of the package assembles the record from the products of
\code{production/paper.sh}.  The model cards of the four models are also included directly in the
source repository under \code{results/13600/rwt-bdt17/}.

%% file: sec-app-conventions.tex
\section{Conventions}
\label{app:conventions}

\paragraph{Weight identifiers} (\code{cards/pwg-rwl-pdf4lhc21.dat}, \nWeights{} weights):
\begin{center}
\small
\begin{tabular}{@{}lp{0.64\columnwidth}@{}}
\toprule
\code{1001}--\code{1009} & $(\muR,\muF)/\muz$ = $(1,1)$, $(1,2)$, $(1,\tfrac12)$, $(2,1)$, $(2,2)$, $(2,\tfrac12)$, $(\tfrac12,1)$, $(\tfrac12,2)$, $(\tfrac12,\tfrac12)$; PDF4LHC21 \\
\code{3000}--\code{3042} & PDF4LHC21\_40\_pdfas members 0--42 (41, 42: $\as=0.117$, $0.119$) \\
\code{2000}--\code{2102} & NNPDF3.1 members 0--102 (101, 102: $\as=0.116$, $0.120$) \\
\bottomrule
\end{tabular}
\end{center}

\paragraph{Event-weight normalization.}
For a sample of $N$ generated events with signed weights $w_i$, the cross section is evaluated as
\begin{equation}
  \sigma=\frac{1}{N}\sum_{i=1}^{N}w_i,\qquad
  \delta\sigma_{\mathrm{MC}}=\frac{1}{N}\left(\sum_{i=1}^{N}w_i^2\right)^{1/2}.
\end{equation}
For Born-suppressed samples the event-weight magnitude contains the compensating factor $1/f$.
Event weights are stored to five significant figures.
\paragraph{Shower weights.}  \PYTHIA{} names: \code{isrRedHi}, \code{fsrRedHi},
\code{isrRedLo}, \code{fsrRedLo} ($\mu_R$ factor $1/\sqrt2$, $\sqrt2$), \code{isrDefHi},
\code{fsrDefHi}, \code{isrDefLo}, \code{fsrDefLo} ($\tfrac12$, 2); in \RIVET{} the LHE weights
appear as \code{MUR1.0\_MUF2.0}, \ldots{} and the PDF weights by their identifier.

\paragraph{Fixed-order normalization.}  Standalone \ggxy{} outputs differential distributions per
lepton flavor in nanobarns; yields are scaled by 3 to represent $\ell=e,\mu,\tau$.  Seeds with
identical sample sizes are combined via unweighted averaging, with statistical uncertainties
evaluated from the standard error of the mean.  Fixed-order \POWHEG{} cross sections represent the
variance-weighted mean of stage-2 integration results across all seeds.

\paragraph{STXS event-record conventions.}
The Higgs boson is identified as its last copy before decay.  The hard-process $Z$ boson is
followed to its last copy before decay and defines $\ptV$.  Jets are built from stable final-state
particles after removing descendants of the Higgs boson and of the decaying $Z$ boson, including
products of $\tau$ decays.  No jet rapidity requirement is applied.  The $30\GeV$ jet threshold
is used throughout; the alternative $25\GeV$ threshold available in the STXS routine is not used.
The routine treats $Z\to q\bar q$ separately, while charged-lepton and neutrino decays enter the
gluon-fusion bins used here.